# DNA Fragments in Crude Oil Reveals Earth's Hidden History

Wan-Qian Zhao[1,3*], Zhan-Yong Guo[2], Yu-Qi Guo[3*], Mei-Jun Li[4*], Gang-Qiang Cao[1], Zeng-Yuan Tian[1], Ran Chai[5], Li-You Qiu[6*], Jin-Hua Zeng[7], Xin-Ge Zhang[8], Tian-Cang Qin[9], Jin-Yu Yang[10], Ming-Jie Chen[11], Mei-Rong Song[12], Fei Liang[13], Jun-Hui Geng[1], Chun-Yan Zhou[14,15], Shu-Jie Zhang[3], Li-Juan Zhao[3]

**Affiliations:**
[01]School of Agricultural Sciences, Zhengzhou University, Zhengzhou, China
[02]College of Agronomy, Henan Agricultural University, Zhengzhou, China
[03]School of Life Sciences, Zhengzhou University, Zhengzhou, China
[04]National Key Laboratory of Petroleum Resources and Engineering, College of Geosciences, China University of Petroleum (Beijing), Beijing, China
[05]Yellow River Conservancy Technical Institute, Zhengzhou, China
[06]College of Life Sciences, Henan Agricultural University, Zhengzhou, China
[07]School of Pharmaceutical Sciences, Zhengzhou University, Zhengzhou, China
[08]College of Tobacco, Henan Agricultural University, Zhengzhou, China
[09]Zhengzhou ToYou Bioengineering Co., Ltd, Zhengzhou, China
[10]Haining BoShang Biotechnology Co., Ltd, Haining, China
[11]NewCore Biotechnology Co., Ltd, Shanghai, China
[12]College of Sciences, Henan Agricultural University, Zhengzhou, China
[13]College of Paleontology, Shenyang Normal University, Shenyang, China
[14]Jiesai Biotechnology Co., Ltd, Haining, China
[15]Jiangsu Jiayu Biomedical Technology Co., Ltd, Jiangyin, China

***Corresponding authors**
Emails: wqzhao@zzu.edu.cn, guoyuqi@zzu.edu.cn, meijunli@cup.edu.cn, qliyou@henau.edu.cn

**Abstract:** This groundbreaking research extracted DNA from petroleum using nanoparticle affinity bead technology, yielding 3,159,020 petroleum DNA (pDNA) sequences, primarily environmental DNA. While most original *in situ* DNA (oriDNA) was lost, ancient DNA (aDNA) from petroleum offers an important source of ecological and evolutionary information, surpassing traditional fossils. This study reveals that oil, mainly sourced from algae and lower aquatic plants, now serves as a new type of fossil, providing detailed insights into Earth's hidden history, including unclassified species and ancient events, revolutionizing petroleum geology and paleontology.

Keywords： nanoparticle affinity bead technology, petroleum DNA, ancient DNA, original *in situ* DNA

## Introduction

Petroleum is a fossil fuel derived from fossilized organic material, such as plankton and bacteria. The remains of these organisms were deposited with sediments to oceans or lakes and were converted to petroleum under thermal stress[1,2]. GMs, also known as molecular fossils or biomarkers, are organic molecules found in geological samples that provide information about the biological and environmental conditions of the past. They are typically preserved in sediments, rocks, crude oils, and coals. They show little or no change in structure from their parent organic molecular in living organisms. Therefore, they can provide reliable geological interpretations to help solve exploration, development, production, environmental, and archeological or paleontological problems[3]. Previous studies show that the source rocks of the Biyang Sag oil field in the Nanxiang Basin are mainly located in the third member (Eh3) of the Paleogene Hetaoyuan Formation[4,5,6,7,8]. Research suggests that the organic matter within the Eh3 shales contains bio-precursors mainly from phytoplankton, bacteria, and benthic algae, with relatively small contributions from animal and higher plant materials[6]. Furthermore, the presence of marine organisms 24-*n*-propylcholestanes indicates the occurrence of marine incursions in the Nanxiang Basin[6,7,8].

To date, paleontological research can successfully obtain original (*in situ*) aDNA, the oriDNA, from certain fossils dating back to 1 million years ago (Ma)[9]. However, fossil distributions are relatively scarce, geographically dispersed, and temporally discontinuous. Obtaining more sources of aDNA could lead to a great breakthrough in the discipline. Many experiments have shown that vegetable oil refined at 400°C contains amplifiable DNA fragments[10,11,12], suggesting that a certain amount of DNA can be stored and protected in the deoxygenated environment of oil-phase liquids. Organic matter begins to mature and hydrocarbons form when the hydrocarbon source rock reaches a certain temperature (threshold temperature) in the formation conditions of petroleum. The main stages of oil production generally begin at temperatures no lower than 60°C and end at temperatures no higher than 180°C[1]. This raises the question of whether amplifiable DNA can be extracted from petroleum. If the reply is yes, it would be a major advance in the petroleum and paleontological belt, especially in revealing the genetic information stored within petroleum during its formation, migration, accumulation, and preservation. To solve this problem, we successfully extracted DNA from crude oil samples from the Nanyang Oilfield, using a new technology called "nanoparticle affinity bead DNA extraction technology"[13], constructed DNA libraries, and sequenced them. A pioneering resource, the pDNA dataset, was created by screening and analyzing pDNA sequences, which revealed many previously undiscovered phenomena. Because GMs and DNA have different physicochemical properties, the two types of molecules are accessed, degraded, lost, and stored in reservoirs in completely different ways.

In this study, 3,159,020 DNA fragments were extracted from crude oil samples from the Nanyang Oilfield in central China, categorized, and analyzed using the "Mega screening method". The results showed that most of the preDNA, mainly bacterial fragments, and species involved in human activities. A small amount of aDNA, including paeDNA, has been discovered and is used to validate important local historical events, such as the local marine invasion, activities of *Homo erectus*, and the presence of ancient local bird species. Surprisingly, the oriDNA is nearly absent, making it challenging to determine the species origin of the petroleum, based on the pDNA data. However, this discovery opens up a new resource for collecting aDNA fragments, which work as DNA fossils. It provides an exciting opportunity to trace the evolutionary path of ancient local life in the distant past within the context of local paleontology and geological evolution. This study emphasizes the need to create a global pDNA database, serving as a new platform for research in interdisciplinary fields, such as geology, industrial genomics, and paleontology.

## Results and Discussion

### Petroleum Geology of the Nanyang area

In the early-middle Triassic, the area where the Nanyang field is located was a trough in the Qinling region of southern China. It formed into a complete land mass after the collision of the Sino-Indian plate in the middle-late Triassic[14]. These oil wells are situated in the Biyang Depression between the North China Plate and the Yangtze Plate. Analysis of the Eh3 shales through geochemical studies has revealed an average organic matter abundance of 2.96 wt%, indicating low to mature stages of maturity, with prevalent type I to type II kerogen dominance. The $Eh_3$ stratum in the Biyang Sag has been proven to be the most favorable hydrocarbon source rocks for the discovered oil in the Nanyang Oilfield[15]. The reservoirs are buried at depths of 90 to 1000 m, with oil layer thicknesses ranging from 1 to 4 m and assemblage thicknesses ranging from 2 to 15 m. The petroleum reservoirs in the Nanyang Oilfield experienced a complex filling history, including an early oil filling, significant uplift, a second phase of filling, and late-stage biodegradation. Therefore, the oils with a burial depth of less than 1100 m are characterized by higher density, higher viscosity, and depletion of light hydrocarbon fraction[16,17,18]. The significant uplift of stratum and erosion made the reservoired oil susceptible to be incorporative to the earth's surface organisms. Fossil evidence suggests that the region was inhabited by *Homo erectus* and a variety of mammals during the Middle Pleistocene about 0.5 Mya ago[19]. It is generally believed that the skeletal morphology of *Homo erectus* is characteristic of both modern humans and apes. By now, the region has a humid and semi-humid continental climate, characterized by abundant rainfall, dense forests, and numerous rivers. It has been densely populated since the arrival of modern agricultural civilization.

The oil samples underwent geochemical analysis to assess their maturity, secondary alteration, sedimentary environment, and organic matter input of their relative source rocks. The results show that these oils and related source rocks have a low thermal maturity. The organic matter source of source rocks for the crude oil is dominant algae and the contributions of bacteria and terrigenous higher plants are minor. The source rocks of these studied crude oil samples were deposited under a high-salinity and strong-reduced environment with a stratified water column.

### The GMs work as indicators of life

In this study, the following chemical molecules, including *n*-alkanes, acyclic isoprenoids (pristane and phytane), β-carotene, tricyclic terpanes, hopanes, gammaceranes, and steranes have been detected as the petroleum biomarkers (Fig S1). Generally, these GMs are difficult to degrade in the environment and are soluble in oil but insoluble in water. It can be assumed that from the source rocks in the Paleogene ages to the subsequent migration, charging, and accumulation of oil reservoirs, it is difficult for the GMs to return to the environmental water. This is because the adsorption capacity of the solid phase of the soil and rock is very limited to these molecules, so most of them are still preserved in the oil. Moreover, when crude oil encounters organic residues in the environment, the GMs can be absorbed into the reservoir.

GMs originating from different species show an overlap distribution pattern; however, GMs originating from specific species indicate specific depositional environments. For example, β-carotene indicates cyanobacteria or algae origin, and 3β-methylhopanes represent the biological source of methanotrophic bacteria. In comparison, gammacerane in oil and sedimentary organic matter was derived from tetrahymanol in ciliates feeding on bacteria. The lower pristane/phytane ratios and higher relative abundance of gammacerane are generally related to anoxic, saline, and stratified water bodies[3]. The lower pristane-to-phytane ratio, higher cammacerane index, and the distribution of other GMs for the oils from the Nanyang Oilfield in this study indicate dominant input of aquatic organisms with less contribution of higher plants (Fig. S1) in the saline lacustrine basin. Our results are in general agreement with the findings of previous authors[6,7].

The distribution of tricyclic terpanes (TT) with the high frequency of the $C_{23}$TT indicates a marine or saline lacustrine depositional environment[20]. The distribution of other GMs confirmed that the oils are mainly derived from lower aquatic organisms with less contribution from higher plants (Fig. S1). Our results are in general agreement with the findings of previous authors[6,7]. To further illustrate these findings, relevant academic papers and paleogeographic maps were referenced in Fig. S2. It is crucial to understand that most GMs in oils suggest their domain or Kingdom but may not differentiate their specific species. For instance, the hopanes indicate the source of prokaryotes, while steranes are the origin of eukaryotes. Hence, it is important to gather solid evidence, such as original (*in situ*) DNA from marine organisms in the source rocks, while ruling out the possibility of paeDNA and preDNA contamination. This will confirm whether local marine intrusions were involved in source rock formation.

The prevailing belief in the academic community is that most petroleum molecules originate from lower plants and plankton, a belief consistent with our findings[1,2,3,21]. Since the early Paleogene before the formation of Hetaoyuan Eh3 shales, angiosperms have been the dominant plant group. Approximately three thousand years ago, the surface of the oil field was transformed into farmland. Today, the once expansive local lake area has nearly vanished, and the prevailing vegetation primarily consists of terrestrial higher plants. Our results suggest that most of the GMs detected belong to the original (*in situ*) GMs of the source rock (oriGMs). Given the lower contribution of higher plants over time, a question arises: Are these higher plant GMs from the source rocks or later-age species, particularly recent/present environmental ones? We suggest that there are three potential sources for these GMs: original (*in situ*) organisms in the source rocks, ancient/post-depositional environmental species (paeGMs), and recent/present environmental organisms (preGMs), such as terrestrial higher plants in the current environment, all of which could potentially enter the oil reservoir. However, the GMs derived from higher plants do not dominate the total GMs, so they enter the reservoir very slowly. Consequently, the GMs exhibit a general trend of "more original molecules and fewer later ones". As a result, we have developed hypothetical patterns representing changes in petroleum genetic markers (Fig. 1)

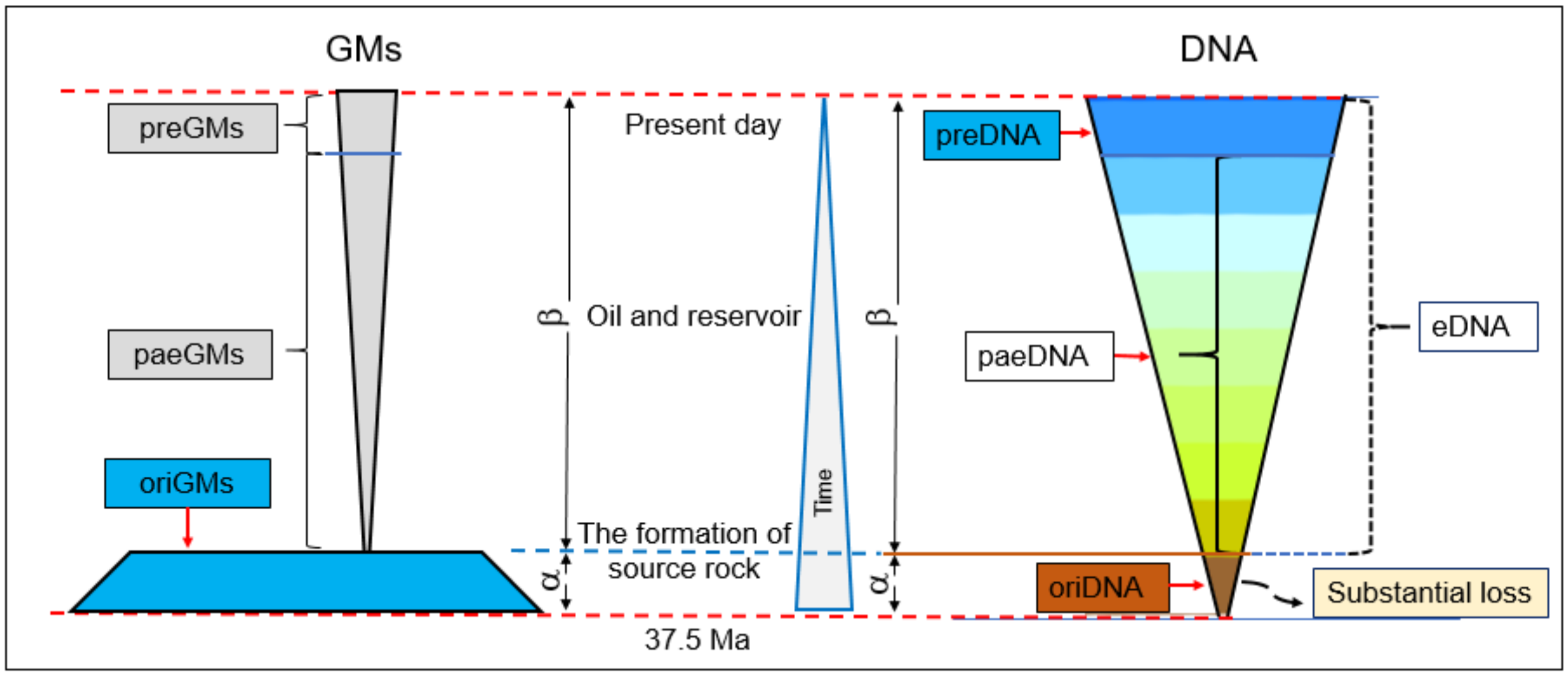

Fig. 1 The preservation patterns of two types of biological indicators

**Extraction of DNA information by the "Mega screening method"**

The 3,159,020 pDNA sequences were categorized and further analyzed using the "Mega screening method" to extract biological information (see Materials and Methods). They originated from different species and sources and can be broadly divided into oriDNA, paeDNA, and preDNA.

**(1) Establishing lineage subsets of pDNA sequences**

Non-discriminatory alignment without predefined target genomes: Using the "minimum E-value model," over 3.15 million sequences (TS) were aligned with the NCBI database (see Materials and Methods). They were then divided into subsets by lineage, with 51.29% (1,620,178 fragments) assigned to the bacterial genome and 33.16% (1,047,532 fragments) to the human genome. 3.85% (121,735 fragments) were assigned to the fungal genome and 0.29% (9,252 fragments) to the avian genome. Non-primate mammals and non-human primates accounted for 0.15% (4785 entries) and 1.11% (35108 entries), respectively. Algae entries accounted for just 5, with a negligible proportion of the total, and no red or brown algae were found (Fig. 2). These results are not entirely consistent with the expected findings of previous authors on this point[6,7]. The GM markers in Fig. S1 suggest that Nanyang crude oil is primarily formed by the decomposition of lower plants and plankton, aligning with the prevailing view of the current academic community[1,2,3,21]. If we consider the DNA of lower plants and plankton as the main source of petroleum oriDNA and the DNA of other species as the main source of eDNA, then the pDNA data above does not highlight this view.

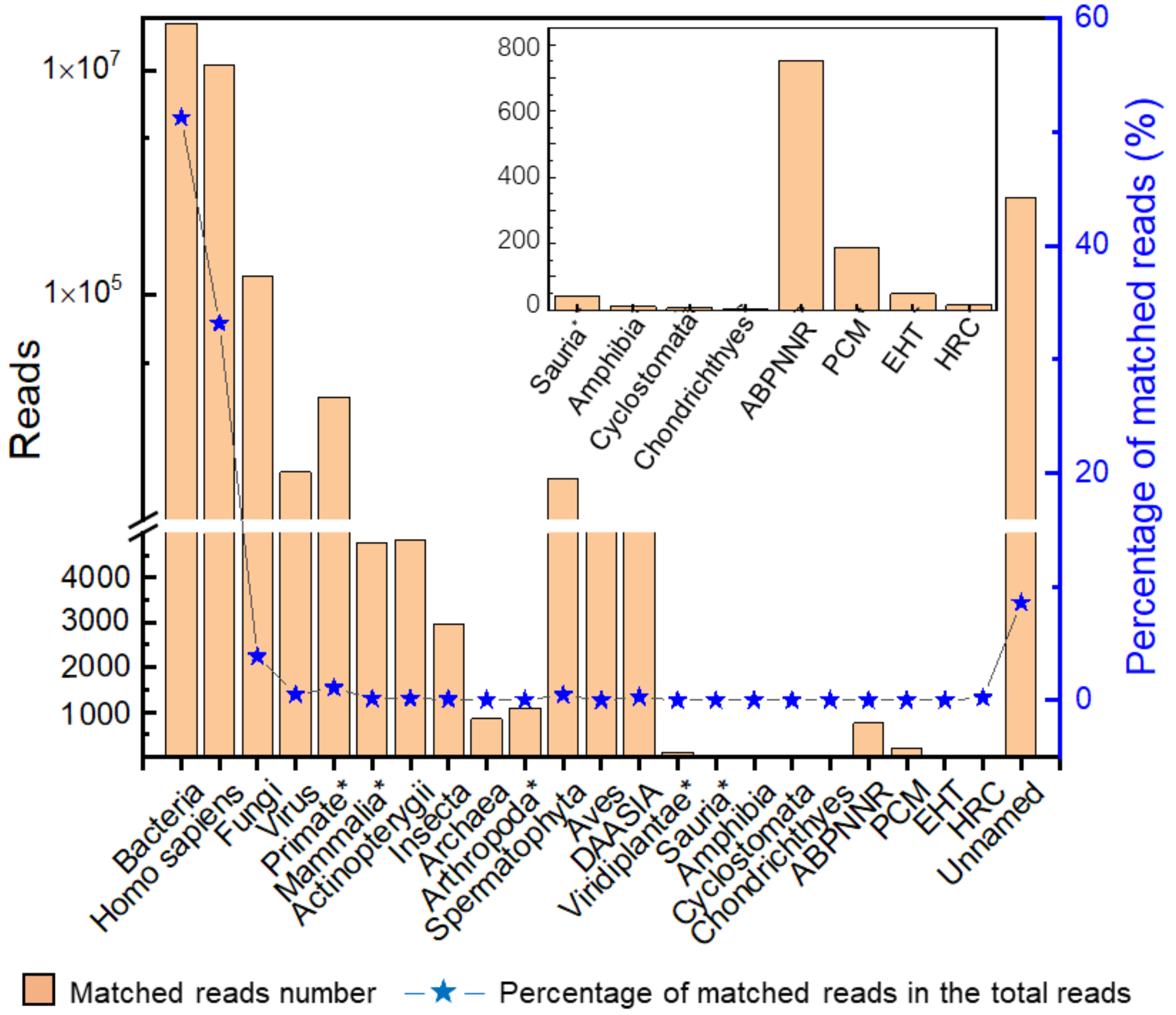

**Fig. 2 displays an analysis of pDNA sequences that correspond with different lineages.** The bars represent the number of pDNA sequences that match a lineage, while the lines represent the percentage of pDNA fragments in that lineage compared to the total number of pDNA fragments. The abbreviations used are as follows: ABPNNR (Rotifera, Nemertea, Nematoda, Platyhelminthes, Annelida, and Bryozoa), PCM (Porifera, Cnidaria, and Mollusca), EHT (Hemichordata and Tunicata), HRC (Haptophyta, Rhodophyta, Cryptophyceae), and DAASIA (Discoba, Apusozoa, Amoebozoa, Sar, Ichthyosporea, and Apicomplexa). It's important to note that Viridiplantae* excludes Spermatophyta, Arthropoda* excludes Insecta, Sauria* excludes Aves, Primates* excludes *Homo*, and Mammalia* excludes Primates*, as detailed in Table S1.

**(2) Identify the unique lineage source of each sequence**

We employed the "MS mode" (see Materials and Methods) to determine the specific genome(s) that each pDNA sequence corresponded to and identify its distinct lineage origin. Due to the lack of genome data for certain lower species, some pDNA sequences could not be genealogically determined. Through this approach, 402 pDNA sequences have been successfully identified and categorized into nine groups based on their morphological characteristics and habitat (Table 1). These sequences originated from lower plants (Group G), marine organisms (Group M), primates (Groups C, P, and T), and birds (Group L). Although these sequences cannot trace major oil-forming species, they can provide valuable data reflecting local geology, history, and evolution. They represent an unexplored reservoir for investigating local geological and species evolution.

**Table 1 Summary of the pDNA sequences matched genomes**

| Group | Target | ID | Lineages | pDNA Count | Affinity | |
|---|---|---|---|---|---|---|
| | | | | | Average | SD |
| G | Plastid + Nuclear DNA | 1G-8G | *Chlorophyta, Streptophyta* | 8 | 90.23% | 16.05% |
| H | Primate mtDNA | 1H-40H | *Home sapiens* | 40 | 96.12% | 15.44% |
| N | Non-primate mtDNA | 1N | *Gallus gallus, Galliformes* | 1 | 99.32% | - |
| | | 2N | *Sus scrofa, Artiodactyla* | 1 | 99.48% | - |
| | | 3N-5N | *Mylopharyngodon piceus, Cypriniformes* | 3 | 100.00% | - |
| T | Primate nuclear gene | 1T-275T | *Pan troglodytes* | 275 | 96.91% | 4.47% |
| P | | 21P-28P | *Pan paniscus* | 8 | 98.23% | 2.15% |
| | | 29P-36P | *Gorilla gorilla* | 8 | 98.42% | 2.03% |
| | | 6P-20P | *Cercopithecidae* (Old World monkeys) | 15 | 94.31% | 2.42% |
| | | 2P-5P | *Platyrrhini* (New World monkeys) | 4 | 94.31% | 5.80% |
| | | 1P | *Carlito syrichta* (*Tarsiiformes*) | 1 | 70.67% | - |
| C | | 1C-9C | Primate protein-coding sequences | 9 | 92.11% | 4.04% |
| A | Nuclear gene of non-indigenous organism | 1A-7A | *Bos mutus, Artiodactyla* | 7 | 93.67% | 5.51% |
| | | 8A-15A | *Meleagris gallopavo, Galliformes* | 8 | 92.06% | 8.02% |
| M | Nuclear gene of marine organisms | 1M | *Scyliorhinus canicula, Chondrichthyes* | 1 | 50% | - |
| | | 2M-5M | *Mercenaria mercenaria, Mollusca* | 4 | 70.89% | 17.23% |
| | | 6M-10M | Marine fishes, *Actinopterygii* | 5 | 57.79% | 18.61% |
| L | Others | 1L-4L | *Matches with very low similarity* | 4 | < 50% | - |

Note: The sequences between 7C and 28P, and between 8C and 4P, are identical.

**(3) Explore local geological changes and species evolution**

The analysis revealed three pDNA matches to Chlorophyta, and five pDNA matches to the Streptophyta genome, with two matching the Klebsormidium genome and three matching the Bryophytes genome from the pDNA data (Table S1). Notably, sequences ID_1G, ID_2G, and ID_3G appear distinct from known genomes and may belong to aDNA. With numerous unsequenced species of algae, it is currently not feasible to determine their specific species affiliation. Furthermore, the rapid evolution of genomes in lower organisms makes it difficult to determine if these sequences originated from ancient times. As a result, it remains uncertain whether they belong to oriDNA, paeDNA, or preDNA derived from post-source rock formation. Nevertheless, these findings strongly suggest that DNA from the original lower species that formed the hydrocarbon source rock has been largely lost. Scholars found morphological evidence of marine invasion in the core samples from the Hetaoyuan Formation (37.5 Ma) in the Biyang Depression, which contain remains of red algae (Rhodophyta) and brown algae (Ochrophyta)[6,7]. Despite this, our data did not yield any DNA from these algae. One possible explanation is that the organisms are too small, which makes them susceptible to rapid dissolution and rupture upon entering the reservoir, releasing DNA. In addition, it is known that pDNA in the reservoir is always lost over time.

The analysis identified DNA sequences containing genetic signatures of marine organisms, including Chondrichthyes, Actinopterygii, Tunicata, and Mollusca, as shown in Table S2, indicating the occurrence of local marine transgressions. Four sequences correspond to ray-finned fish species, with two sequences attributed to Scombriformes and two to Eupercaria. These fish species emerged during the mid-to-late Cretaceous period[22], suggesting that at least one marine invasion occurred after this era. However, pDNA evidence can neither pinpoint the timing of these transgressions, such as whether they occurred during the Paleocene Epoch (approximately 37.5 Ma) or some other time, nor the location and stratigraphy of the hosts of the DNA fragments. Consequently, we cannot be sure whether they are oriDNA or paeDNA from other sources. To reach a conclusive determination, it is imperative to collect stratigraphic DNA data. The amount and source of DNA data obtained are directly related to the sufficiency of the evidence available for analysis. Our current findings only partially support the suggestion made by Xia[6].

Furthermore, sequences associated with non-native species, such as yaks in the northern Himalayan foothills, turkeys in the Americas, and kiwi birds in the South Pacific Islands, were discovered (Table S3). It is hypothesized that these species are descendants of ancient Pangaea ancestors who migrated and reproduced in their local habitats due to continental drift. Their eventual disappearance left molecular evidence underground, offering new insights into the theory of continental drift.

Generally, aDNA with lower-affinity values (see Materials and Methods) can serve as DNA fossils, providing valuable information about species evolution. For example, the ID_1L sequence shows some degree of similarity to several bird genome sequences, with an affinity value of 61.91 to the mRNA transcript sequence of *Calidris pugnax*. This suggests that the sequence may have originated from its ancient avian ancestor, indicating its ancient identity (Table S4). It comprises the protein-coding sequence of exons 3 and 4, along with a portion of exon 5, without any intronic insertions between the exons, hinting that it may function as a pseudogene. Moreover, the sequence can be divided into two distinct parts: the initial 25 base pairs upstream closely resemble the intergenic sequence of the non-native reptile *Podarcis lilfordi* and the remaining part bears similarity to various bird species (Fig. 3). The "composite" nature of this sequence suggests that the avian and reptilian genes originate from the LCA genome. Our analysis indicates that this DNA fragment does not align with any known bird DNA sequence, suggesting that the host organism may have become extinct or that its descendants have undergone significant genomic divergence due to long-term evolution.

The Late Mesozoic strata beneath the Hetaoyuan Formation are believed to have existed during the period of dinosaur breeding and the emergence of primitive birds, likely during the Cretaceous period or later. Although DNA fossils can reveal the time of origin of DNA sequences, they cannot provide evidence of the historical existence of the host organism. Therefore, it is challenging to determine whether this sequence belongs to the oriDNA or eDNA from a certain period. To address this question, additional information is needed, such as the host's life cycle, the location of the fossilized body, and the specific stratum in which it was found. Importantly, the pDNA sequence exhibits both dinosaurian and avian characteristics, which could provide the first molecular evidence of transitional species within Darwin's proposed evolutionary spectrum. This finding also supports the theory that the Nanyang Basin served as an ancestral habitat for birds.

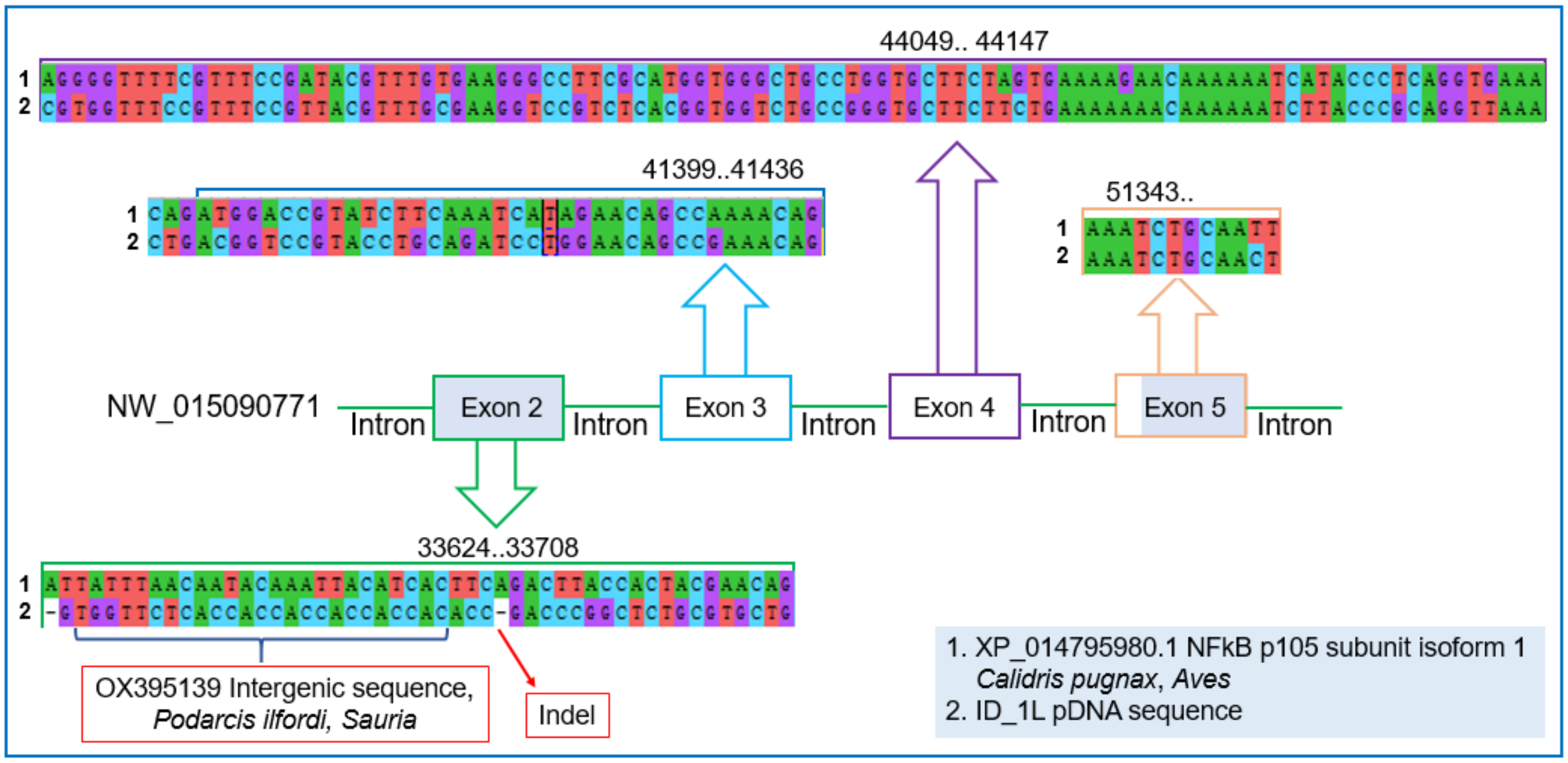

**Fig. 3 The Schematic diagram of *NFκB* p105 subunit-like DNA sequence**

We used MS mode to analyze 1,047,532 sequences in a subset of humans and identify sequences that differed from the human mitogenome and nuclear genome (Fig. 2). Furthermore, we discovered five non-primate mitogenomic sequences in the dataset (Table S5), with the absence of non-human primate mitogenomic sequences. Among these, 846 sequences (affinity value: 100%) were identified to correspond to the human mitogenome and were not subject to further discussion. Additionally, there were 41 sequences displaying an affinity range of 91.48% to 99.5% with the human mitogenome. Given that NCBI houses comprehensive data on the entire human mitogenome, it can be inferred that these sequences are likely of ancient human origin. Nevertheless, potential individual misclassifications stemming from the approximate existence of a few unknown loci differences in modern humans cannot be ruled out (Table S6). Upon scrutiny of two specific sequences, ID_17H and ID_20H, it was discerned that they exhibited 100% coverage of each other, with distinct identities of 96.98% and 99.50%, respectively, implicating that they originate from disparate maternal lineages. The pDNA fragment ID_18H displayed an affinity value of 96.80%, and phylogenetic analyses indicated its positioning on a node with *Homo sapiens*, hinting at the possibility that its host may pertain to a previously unidentified species within the genus *Homo* (Fig. S3, above). Fragment ID_19H, with an affinity value of 83.85%, aligned with human ND4. Phylogenetic analysis suggested that ID_19H lies between the human and chimpanzee genera, potentially representing the earliest locally occurring species within the genus *Homo*; RelTime-ML analysis indicated a divergence time of 0.49 Ma from human (Fig. S3, middle & bottom). Notably, the above pDNA sequences do not appear in the modern human genome, indicating that the ancient humans carrying these fragments are likely extinct.

Many pDNA sequences were found to match nuclear genome sequences of non-human primates at the following frequencies: 1 in the Tarsiidae family, 4 in Platyrrhini (New World monkeys), 15 in Macaca (Old World monkeys), 8 in Gorillas, 8 in pygmy chimpanzees (Table S7), and 271 in chimpanzees (Table S8). Among the 271 sequences for chimpanzees, the second hit species were modern humans in 241 (88.93%), other primates in 29 (10.70%), and non-primates in the remainder. These sequences should belong to the ancient hominins. The evidence derived from pDNA establishes a genomic link between modern humans and chimpanzees, consistent with scholars' traditional view that the two species are closely related.

Human burial practices effectively prevented DNA decomposition, leading to its accumulation in the environment at burial sites. Although other hominids also engaged in burial activities for their companions, the frequency, duration, and depth of these burials were not as significant as those observed in the species of *Homo*. Researchers have discovered *Homo erectus* fossils dating back to 0.5 Ma in Nanzhao County near the oil field[19]. Consequently, it is inferred that the ID_19H pDNA fragment discussed in this paper, along with most of the 271 pDNA fragments mentioned earlier, likely originated from unsequenced *Homo erectus*. Since these DNA fragments are significantly "younger" than the age of the source rock (37.5 Ma), it can be concluded that these sequences do not belong to oriDNA but to paeDNA.

The study suggests that the proposed *Homo erectus* shared some nuclear genes with apes. The difference between the mitogenomic sequences of the *Homo erectus* and modern humans is much smaller than that between the *Homo erectus* and chimpanzees. This indicates a lack of close mitogenomic connection between the *Homo erectus* and chimpanzees, suggesting that during hominin evolution, the mitogenome led the transition from apes to humans and the nuclear genome underwent a slower evolutionary process. Based on extensive research, it is widely accepted that mitochondrial genomic sequences exhibit a higher level of conservation compared to cell nuclear sequences. Previous studies revealed that ancient humans experienced severe population bottlenecks and limited mitochondrial haplogroup diversity[23,24,25]. Based on all the information, we proposed that early species of the genus *Home* may originate from a small group of chimpanzees, possibly with a small number of females. They constituted small clan tribes with a female-centered structure and kinship ties. Later, reproductive isolation may have occurred due to karyotypic changes, preventing descendants from acquiring mitochondrial genomes from apes. Genetic isolation had likely happened before the time of the *Homo erectus*. Determining the exact time when this reproductive isolation occurred, while humans and apes separated, is a primary scientific question that always calls for further studies. The logical steps involved in utilizing the "Mega screening method" to search for ancient *Homo* DNA are illustrated in Fig. S4.

Because DNA is primarily soluble in water and only slightly lipid-soluble, the lipid envelope surrounding the cells and organelles of living organisms aids in their entry into the oil. When the lipid envelope ruptures, DNA mixes with alkane molecules. After encountering water, the DNA leaves the oil phase because of water extraction on the water-oil interface. It is hypothesized that tissues, cells, or casein fragments of remains can enter the oil reservoir. In larger remains (e.g., animal and plant), the body's multiple layers of tissue can be encapsulated by the oil, resulting in a phenomenon, "Oil immersion preservation". This process allows the gradual release of DNA from tissue adjacent to the oil phase. When one layer of tissue is degraded, cells in other layers continue to release DNA until the organism is completely dissolved. Moreover, the fat tissue of the organism contains DNA, and when the fat liquefies, the DNA can flow into the oil reservoir, leading to a long-term, slow, and continuous release of DNA from the organism. Additionally, small organisms such as algae and plankton have cells that can rupture quickly, releasing their DNA into the reservoir. As a result, pDNA is continuously lost over time (Fig. 1). Here arises the interesting question: Are eDNA (paeDNA and preDNA) hosts also involved in oil formation? While their contribution may seem small, the answer is positive.

Some scholars have observed a decrease in original (*in situ*) DNA and an increase in eDNA, mainly from microorganisms such as bacteria, in bone fossils[26, 27]. It is challenging to determine the exact source of paeDNA in the distant period before and after the formation of hydrocarbon source rocks due to the lack of knowledge about the paleoenvironment at ancient times. During the reservoir formation process, the oriDNA molecules undergo continuous degradation and loss, resulting in only a small amount of their remaining, while eDNA gradually infiltrates the oil. Human urban habitation, burial practices, and widespread cultivation of angiosperm crops have contributed to the ongoing introduction of DNA from human-associated species into the environment, serving as the primary source of preDNA. Utilizing the data in Fig. 2 and adding in our assumptions, the pattern of pDNA changes in oil was depicted (Fig. 1).

Below are a few instances of the "More new molecules and fewer old ones" phenomenon. First, the DNA from lower-life species that contributed to the formation of crude oil has nearly disappeared, as the DNA representing modern species has steadily increased (Fig. 2). Second, out of 886 human mitogenome sequences, 845 show a 100% affinity value with the human genome (preDNA, modern humans), while 41 sequences, possibly ancient human mitogenome sequences (paeDNA), exhibit variant divergences from the human mitogenome (Table S6). The sequence ID_19H displays the most significant divergence from the human mitogenome (Fig. S3) and is likely derived from *Homo sp*.

In our study, it was interesting to note that the analysis of the pDNA sequences did not show the typical "deamidation reaction" often found in aDNA from traditional fossils[28]. We hypothesize that two factors may contribute to this phenomenon. Firstly, the entry of eDNA to the reservoir necessitates specific conditions, such as encapsulation in lipids, leading to an anaerobic and anhydrous environment within the oil-forming materials and the reservoir. Secondly, the DNA sequences undergoing the deamidation reaction may not have been selectable for the “MS mode”.

## Methods

### The geochemical survey

The oils were deasphalted with petroleum ether and then fractionated by column chromatography using a 2:1 (v/v) mixture of silica gel and alumina. The fractions were separated into saturated, aromatic, and resin components by sequential elution with 60 mL of petroleum ether, 40 mL of a dichloromethane/petroleum ether mixture (2:1, v/v), and 30 mL of a dichloromethane/methanol solution (93:7, v/v), respectively.

Gas chromatography (GC) analysis was conducted on the saturated fractions using an Agilent Model 6890 gas chromatograph, with a silica column (HP-1, 30 mm × 0.25 mm i.d.) installed. The oven temperature was programmed to increase from 100°C to 300°C for 10 minutes at a rate of 4°C/min. Helium was employed as the carrier gas.

Gas chromatography-mass spectrometry (GC-MS) analysis of the saturated and aromatic fractions was performed using an Agilent Model 6890 gas chromatograph coupled to a DB-5MS capillary column. The temperature program for the saturated fraction was as follows: the initial temperature was set at 100°C for 1 minute, followed by a ramp of 4°C per minute to 220°C, and then increased to 300°C at a rate of 3°C per minute and held for 15 minutes. The mass spectrometer was operated in electron impact mode, with an ionization energy of 70 eV and a scan range of 50-600 Da. The biomarker parameters of the saturated and aromatic hydrocarbons were determined by integrating the peak areas on the mass chromatograms.

**DNA extraction, DNA library construction, sequencing, and NCBI nucleotide BLAST**

**The wet lab conditions**: The procedure was conducted in a BSL-2 industrial laboratory (Biosafety Level 2), following rigorous ancient DNA extraction protocols[29]. The process involved UV sterilization of all items used and the utilization of Geobio® DNA Remover (Jiaxing Jiesai Biotechnology Co., Ltd., Jiaxing, China) on clean benches and solid equipment. It is important to note that this laboratory has never been exposed to fish, non-human primates, or plant samples before this work.

**To extract DNA** from petroleum samples, we followed the recommended protocol of the Geobio® DPPS-PMNP DNA extraction kit developed by Jiaxing Jiesai Biotechnology Co[30]. Due to various long-term factors, some pDNA molecules are cross-linked with other macromolecules, forming complexes. To estimate the extract concentration, we measured the optical density (OD) at the UV 260 nm peak of the DNAmix, containing DNA complexes and free DNA molecules. We performed the extraction experiments on the samples five times, and the results were consistent (Table S9), providing sufficient DNA (Table S10). We used NEB PreCR®Mix for DNA repair.

**The NGS DNA library construction** followed the SSDLSP standard procedure (Sangon Biotech, CORP). Sequencing was conducted on Illumina NovaSeq 6000 (S-4 Reagent Kit) using the "Dual index sequencing" standard procedure, and the reads met the quality control standards with satisfactory quality distribution. Sometimes, it's necessary to manually remove extra adapter sequences at both ends of the read to obtain accurate information. After merging the forward and reverse reads, we performed local NCBI Nucleotide BLAST, Size 11, using the database (Version 5, Nucleotide sequence Database), identifying the amount of 3,159,020 valid sequences. The sequences were classified according to their biological characteristics, and their lengths were measured (Table S11). It is important to note that the processes of secondary oil collection may inadvertently lead to the loss of many aDNA fragments.

**The "Mega screening method"**: In the first step, we employed a "minimum E-value mode" to group sequences by aligning fossil DNA with all known sequences from the NCBI database (Version 5, Nucleotide Sequence Database). We applied an E-value cutoff below 1E-07 to filter for qualified sequences (QS) and identify the best match at a specific taxonomic level. In the second step, we used the "MS mode" by selecting "exclude" option to conduct a search that omits the top result. If the E-value difference between two search results at the same taxonomic level (species, genus, family, order, and class) exceeds 1E-02, it suggests that the test sequence (query) belongs to the species identified in the first search result, suggesting a unique origin. Conversely, if the difference does not surpass the established threshold (1E-02), the sequence is shared between two taxa, rather than originating from a single one. In this scenario, we cannot determine the true source of the sequence, making it impossible to confirm whether the fragment belongs to the target species or irrelevant environmental species. Thus, we had to forgo any further analysis of the sequence.

Additionally, the "mega screen method" can be employed to reveal relationships among similar sequences discovered across different genomic organisms, providing valuable insights into the patterns of genome evolution.

The whole process involves the following steps:
1) Nucleotide Blast is employed to align sequences to the entire NCBI database without any limitation.
2) Creation of subsets for sorting the sequences: Using the "minimum E-value mode," results are matched and sequences are grouped into the appropriate subsets, determining the total number of sequences (TS). Subsequently, an E-value threshold of < 1E-07 is established to filter for qualified sequences (QS).
3) The subset primarily containing ancient DNA (aDNA) is selected.
4) Each sequence within this subset is screened individually using the "MS mode" to identify those originating from a single lineage, categorizing the "unique" sequences.
5) Considering various factors, including local climate shifting, geological changes, and evolutionary principles of the host species, we can ascertain if the sequence is oriDNA.

**Sequence affinity (Affinity)**: The metrics we set up to assess the similarity relationship between subject sequences and hit genomes were obtained by multiplying the two values of Identity and Cover obtained from NCBI BLAST and then converted to percentages (Identity × Cover × 100%). Mutations in the genomic and mitogenomic sequences occur continuously during the evolutionary process, leading to significant sequence differences between ancient and modern species. However, this variation is smaller in conserved sequences than in non-conserved sequences. A high-affinity value indicates that the sequence is highly coherent and related to the genome of the modern species, which is likely to be either a conserved sequence or a modern eDNA. A low-affinity value indicates a low similarity, and the sequence may be from either a distantly related species or other modern species that have not yet been sequenced. It is necessary to consider various factors, including the sequence composition and conservation, host species information, host genome sequencing data, etc., rather than relying on the Affinity value alone, when determining whether a sequence is an oriDNA.

**Data and software availability**

The sequencing data have been deposited in the NCBI BioProject (PRJNA1091869). The pDNA sequences utilized in this manuscript are listed in the Appendix (Table S12). In this study, MEGA11 was used to construct the RelTime-ML local timetree[31].

**Acknowledgments:** We extend our gratitude to the "Chao-Xiang Talent Program" foundation of Haining City, the support that covers the costs of petroleum DNA extraction and DNA sequencing from 2018 to 2022. The collection of petroleum samples from 2018 to 2021 received support from the Key Research & Development Special Program of Henan Province (231111111200), with the indispensable help of numerous workers and administrative staff from Nanyang oilfield (Sinopec Corp.), including Ms. Yu Liu, Mr. Pin-Lei Cai, and Mr. Peng Wei. Special thanks to Haining BoShang Biotechnology Corporation and Sangon Biotech (Shanghai) Corporation for their collaborative support in the success of this project, especially in the construction of the NGS DNA library using the SSDLSP standard procedure and sequencing. We also appreciate the contributions of Mr. HAN-Yu Zhu in creating a paleogeographic map and the assistance of Ms. Yun-Li Wang and Ms. Yue Zhang in processing the data. We extend our heartfelt appreciation to the following individuals for their invaluable contributions to this project: Dr. Hua-Jun Zheng from the Shanghai Institute for Biomedical and Pharmaceutical Technologies, Dr. Jun-Ye Ma from the Nanjing Institute of Geology and Paleontology, at the Chinese Academy of Sciences, Dr. Alida M. Bailleul from the Institute of Vertebrate Paleontology and Paleoanthropology, at the Chinese Academy of Sciences, Dr Shun-Ping He from the Institute of Hydrobiology, also at Chinese Academy of Sciences. Their stimulating discussions regarding the manuscript have greatly enriched our work.

**Author contribution**: The study was conceptualized and designed by W.Q.Z. and Y.Q.G., Y.Q.G., T.C.Q., and W.Q.Z. supervised the transportation of oil samples. M.J.L. performed the geochemical marker analysis. Geographical and paleogeological data analysis, as well as map creation, were conducted by F.L. and H.Y.Z. Petroleum DNA extraction from the samples was carried out by W.Q.Z., C.Y.Z., and J.Y.Y. Alignment between the sequences and present genomes was completed by Z.Y.G. and M.J.C. DNA laboratory analysis, interpretations, taxonomic profiling, and annotation were conducted by Z.Y.G., Z.Y.T., S.J.Z, L.J.Z and W.Q.Z. The statistical analyses were performed and finished by G.Q.C., X.G.Z., R.C., and J.H.Z. Phylogenetic analyses of mitogenomic DNA sequences were performed by J.H.G., under the supervision of G.Q.C. and W.Q.Z. Figures were designed and finished by R.C., J.H.Z., L.Y.Q., and M.R.S. The manuscript was written by W.Q.Z. and corrected by Z.Y.T., M.J.L., and L.Y.Q. The Project Coordinator was T.C.Q.

**Competing interests:** Authors declare that they have no competing interests.

**Data and materials availability:** All data are available in the main text or the supplementary materials.

**Supplementary Materials**

Figs. S1 to S4

Tables S1 to S12

Supplementary Materials for

# DNA Fragments in Crude Oil Reveals Earth's Hidden History

Wan-Qian Zhao[1,3*], Zhan-Yong Guo[2], Yu-Qi Guo[3*], Mei-Jun Li[4*], Gang-Qiang Cao[1], Zeng-Yuan Tian[1], Ran Chai[5], Li-You Qiu[6*], Jin-Hua Zeng[7], Xin-Ge Zhang[8], Tian-Cang Qin[9], Jin-Yu Yang[10], Ming-Jie Chen[11], Mei-Rong Song[12], Fei Liang[13], Jun-Hui Geng[1], Chun-Yan Zhou[14,15], Shu-Jie Zhang[3], Li-Juan Zhao[3]

Corresponding author: wqzhao@zzu.edu.cn; guoyuqi@zzu.edu.cn; meijunli@cup.edu.cn; qliyou@henau.edu.cn

**The PDF file includes:**

Figs. S1 to S4
Tables S1 to S12

**A**

| Physical properties of the petroleum fluids | | | | | | |
|---|---|---|---|---|---|---|
| Oilfields | Sample Number | Density ($g/cm^3$) | Colloid Asphaltene (%) | Wax (%) | Sulfur (%) | Viscosity (cP) |
| Gucheng Well number：51215 | 1 | 0.863 | 39.3 | 14.5 | 0.084 | 15.5 |
| | 2 | 0.865 | 39.9 | 15.4 | 0.085 | 15.7 |
| | 3 | 0.867 | 38.1 | 14.8 | 0.083 | 15.2 |
| | 4 | 0.871 | 39.1 | 15.2 | 0.084 | 15.4 |
| Storage depots 20210301 | Gucheng | NA | NA | NA | NA | NA |
| | Jinglou | NA | NA | NA | NA | NA |
| | Shuanglou | NA | NA | NA | NA | NA |
| | Shuanghe | NA | NA | NA | NA | NA |

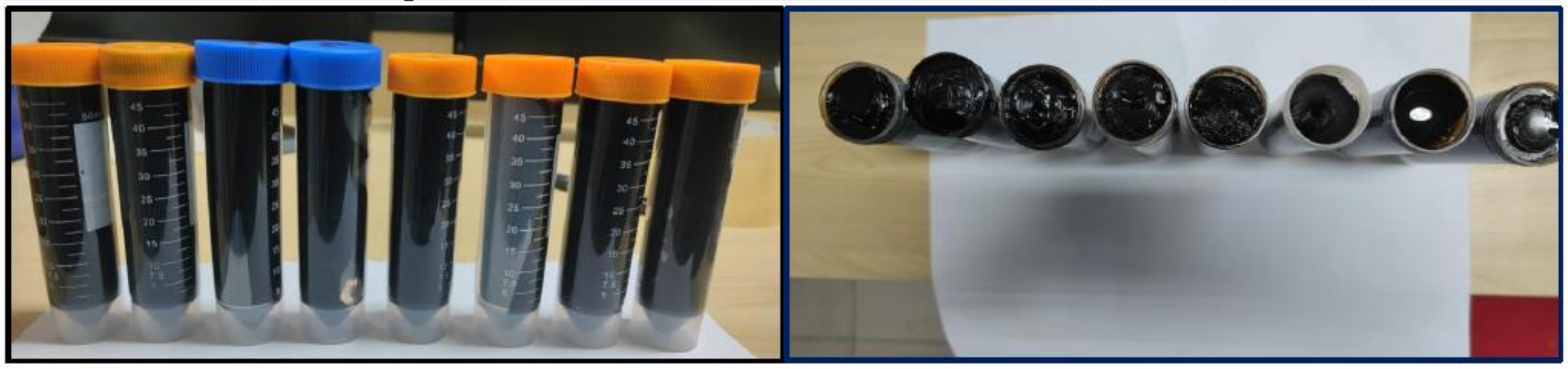

**B**

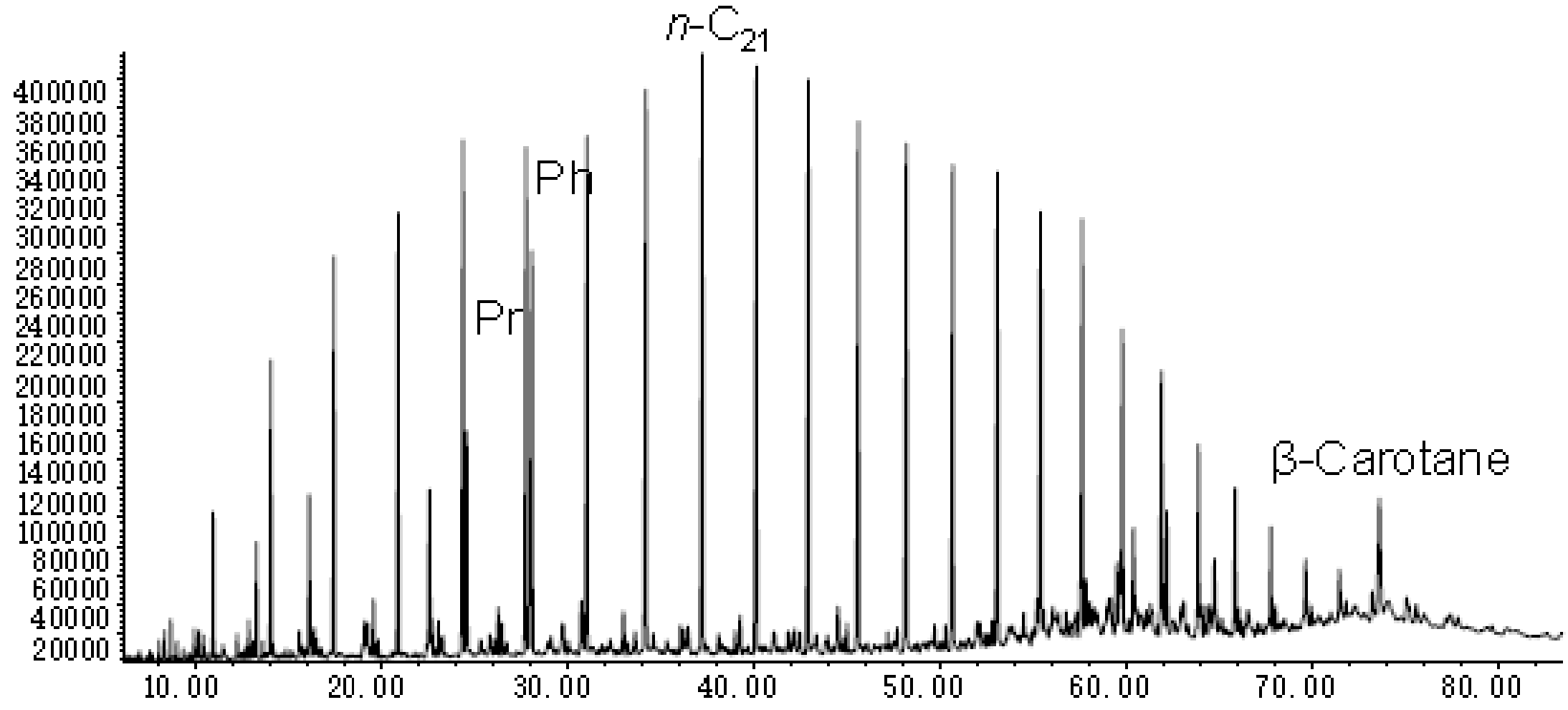

**C**

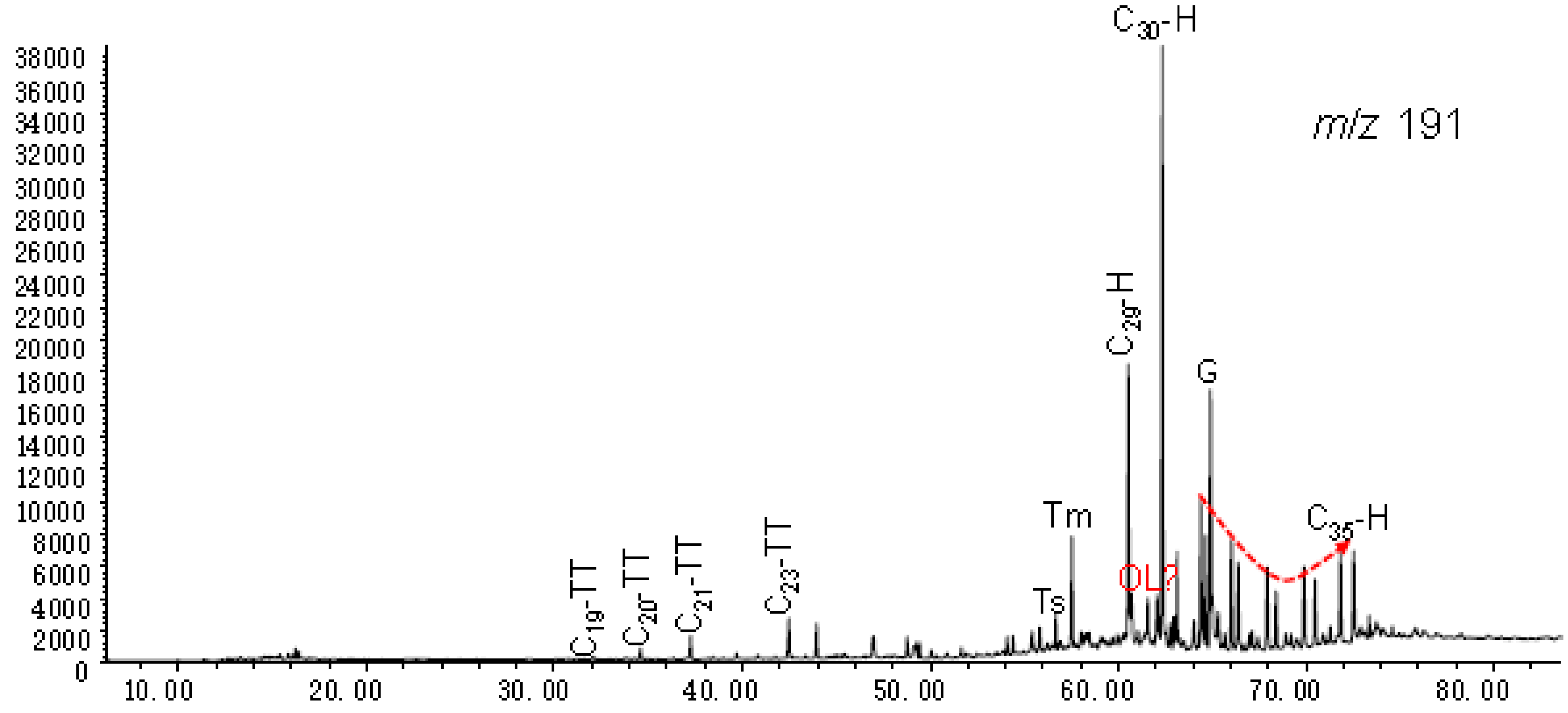

**D**

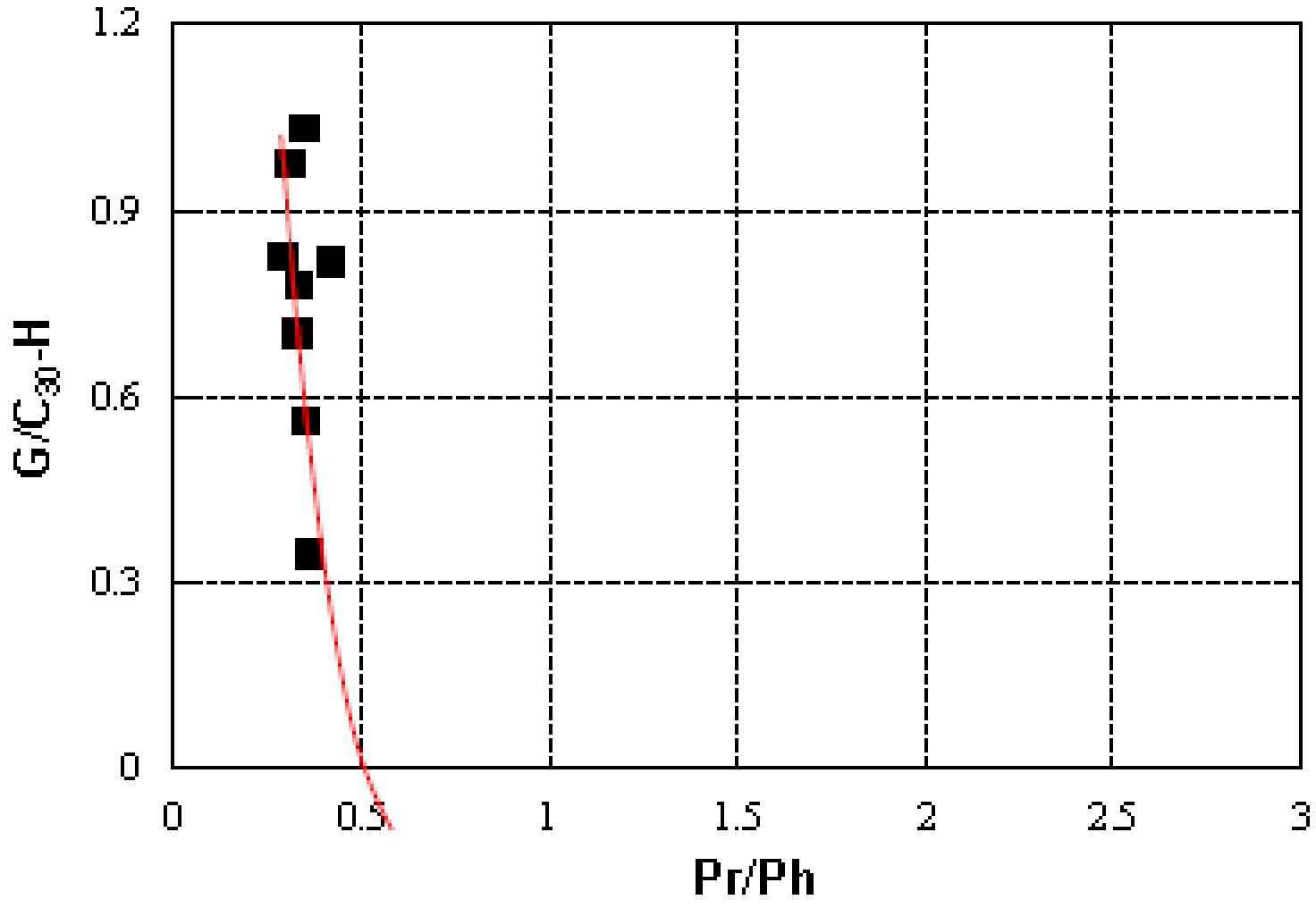

**E**

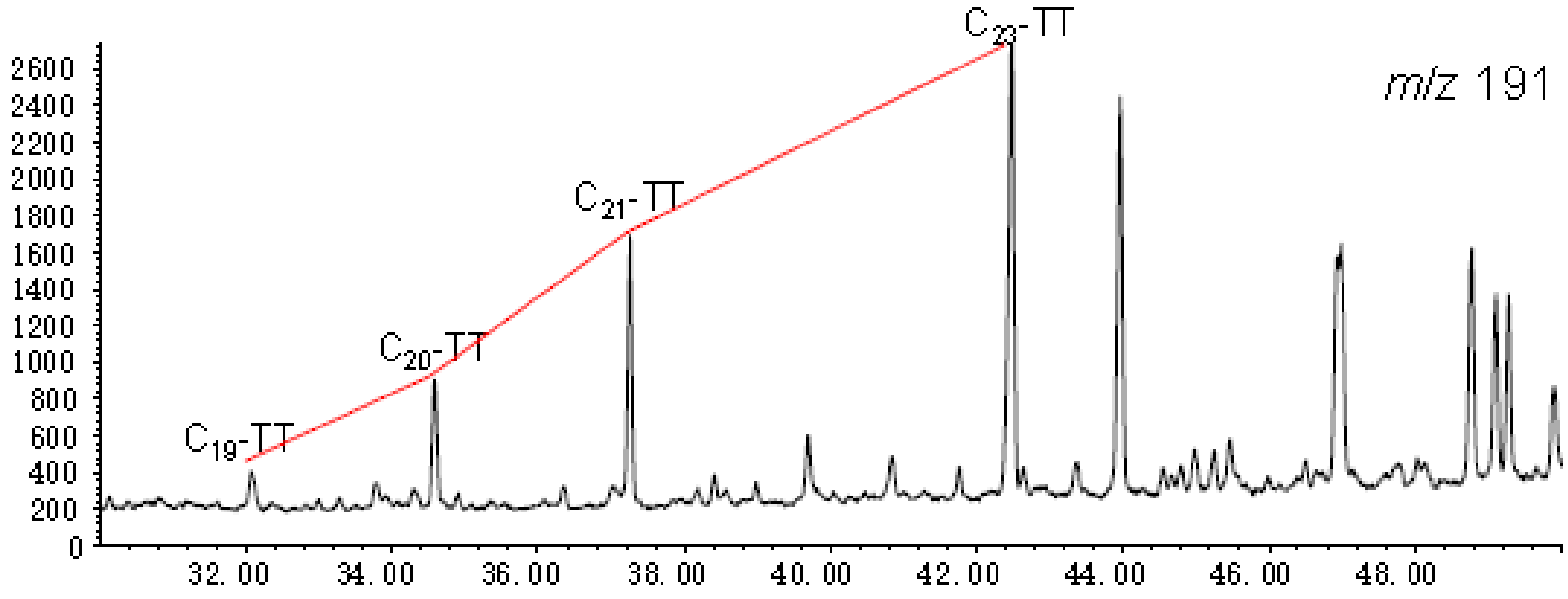

**F**

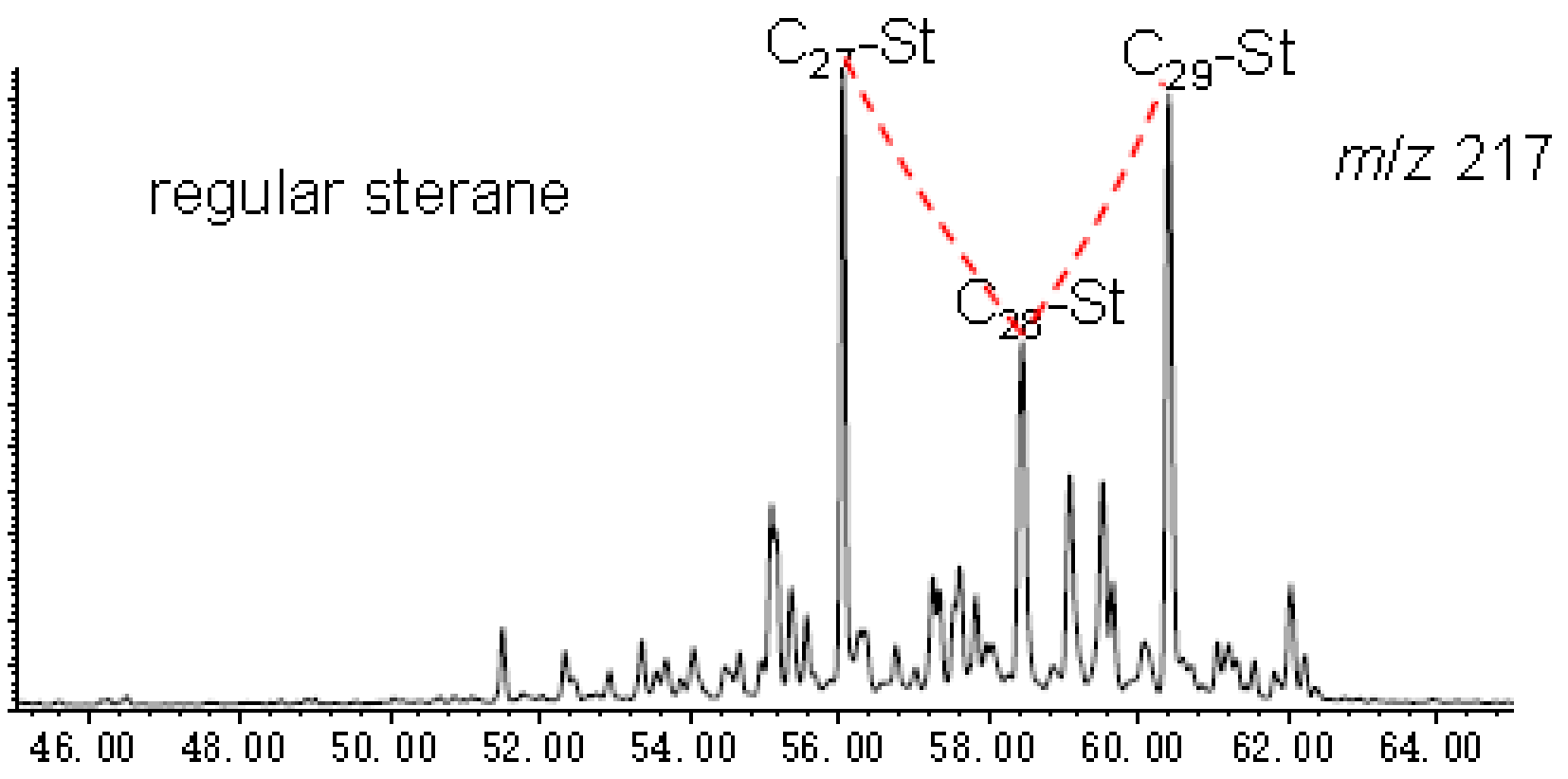

**G**

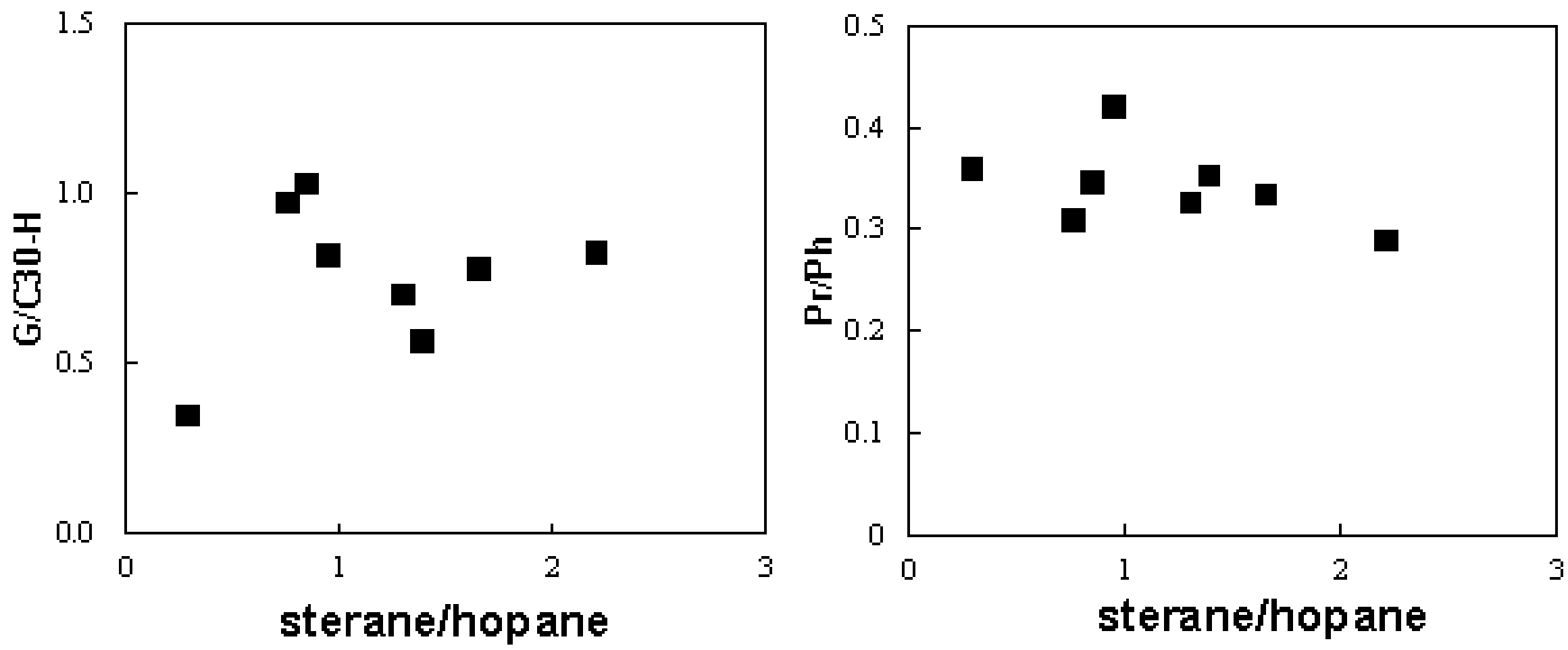

**Fig S1. The geochemical analysis of petroleum samples**

**A**, The buk geochemical characteristics of the petroleum used in this study. **B**, The distribution of *n*-alkanes and isoprenoids in the petroleum used in this study. **C**, The distribution of hopanes in the petroleum used in this study. D, The distribution of hopanes-gammacerane in the petroleum used in this study. E, The distribution of tricyclic terpanes in the petroleum used in this study. **F**, The distribution of sterane in the petroleum used in this study. **G**, The distribution of sterane in the petroleum used in this study.

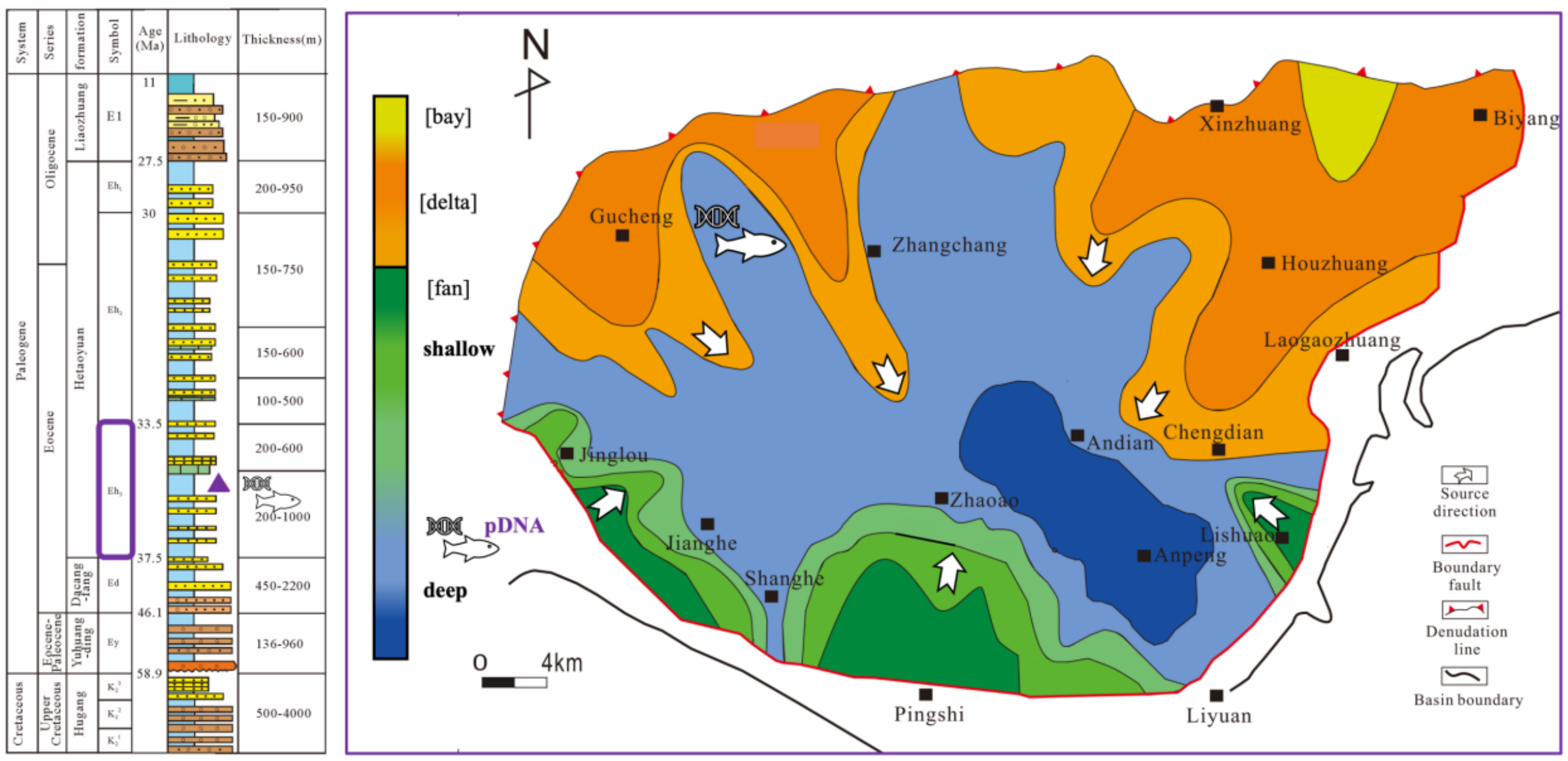

**Fig S2.** The lithostratigraphy of the Hetaoyuan Formation (**left**) and corresponding paleogeography in the Eocene ages (**right**) at the sample position. The paleogeographic map is modified from Li's work[1-3], based on the corrective information from Dong's work[4] and Xia[5].The samples are located in Member 3 of the Hetaoyuan Formation.

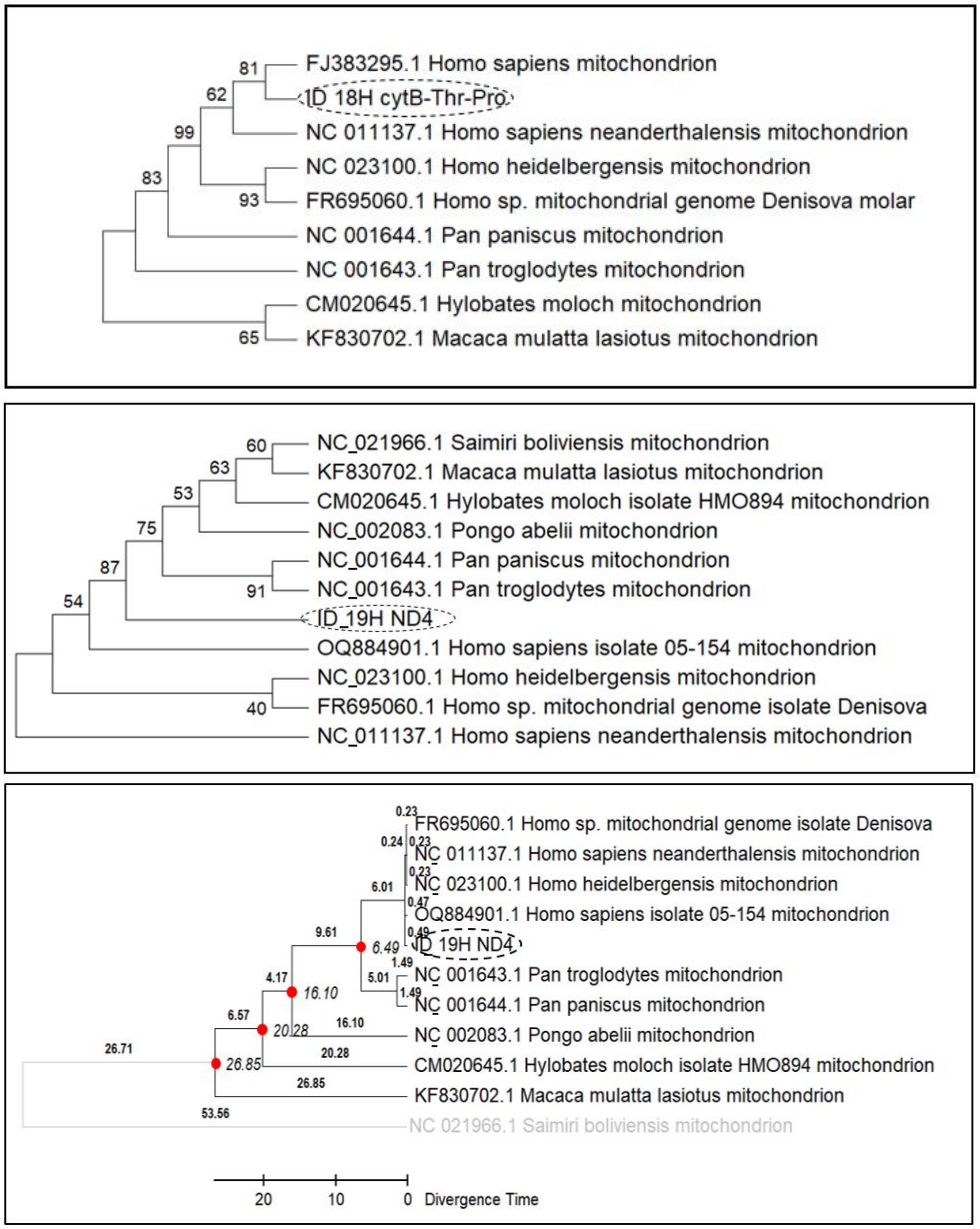

**Fig S3. The Phylogenetical analysis of the pDNA sequences**

The phylogenetic analyses are established using primate mitogenomic sequences as references. The data is being examined utilizing the Maximum Likelihood method in MEGA11 on a locally installed computational platform. The divergences are evaluated in Bootstrap consensus trees through specific calculations with 1000 bootstrap replicates under the HKY+G model and a Timetree under RelTime-ML mode. In each context, ID_18H (shown in the top) and ID_19H (shown in the middle and the bottom) correspond to the human cytB+Thr+Pro and ND4 regions, respectively (OQ884901.1). The Bootstrap value of the node on the Bootstrap consensus tree and the divergence time near the node on the Timetree are numerically represented and displayed for analysis.

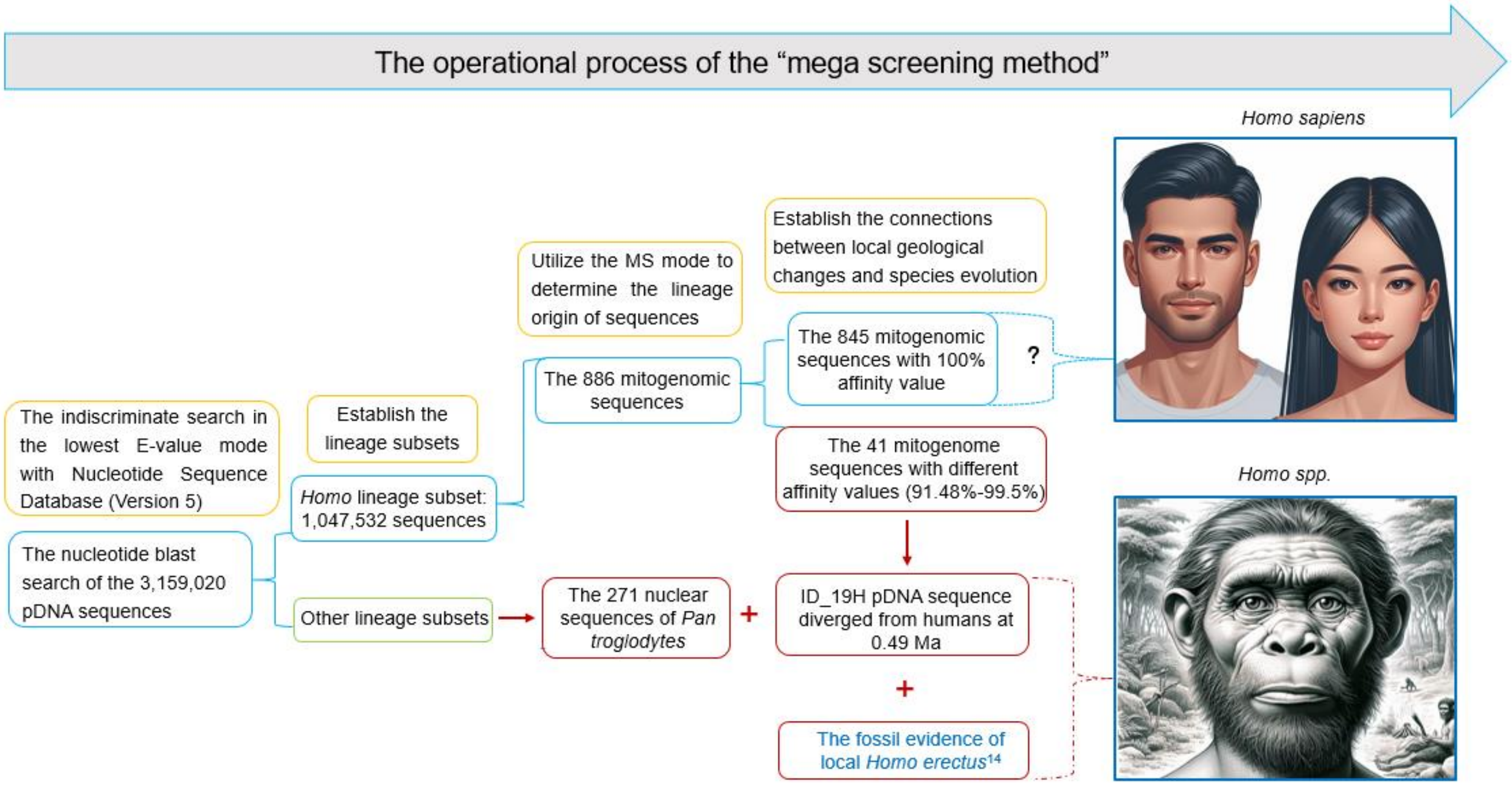

**Fig S4. The molecular evidence for the presence of *Homo spp.***

Table S1 The 8 pDNA Sequences Match Lower Plant Genomes

| ID | Len | Taxo | Nucleotide ID | Target Description | Cover | Identity | E-value | Affinity |
|---|---|---|---|---|---|---|---|---|
| 1G | 75 | *Chlorella desiccata , Chlorophyta* | OK569792.1 | Mitochondrion | 74% | 100% | 1.00E-18 | 74.00% |
| 2G | 103 | *Picochlorum sp., Chlorophyta* | CP120464.1 | SENEW3 chromosome 10 | 99% | 91.18% | 6.00E-30 | 90.26% |
| 3G | 179 | *Rhytidiadelphus loreus; Streptophyta; Bryophyta; Hylocomiaceae* | OX344761.1 | Genome, chromosome 1 | 96% | 83% | 9.00E-38 | 79.32% |
| 4G | 83 | *Ulva expansa, Chlorophyta* | MH730161.1 | 28S rDNA | 100% | 100% | 3.00E-32 | 100.00% |
| 5G | 100 | *Klebsormidium sp.; Streptophyta; Klebsormidiaceae; Klebsormidium* | OR168110.1 | Plastid DNA | 100% | 100% | 2.00E-41 | 100.00% |
| 6G | 141 | *Klebsormidium sp., Streptophyta; Klebsormidiaceae; Klebsormidium* | OR168110.1 | Plastid DNA | 100% | 100% | 2.00E-63 | 100.00% |
| 7G | 153 | *Marchantia polymorpha, Streptophyta; Embryophyta; Marchantiophyta* | AP019866.1 | Chloroplast DNA | 100% | 99% | 4.00E-67 | 99.35% |
| 8G | 155 | *Marchantia polymorpha, Streptophyta; Embryophyta; Marchantiophyta* | AP020202.1 | Chloroplast DNA | 100% | 99.35% | 3.00E-69 | 99.35% |

Table S2. The pDNA Sequences Corresponding to the Genomes of Marine Organisms

| pDNA | | Results of the Nucleotide Blast | | | |
|---|---|---|---|---|---|
| ID | Len | Taxo | Nucleotide | E-value | Affinity |
| 1M | 78 | *Scyliorhinus canicula, Chondrichthyes* | LR744051.1 | 2.00E-06 | 50% |
| 2M | 80 | *Mercenaria mercenaria, Mollusca* | XM_045350734.2 | 9.00E-18 | 83% |
| 3M | 92 | *Nucula nucleus, Mollusca* | LC123061.1 | 5.00E-22 | 78% |
| 4M | 112 | *Ruditapes philippinarum, Mollusca* | KU682310.1 | 2.00E-16 | 77% |
| 5M | 184 | *Phorcus lineatus, Mollusca* | OV121097.1 | 2.00E-07 | 45% |
| 6M | 253 | *Lateolabrax maculatus, Actinopterygii* | CP027279.1 | 2.00E-52 | 68% |
| 7M | 203 | *Hyperoplus immaculatus, Actinopterygii* | OX442354.1 | 1.00E-29 | 44% |
| 8M | 187 | *Thunnus maccoyii, Actinopterygii* | OU343202.1 | 8.00E-19 | 70% |
| 9M | 83 | *Thunnus maccoyii, Actinopterygii* | OU343200.1 | 9.00E-12 | 72% |
| 10M | 131 | *Aplidium turbinatum, Tunicata* | OU964919.12 | 3.00E-09 | 35% |

Table S3. The 15 pDNA Sequences Corresponding to the Genomes of Non-Indigenous Organisms

| pDNA | | Results of the Nucleotide Blast | | | |
|---|---|---|---|---|---|
| ID | Len | Taxo | Nucleotide ID | E-value | Affinity |
| 1A | 117 | *Bos mutus* | CP027076.1 | 3.00E-46 | 97.44% |
| 2A | 149 | *Bos mutus* | CP027076.1 | 8.00E-55 | 92.98% |
| 3A | 144 | *Bos mutus* | CP027076.1 | 2.00E-61 | 96.62% |
| 4A | 97 | *Bos mutus* | CP027076.1 | 5.00E-23 | 81.66% |
| 5A | 162 | *Bos mutus* | CP027078.1 | 3.00E-67 | 95.92% |
| 6A | 119 | *Bos mutus* | CP027078.1 | 3.00E-46 | 96.64% |
| 7A | 140 | *Bos mutus* | CP027087.1 | 9.00E-54 | 94.45% |
| 8A | 223 | *Meleagris gallopavo* | OW982293.1 | 4.00E-93 | 92.34% |
| 9A | 151 | *Meleagris gallopavo* | OW982293.1 | 2.00E-49 | 95.26% |
| 10A | 117 | *Meleagris gallopavo* | OW982293.1 | 8.00E-41 | 97.20% |
| 11A | 198 | *Meleagris gallopavo* | OW982293.1 | 5.00E-53 | 94.28% |
| 12A | 288 | *Meleagris gallopavo* | OW982293.1 | 9.00E-128 | 97.20% |
| 13A | 92 | *Meleagris gallopavo* | OW982297.1 | 2.00E-20 | 72.90% |
| 14A | 154 | *Meleagris gallopavo* | OW982293.1 | 4.00E-53 | 95.26% |
| 15A | 103 | *Apteryx mantelli* | LK358553.1 | 2.00E-08 | 39.05% |

Table S4. The pDNA Sequences Exhibiting Low Affinity Values

| pDNA | | Results of the Nucleotide Blast | | | | |
|---|---|---|---|---|---|---|
| ID | Len | Taxo | Loci/Protein | Nucleotide | E-value | Affinity |
| 1L | 250 | *Calidris pugnax Aves* | NFKB1 | XM_014940504.1 | 2.00E-32 | 61.91 |
| 2L | 250 | *Myotis daubentonii Mammalia* | Chromosome 13 | OY725372.1 | 3.00E-38 | 59.8 |
| 3L | 273 | *Aquila chrysaetos Aves* | Chromosome 13 | LR606193.1 | 7.00E-19 | 45.46 |
| 4L | 250 | None | | | | |

Table S5. The 5 pDNA Sequences Corresponding to Non-Primate Mitogenomes

| pDNA | | Results of the Nucleotide Blast | | | | |
| --- | --- | --- | --- | --- | --- | --- |
| ID | Len | Taxo | Nucleotide | Range/Area | E-value | Affinity |
| 1N | 146 | *Gallus gallus; Galliformes* | OP740918.1 | 503-648; 12S | 2.00E-66 | 99.3200% |
| 2N | 193 | *Sus scrofa; Artiodactyla* | MN124249.1 | 344-536; COX1 | 2.00E-92 | 99.4800% |
| 3N | 81 | *Mylopharyngodon piceus; Cypriniformes* | MF687133.1 | 13586-13666; ND5 | 3.00E-32 | 100.0000% |
| 4N | 203 | *Ctenopharyngodon idella; Cypriniformes* | MG827396.1 | 10382-10582; ND4 | 6.00E-97 | 100.0000% |
| 5N | 113 | *Ctenopharyngodon idella; Cypriniformes* | MG827396.1 | 400-288; 12S | 2.00E-48 | 100.0000% |

Table S6. The 41 pDNA Sequences Corresponding to Human Mitogenome

| pDNA | | Results of the Nucleotide Blast | | | | |
|---|---|---|---|---|---|---|
| ID | Len | Taxo | Nucleotide ID | Mito Range/Area | E-value | Affinity |
| 1H | 249 | *Home sapiens* | OR182493.1 | 13856-14104; ND5 | 2.00E-125 | 99.60% |
| 2H | 238 | *Home sapiens* | ON422324.1 | 8347-8584; ATP8+ATP6 | 2.00E-119 | 99.58% |
| 3H | 216 | *Home sapiens* | MH449153.1 | 9787-10002; COX3+trnG | 3.00E-107 | 99.54% |
| 4H | 215 | *Home sapiens* | OQ388438.1 | 47-261; D-loop | 1.00E-106 | 99.53% |
| 5H | 212 | *Home sapiens* | ON946746.1 | 765-554; D-loop | 5.00E-105 | 99.53% |
| 6H | 194 | *Home sapiens* | MH449153.1 | 9974-9781; COX3 | 5.00E-95 | 99.48% |
| 7H | 186 | *Home sapiens* | MG660585.1 | 8790-8975; ATP6 | 1.00E-90 | 99.46% |
| 8H | 176 | *Home sapiens* | JQ664512.1 | 4404-4578; trnM+ND2 | 5.00E-85 | 98.44% |
| 9H | 164 | *Home sapiens* | MN503417.1 | 11624-11787; trnH+trnS+trnL+ND5 | 2.00E-76 | 99.39% |
| 10H | 153 | *Home sapiens* | MN207699.1 | 652-804; D-loop: | 2.00E-72 | 99.35% |
| 11H | 111 | *Home sapiens* | OR182493.1 | 13525-13415; ND5 | 3.00E-49 | 99.10% |
| 12H | 200 | *Home sapiens* | LS997688.1 | 28-227; D-loop: | 1.00E-96 | 99.00% |
| 13H | 280 | *Home sapiens* | MF362729.1 | 11759-12038; ND4 | 7.00E-140 | 98.93% |
| 14H | 245 | *Home sapiens* | OQ707209.1 | 8658-8902; ATP6 | 6.00E-120 | 98.78% |
| 15H | 199 | *Home sapiens* | MG660585.1 | 7993-8191; COX2: | 1.00E-92 | 98.49% |
| 16H | 189 | *Home sapiens* | KF162400.1 | 8394-8206; ATP8 | 2.00E-89 | 98.41% |
| 17H | 199 | *Home sapiens* | OR182493.1 | 13537-13735; ND5 | 2.00E-89 | 96.98% |
| 18H | 250 | *Home sapiens* | FJ383295.1 | 16068-15820; CYTB+trnT+trnP | 1.00E-113 | 96.80% |
| 19H | 177 | *Home sapiens* | OQ884901.1 | 11485-11325; ND4 | 5.00E-60 | 83.85% |
| 20H | 199 | *Home sapiens* | OR182493.1 | 13537-13735; ND5 | 8.00E-98 | 99.50% |
| 21H | 189 | *Home sapiens* | OR182493.1 | 15872-15684; CYTB | 3.00E-92 | 99.47% |
| 22H | 188 | *Home sapiens* | OR159678.1 | 3294-3481; trnL-ND1 | 1.00E-91 | 99.47% |
| 23H | 179 | *Home sapiens* | OR182493.1 | 14442-14620; ND6 | 1.00E-86 | 99.44% |
| 24H | 175 | *Home sapiens* | OR159678.1 | 4656-4482; ND2 | 2.00E-84 | 99.43% |
| 25H | 137 | *Home sapiens* | OR159678.1 | 3223-3359; 16S+trnL+ND1 | 2.00E-63 | 99.27% |

| | | | | | | | |
|---|---|---|---|---|---|---|---|
| 26H | 128 | | *Home sapiens* | OQ707209.1 | 10974-10847; ND4 | 1.00E-58 | 99.22% |
| 27H | 128 | | *Home sapiens* | LC733705.1 | 10847-10974; ND4 | 1.00E-58 | 99.22% |
| 28H | 127 | | *Home sapiens* | OR182493.1 | 11643-11769; ND4 | 5.00E-58 | 99.21% |
| 29H | 123 | | *Home sapiens* | OR182493.1 | 3674-3796; ND1 | 8.00E-56 | 99.19% |
| 30H | 118 | | *Home sapiens* | OR182493.1 | 3430-3547; ND1 | 5.00E-53 | 99.15% |
| 31H | 106 | | *Home sapiens* | OR182493.1 | 14567-14672; ND6 | 2.00E-46 | 99.06% |
| 32H | 199 | | *Home sapiens* | OR182493.1 | 15529-15727; CYTB | 4.00E-96 | 98.99% |
| 33H | 99 | | *Home sapiens* | OR182493.1 | 13226-13128; ND5 | 1.00E-42 | 98.99% |
| 34H | 83 | | *Home sapiens* | OR182493.1 | 11867-11785; ND4 | 8.00E-34 | 98.80% |
| 35H | 173 | | *Home sapiens* | OR182493.2 | 3320-3148; 16S+trnL+ND1 | 4.00E-80 | 98.27% |
| 36H | 200 | | *Home sapiens* | OR182493.1 | 12487-12288; trnL+ND5 | 2.00E-93 | 98.00% |
| 37H | 259 | | *Home sapiens* | OR182493.1 | 14124-14382; trnL+ND5 | 5.00E-121 | 97.30% |
| 38H | 143 | | *Home sapiens* | OR182493.1 | 13614-13754; ND5 | 2.00E-67 | 98.00% |
| 39H | 152 | | *Home sapiens* | MG660585.1 | 7949-8040/COX2: | 4.00E-65 | 96.71% |
| 40H | 225 | | *Home sapiens* | ON422324.1 | 8367-8589/ATP8-ATP6 | 5.00E-81 | 90.57% |
| 41H | 123 | | *Home sapiens* | OR182493.1 | 3674-3796/ND1 | 8.00E-56 | 99.19% |

Table S7. The 36 pDNA Sequences Aligning with Non-Homo Primate Nuclear Sequences

| pDNA | | Results of the Nucleotide Blast | | | |
|---|---|---|---|---|---|
| ID | Len | Taxo | Nucleotide | E-value | Affinity |
| 1P | 150 | *Carlito syrichta* | XM_021713054.1 | 3.00E-20 | 70.67% |
| 2P | 122 | *Sapajus apella* | XR_004259191.1 | 8.00E-41 | 92.62% |
| 3P | 63 | *Callithrix jacchus* | AC188585.3 | 5.00E-19 | 96.83% |
| 4P | 160 | *Sapajus apella* | XM_032273219.1 | 1.00E-47 | 84.12% |
| 5P | 164 | *Aotus nancymaae* | XM_012443881.1 | 1.00E-68 | 95.96% |
| 6P | 166 | *Macaca thibetana* | XM_050789932.1 | 4.00E-66 | 95.18% |
| 7P | 187 | *Macaca mulatta* | AC188451.1 | 1.00E-86 | 99.47% |
| 8P | 93 | *Macaca mulatta* | AC187507.2 | 1.00E-29 | 93.55% |
| 9P | 178 | *Macaca mulatta* | AC214608.1 | 3.00E-67 | 93.82% |
| 10P | 121 | *Macaca mulatta* | AC279264.1 | 4.00E-45 | 95.87% |
| 11P | 150 | *Macaca mulatta* | AC211796.4 | 3.00E-61 | 96.67% |
| 12P | 262 | *Macaca mulatta* | AC204076.9 | 1.00E-99 | 92.37% |
| 13P | 118 | *Macaca mulatta* | OM502028.1 | 3.00E-40 | 97.17% |
| 14P | 139 | *Macaca thibetana* | XM_050772658.1 | 2.00E-48 | 90.72% |
| 15P | 149 | *Macaca nemestrina* | XR_003018358.1 | 7.00E-56 | 94.63% |
| 16P | 118 | *Macaca mulatta* | AC280286.1 | 5.00E-43 | 95.00% |
| 17P | 196 | *Macaca mulatta* | AC197276.3 | 6.00E-71 | 91.33% |
| 18P | 140 | *Macaca mulatta* | AC202605.16 | 3.00E-53 | 95.00% |
| 19P | 250 | *Macaca mulatta* | AB128049.1 | 4.00E-94 | 91.67% |
| 20P | 121 | *Macaca nemestrina* | XR_996760.2 | 7.00E-42 | 92.24% |
| 21P | 71 | *Pan paniscus* | XM_034938079.2 | 1.00E-21 | 95.77% |
| 22P | 146 | *Pan paniscus* | AC280336.1 | 3.00E-66 | 100.00% |
| 23P | 141 | *Pan paniscus* | AC280342.1 | 2.00E-63 | 100.00% |
| 24P | 163 | *Pan paniscus* | XM_055115310.1 | 6.00E-70 | 97.55% |
| 25P | 148 | *Pan paniscus* | AC280341.1 | 3.00E-67 | 100.00% |
| 26P | 153 | *Pan paniscus* | XM_008974855.5 | 3.00E-67 | 98.71% |

| | | | | | | |
|---|---|---|---|---|---|---|
| 27P | 173 | | *Pan paniscus* | AC280344.1 | 4.00E-79 | 99.42% |
| 28P | 139 | | *Pan paniscus* | XM_034963660.2 | 1.00E-52 | 94.37% |
| 29P | 97 | | *Gorilla gorilla* | AC270505.1 | 3.00E-38 | 98.97% |
| 30P | 97 | | *Gorilla gorilla* | AC270505.1 | 5.00E-41 | 100.00% |
| 31P | 135 | | *Gorilla gorilla* | AC275387.1 | 3.00E-60 | 99.26% |
| 32P | 103 | | *Gorilla gorilla* | AC145355.3 | 2.00E-44 | 100.00% |
| 33P | 209 | | *Gorilla gorilla* | AC145852.3 | 3.00E-81 | 93.78% |
| 34P | 163 | | *Gorilla gorilla* | AC280153.1 | 9.00E-71 | 97.55% |
| 35P | 150 | | *Gorilla gorilla* | AC275724.1 | 6.00E-67 | 98.67% |
| 36P | 114 | | *Gorilla gorilla* | AC280262.1 | 1.00E-48 | 99.12% |

Table S8. The 271 pDNA Sequences Matching Pan troglodyte Nuclear Sequences

| pDNA | | Results of the Nucleotide Blast | | | |
|---|---|---|---|---|---|
| ID | Len | Taxo | Accession | E-Value | Affinity |
| 0 | 109 | *Pan troglodytes* | AC183379.3 | 1.00E-44 | 99.00% |
| 2T | 129 | *Pan troglodytes* | AC183666.2 | 2.00E-55 | 99.22% |
| 3T | 110 | *Pan troglodytes* | AC194549.3 | 7.00E-47 | 100.00% |
| 4T | 127 | *Pan troglodytes* | AC190233.5 | 5.00E-56 | 100.00% |
| 5T | 182 | *Pan troglodytes* | AC188551.3 | 3.00E-75 | 96.15% |
| 6T | 210 | *Pan troglodytes* | AC192764.3 | 1.00E-93 | 97.62% |
| 8T | 145 | *Pan troglodytes* | AC186180.3 | 3.00E-60 | 97.28% |
| 9T | 164 | *Pan troglodytes* | AC142339.1 | 3.00E-74 | 99.39% |
| 10T | 146 | *Pan troglodytes* | AC142339.1 | 2.00E-69 | 99.32% |
| 12T | 86 | *Pan troglodytes* | BS000075.1 | 5.00E-34 | 100.00% |
| 13T | 153 | *Pan troglodytes* | AC157217.2 | 3.00E-64 | 96.18% |
| 14T | 150 | *Pan troglodytes* | AC161061.2 | 1.00E-66 | 99.33% |
| 15T | 154 | *Pan troglodytes* | AC183961.2 | 2.00E-70 | 100.00% |
| 16T | 225 | *Pan troglodytes* | AC183961.2 | 4.00E-110 | 99.56% |
| 17T | 146 | *Pan troglodytes* | AC183768.2 | 9.00E-68 | 98.63% |
| 18T | 247 | *Pan troglodytes* | AC146269.3 | 2.00E-99 | 94.33% |
| 19T | 150 | *Pan troglodytes* | AC211714.3 | 2.00E-68 | 96.67% |
| 20T | 225 | *Pan troglodytes* | AC183684.3 | 5.00E-115 | 99.56% |
| 21T | 286 | *Pan troglodytes* | AC192151.3 | 7.00E-85 | 86.23% |
| 22T | 91 | *Pan troglodytes* | AC200162.4 | 2.00E-34 | 97.80% |
| 23T | 120 | *Pan troglodytes* | AC275774.1 | 1.00E-33 | 91.74% |
| 24T | 133 | *Pan troglodytes* | AC145864.2 | 4.00E-59 | 99.25% |
| 25T | 217 | *Pan troglodytes* | XM_016920966.2 | 4.00E-112 | 100.00% |
| 26T | 156 | *Pan troglodytes* | AC200710.3 | 1.00E-44 | 89.87% |
| 27T | 163 | *Pan troglodytes* | AC200710.3 | 1.00E-59 | 93.33% |

| | | | | | | |
|---|---|---|---|---|---|---|
| 28T | 171 | | *Pan troglodytes* | BS000063.1 | 6.00E-77 | 97.08% |
| 29T | 198 | | *Pan troglodytes* | AC188570.3 | 3.00E-88 | 97.51% |
| 30T | 191 | | *Pan troglodytes* | AC192066.3 | 3.00E-75 | 93.35% |
| 31T | 108 | | *Pan troglodytes* | AC192447.3 | 4.00E-44 | 99.07% |
| 32T | 150 | | *Pan troglodytes* | AC188569.4 | 2.00E-68 | 100.00% |
| 33T | 200 | | *Pan troglodytes* | AC279071.1 | 3.00E-88 | 97.04% |
| 34T | 200 | | *Pan troglodytes* | AC279071.1 | 1.00E-86 | 96.55% |
| 35T | 208 | | *Pan troglodytes* | AC279071.1 | 9.00E-87 | 96.55% |
| 36T | 101 | | *Pan troglodytes* | BS000130.1 | 1.00E-37 | 97.03% |
| 37T | 146 | | *Pan troglodytes* | AC190214.5 | 8.00E-61 | 96.69% |
| 38T | 148 | | *Pan troglodytes* | AC190214.5 | 5.00E-64 | 98.01% |
| 39T | 195 | | *Pan troglodytes* | AC193038.3 | 4.00E-90 | 98.46% |
| 40T | 163 | | *Pan troglodytes* | AC195663.3 | 2.00E-72 | 98.16% |
| 41T | 183 | | *Pan troglodytes* | AC186383.2 | 3.00E-81 | 97.81% |
| 42T | 101 | | *Pan troglodytes* | AC192189.4 | 6.00E-40 | 98.02% |
| 43T | 188 | | *Pan troglodytes* | AC188419.3 | 3.00E-81 | 97.34% |
| 44T | 98 | | *Pan troglodytes* | AC187371.2 | 9.00E-39 | 98.98% |
| 45T | 196 | | *Pan troglodytes* | AC146236.3 | 7.00E-89 | 98.47% |
| 46T | 166 | | *Pan troglodytes* | AC146236.3 | 5.00E-81 | 99.40% |
| 47T | 99 | | *Pan troglodytes* | AC187746.3 | 1.00E-35 | 90.00% |
| 48T | 211 | | *Pan troglodytes* | AC198057.3 | 2.00E-97 | 98.05% |
| 49T | 135 | | *Pan troglodytes* | AC190196.3 | 8.00E-46 | 80.00% |
| 50T | 146 | | *Pan troglodytes* | AC163767.4 | 9.00E-68 | 98.63% |
| 51T | 200 | | *Pan troglodytes* | BS000037.1 | 4.00E-95 | 99.01% |
| 52T | 210 | | *Pan troglodytes* | AC270591.1 | 9.00E-102 | 99.52% |
| 53T | 127 | | *Pan troglodytes* | AC189685.3 | 2.00E-54 | 99.21% |
| 54T | 150 | | *Pan troglodytes* | AC145994.2 | 3.00E-65 | 98.00% |
| 55T | 120 | | *Pan troglodytes* | AC157954.2 | 3.00E-56 | 100.00% |
| 56T | 101 | | *Pan troglodytes* | AC183764.3 | 2.00E-45 | 100.00% |
| 57T | 135 | | *Pan troglodytes* | AC197161.3 | 1.00E-59 | 97.78% |
| 58T | 188 | | *Pan troglodytes* | CT737396.3 | 8.00E-84 | 97.34% |
| 59T | 142 | | *Pan troglodytes* | AC192120.4 | 7.00E-65 | 99.30% |
| 60T | 189 | | *Pan troglodytes* | AC188680.4 | 5.00E-74 | 94.18% |
| 61T | 109 | | *Pan troglodytes* | AL593856.8 | 3.00E-44 | 98.18% |
| 63T | 94 | | *Pan troglodytes* | AC216085.3 | 2.00E-41 | 100.00% |
| 64T | 255 | | *Pan troglodytes* | AC216085.3 | 1.00E-120 | 98.82% |
| 65T | 182 | | *Pan troglodytes* | AC216085.3 | 6.00E-83 | 98.90% |
| 66T | 145 | | *Pan troglodytes* | AC216085.3 | 1.00E-65 | 100.00% |
| 67T | 238 | | *Pan troglodytes* | AC187370.3 | 3.00E-112 | 98.32% |
| 68T | 267 | | *Pan troglodytes* | AC192481.1 | 3.00E-108 | 94.96% |
| 69T | 87 | | *Pan troglodytes* | AC192937.3 | 1.00E-30 | 96.63% |
| 70T | 154 | | *Pan troglodytes* | AC192196.4 | 8.00E-71 | 99.35% |
| 71T | 148 | | *Pan troglodytes* | AC187381.2 | 1.00E-63 | 96.98% |
| 72T | 153 | | *Pan troglodytes* | AC186877.3 | 2.00E-71 | 99.00% |
| 73T | 164 | | *Pan troglodytes* | XR_008537664.1 | 4.00E-69 | 96.95% |
| 74T | 107 | | *Pan troglodytes* | AC190156.3 | 3.00E-43 | 98.15% |
| 75T | 88 | | *Pan troglodytes* | AC193863.2 | 2.00E-33 | 98.86% |
| 76T | 121 | | *Pan troglodytes* | AC193907.3 | 2.00E-47 | 97.52% |
| 77T | 112 | | *Pan troglodytes* | AC193928.3 | 3.00E-45 | 98.21% |
| 78T | 159 | | *Pan troglodytes* | AC183417.2 | 1.00E-71 | 99.37% |
| 79T | 173 | | *Pan troglodytes* | AC183417.2 | 5.00E-78 | 98.84% |
| 80T | 139 | | *Pan troglodytes* | AC188669.3 | 2.00E-62 | 100.00% |
| 81T | 164 | | *Pan troglodytes* | AC142309.1 | 1.00E-65 | 94.12% |
| 82T | 241 | | *Pan troglodytes* | AC113436.28 | 1.00E-106 | 97.10% |
| 83T | 238 | | *Pan troglodytes* | AC190258.1 | 3.00E-113 | 99.16% |
| 84T | 113 | | *Pan troglodytes* | AC200161.3 | 3.00E-39 | 98.02% |
| 85T | 181 | | *Pan troglodytes* | AC194556.3 | 2.00E-82 | 98.90% |
| 86T | 207 | | *Pan troglodytes* | AC190228.3 | 2.00E-97 | 99.52% |
| 87T | 183 | | *Pan troglodytes* | AC186385.2 | 2.00E-84 | 99.45% |
| 88T | 150 | | *Pan troglodytes* | AC186385.2 | 3.00E-61 | 96.67% |
| 89T | 201 | | *Pan troglodytes* | AC187722.3 | 4.00E-80 | 93.60% |
| 90T | 182 | | *Pan troglodytes* | AC160651.2 | 3.00E-90 | 99.45% |
| 91T | 150 | | *Pan troglodytes* | LO018685.1 | 6.00E-36 | 87.06% |
| 92T | 107 | | *Pan troglodytes* | AC187607.2 | 1.00E-43 | 99.07% |
| 93T | 101 | | *Pan troglodytes* | AC146004.3 | 5.00E-42 | 100.00% |
| 94T | 219 | | *Pan troglodytes* | AC192564.2 | 3.00E-100 | 98.17% |
| 95T | 135 | | *Pan troglodytes* | AC092859.26 | 2.00E-56 | 98.52% |
| 96T | 148 | | *Pan troglodytes* | AC192953.2 | 1.00E-58 | 93.59% |
| 97T | 150 | | *Pan troglodytes* | AC192953.2 | 3.00E-65 | 98.00% |
| 98T | 218 | | *Pan troglodytes* | AC191970.3 | 1.00E-67 | 84.94% |
| 99T | 110 | | *Pan troglodytes* | AC214852.3 | 3.00E-45 | 99.09% |
| 100T | 81 | | *Pan troglodytes* | AC190217.3 | 3.00E-31 | 100.00% |
| 101T | 150 | | *Pan troglodytes* | AP023480.1 | 9.00E-42 | 86.38% |
| 102T | 247 | | *Pan troglodytes* | AC270562.1 | 2.00E-109 | 97.17% |
| 103T | 157 | | *Pan troglodytes* | AC214854.3 | 2.00E-70 | 99.36% |
| 104T | 157 | | *Pan troglodytes* | AC200159.3 | 5.00E-68 | 98.68% |
| 105T | 116 | | *Pan troglodytes* | AC200159.3 | 3.00E-52 | 99.14% |
| 106T | 137 | | *Pan troglodytes* | AC196938.3 | 3.00E-59 | 99.28% |
| 107T | 82 | | *Pan troglodytes* | AC211978.4 | 2.00E-26 | 95.12% |
| 108T | 225 | | *Pan troglodytes* | AC211978.4 | 1.00E-111 | 100.00% |
| 109T | 153 | | *Pan troglodytes* | AC193935.3 | 6.00E-67 | 98.04% |
| 110T | 235 | | *Pan troglodytes* | AC161620.5 | 3.00E-117 | 100.00% |
| 111T | 191 | | *Pan troglodytes* | AC184799.2 | 1.00E-79 | 96.34% |
| 112T | 109 | | *Pan troglodytes* | AC192408.3 | 9.00E-76 | 100.00% |
| 113T | 102 | | *Pan troglodytes* | AC186892.2 | 3.00E-38 | 97.06% |
| 114T | 223 | | *Pan troglodytes* | AC149092.6 | 3.00E-87 | 94.17% |
| 115T | 96 | | *Pan troglodytes* | AC149092.6 | 2.00E-39 | 100.00% |
| 116T | 97 | | *Pan troglodytes* | AC190234.6 | 7.00E-40 | 100.00% |
| 117T | 190 | | *Pan troglodytes* | AC186045.5 | 1.00E-90 | 99.47% |
| 118T | 189 | | *Pan troglodytes* | AC193895.4 | 9.00E-88 | 99.47% |
| 119T | 249 | | *Pan troglodytes* | AC145375.5 | 7.00E-116 | 98.02% |
| 120T | 72 | | *Pan troglodytes* | AC186387.2 | 2.00E-18 | 91.67% |
| 121T | 152 | | *Pan troglodytes* | AC190249.3 | 8.00E-68 | 99.34% |
| 122T | 138 | | *Pan troglodytes* | NG_013342.1 | 3.00E-59 | 98.55% |
| 123T | 123 | | *Pan troglodytes* | AC183376.4 | 3.00E-52 | 99.19% |
| 124T | 258 | | *Pan troglodytes* | AC146396.5 | 7.00E-116 | 96.58% |
| 125T | 172 | | *Pan troglodytes* | AC068196.4 | 7.00E-76 | 97.18% |
| 126T | 104 | | *Pan troglodytes* | AC192725.2 | 1.00E-41 | 100.00% |
| 127T | 186 | | *Pan troglodytes* | AC192725.2 | 4.00E-90 | 100.00% |
| 128T | 150 | | *Pan troglodytes* | CP068263.2 | 2.00E-53 | 93.88% |
| 129T | 115 | | *Pan troglodytes* | AC145378.2 | 3.00E-46 | 99.13% |
| 130T | 83 | | *Pan troglodytes* | AC145378.2 | 5.00E-30 | 97.59% |
| 131T | 253 | | *Pan troglodytes* | AC188550.3 | 7.00E-111 | 97.23% |
| 132T | 152 | | *Pan troglodytes* | AC196251.3 | 1.00E-66 | 98.68% |
| 133T | 173 | | *Pan troglodytes* | AC196251.3 | 3.00E-80 | 99.00% |
| 134T | 124 | | *Pan troglodytes* | AC147033.2 | 2.00E-48 | 96.06% |
| 135T | 238 | | *Pan troglodytes* | AC147978.3 | 4.00E-112 | 99.16% |
| 136T | 202 | | *Pan troglodytes* | AC198619.4 | 8.00E-95 | 99.50% |
| 137T | 171 | | *Pan troglodytes* | XM_024357983.2 | 2.00E-73 | 97.08% |
| 138T | 150 | | *Pan troglodytes* | CU464036.1 | 2.00E-68 | 100.00% |
| 139T | 168 | | *Pan troglodytes* | AC193856.2 | 2.00E-78 | 99.40% |
| 140T | 225 | | *Pan troglodytes* | AC146171.2 | 3.00E-75 | 87.56% |
| 141T | 156 | | *Pan troglodytes* | AC220072.5 | 1.00E-71 | 100.00% |
| 142T | 116 | | *Pan troglodytes* | AC193260.2 | 2.00E-46 | 97.41% |
| 143T | 236 | | *Pan troglodytes* | AC193885.4 | 2.00E-104 | 96.61% |
| 144T | 206 | | *Pan troglodytes* | XR_002942333.2 | 3.00E-96 | 98.55% |
| 145T | 146 | | *Pan troglodytes* | AC200342.3 | 5.00E-68 | 100.00% |
| 146T | 138 | | *Pan troglodytes* | AC172375.3 | 3.00E-49 | 94.20% |
| 147T | 93 | | *Pan troglodytes* | BS000211.1 | 7.00E-39 | 100.00% |
| 148T | 203 | | *Pan troglodytes* | AC006582.13 | 2.00E-69 | 86.60% |
| 149T | 135 | | *Pan troglodytes* | AC200459.2 | 1.00E-57 | 98.52% |
| 150T | 288 | | *Pan troglodytes* | AC190143.3 | 2.00E-135 | 97.92% |
| 151T | 124 | | *Pan troglodytes* | AC186185.2 | 7.00E-48 | 90.00% |
| 152T | 253 | | *Pan troglodytes* | AC186185.2 | 3.00E-120 | 98.21% |
| 153T | 150 | | *Pan troglodytes* | AC186185.2 | 2.00E-68 | 100.00% |
| 154T | 99 | | *Pan troglodytes* | AC186185.2 | 3.00E-39 | 98.99% |
| 155T | 210 | | *Pan troglodytes* | AC183773.2 | 2.00E-89 | 94.44% |
| 156T | 81 | | *Pan troglodytes* | AC217027.3 | 3.00E-30 | 97.00% |
| 157T | 125 | | *Pan troglodytes* | NG_009388.1 | 1.00E-44 | 94.09% |
| 158T | 126 | | *Pan troglodytes* | AC229904.2 | 2.00E-54 | 97.62% |
| 159T | 83 | | *Pan troglodytes* | AC147079.4 | 1.00E-31 | 97.59% |
| 160T | 221 | | *Pan troglodytes* | AC213062.4 | 6.00E-104 | 98.64% |
| 161T | 212 | | *Pan troglodytes* | AC192213.3 | 2.00E-99 | 98.58% |
| 162T | 146 | | *Pan troglodytes* | AC183271.3 | 2.00E-66 | 99.32% |
| 163T | 264 | | *Pan troglodytes* | AC159827.1 | 2.00E-128 | 98.86% |
| 164T | 156 | | *Pan troglodytes* | AC159827.1 | 1.00E-68 | 98.08% |
| 165T | 256 | | *Pan troglodytes* | AC182427.3 | 3.00E-112 | 95.51% |
| 166T | 110 | | *Pan troglodytes* | AC092033.4 | 3.00E-43 | 97.30% |
| 167T | 100 | | *Pan troglodytes* | AC192124.3 | 5.00E-41 | 99.00% |
| 168T | 179 | | *Pan troglodytes* | AC161386.4 | 1.00E-75 | 96.65% |
| 169T | 171 | | *Pan troglodytes* | AC188559.3 | 2.00E-78 | 98.83% |
| 170T | 158 | | *Pan troglodytes* | AC147058.3 | 1.00E-54 | 93.04% |
| 171T | 125 | | *Pan troglodytes* | AC191700.3 | 8.00E-55 | 99.20% |
| 172T | 226 | | *Pan troglodytes* | AC192996.2 | 2.00E-104 | 98.23% |
| 173T | 146 | | *Pan troglodytes* | XR_001716146.2 | 2.00E-66 | 99.32% |
| 174T | 214 | | *Pan troglodytes* | XM_055114440.2 | 3.00E-98 | 96.73% |
| 175T | 237 | | *Pan troglodytes* | AC146059.4 | 1.00E-115 | 99.16% |
| 176T | 98 | | *Pan troglodytes* | AC189686.4 | 1.00E-41 | 100.00% |
| 177T | 140 | | *Pan troglodytes* | AC192800.2 | 1.00E-59 | 97.86% |
| 178T | 154 | | *Pan troglodytes* | CT989257.3 | 8.00E-71 | 99.35% |
| 179T | 223 | | *Pan troglodytes* | AC275882.1 | 1.00E-95 | 96.41% |
| 180T | 239 | | *Pan troglodytes* | AC183992.2 | 2.00E-69 | 87.45% |
| 181T | 251 | | *Pan troglodytes* | XM_054669499.1 | 4.00E-126 | 100.00% |
| 182T | 167 | | *Pan troglodytes* | AC146242.3 | 1.00E-49 | 74.43% |
| 183T | 150 | | *Pan troglodytes* | AC146384.4 | 3.00E-65 | 98.00% |
| 184T | 74 | | *Pan troglodytes* | AC213065.4 | 2.00E-28 | 100.00% |
| 185T | 107 | | *Pan troglodytes* | AC213065.4 | 2.00E-45 | 98.13% |
| 186T | 161 | | *Pan troglodytes* | AC159020.3 | 2.00E-76 | 100.00% |
| 187T | 150 | | *Pan troglodytes* | AC187623.2 | 6.00E-67 | 98.67% |
| 188T | 228 | | *Pan troglodytes* | AC187732.3 | 1.00E-111 | 99.56% |
| 189T | 56 | | *Pan troglodytes* | AC188437.4 | 1.00E-18 | 100.00% |
| 190T | 105 | | *Pan troglodytes* | AC186388.3 | 2.00E-45 | 100.00% |
| 191T | 135 | | *Pan troglodytes* | AC196245.2 | 4.00E-58 | 98.52% |
| 192T | 117 | | *Pan troglodytes* | AC146132.3 | 5.00E-52 | 100.00% |
| 193T | 140 | | *Pan troglodytes* | XM_016934978.2 | 7.00E-61 | 96.60% |
| 194T | 151 | | *Pan troglodytes* | AC190144.3 | 2.00E-62 | 96.71% |
| 195T | 150 | | *Pan troglodytes* | AC194763.3 | 3.00E-70 | 100.00% |
| 196T | 112 | | *Pan troglodytes* | AC160143.4 | 4.00E-18 | 62.81% |
| 197T | 244 | | *Pan troglodytes* | AC191942.3 | 2.00E-89 | 93.03% |
| 198T | 69 | | *Pan troglodytes* | AC148317.3 | 3.00E-23 | 98.55% |
| 199T | 131 | | *Pan troglodytes* | AC202278.2 | 3.00E-47 | 93.89% |
| 200T | 97 | | *Pan troglodytes* | AC188802.3 | 5.00E-41 | 100.00% |
| 201T | 122 | | *Pan troglodytes* | AC198462.3 | 1.00E-43 | 93.44% |
| 202T | 198 | | *Pan troglodytes* | AC113435.14 | 1.00E-73 | 92.42% |
| 203T | 200 | | *Pan troglodytes* | AC279053.1 | 1.00E-30 | 77.76% |
| 204T | 150 | | *Pan troglodytes* | AC199715.3 | 1.00E-68 | 99.33% |
| 205T | 156 | | *Pan troglodytes* | AC192734.2 | 2.00E-63 | 96.20% |
| 206T | 195 | | *Pan troglodytes* | AC192834.3 | 8.00E-92 | 98.97% |
| 207T | 131 | | *Pan troglodytes* | AC199235.3 | 8.00E-57 | 98.48% |
| 208T | 193 | | *Pan troglodytes* | AC190232.2 | 8.00E-95 | 98.96% |
| 209T | 98 | | *Pan troglodytes* | AC275872.1 | 1.00E-43 | 100.00% |
| 210T | 145 | | *Pan troglodytes* | AC182425.3 | 5.00E-55 | 93.79% |
| 211T | 242 | | *Pan troglodytes* | AC142323.1 | 1.00E-126 | 100.00% |
| 212T | 249 | | *Pan troglodytes* | AC270583.1 | 3E-120 | 99.60% |
| 213T | 189 | | *Pan troglodytes* | AC192566.2 | 4E-90 | 99.47% |
| 214T | 201 | | *Pan troglodytes* | AC183962.3 | 9E-80 | 91.64% |
| 215T | 201 | | *Pan troglodytes* | AC183962.3 | 9E-87 | 96.52% |
| 216T | 95 | | *Pan troglodytes* | AC190197.3 | 8E-39 | 100.00% |
| 217T | 114 | | *Pan troglodytes* | AC149231.2 | 2E-50 | 100.00% |
| 218T | 153 | | *Pan troglodytes* | AC280209.1 | 1E-47 | 89.54% |
| 219T | 124 | | *Pan troglodytes* | AC182393.2 | 4E-48 | 96.77% |
| 220T | 117 | | *Pan troglodytes* | AC183622.3 | 8.00E-53 | 99.15% |
| 221T | 99 | | *Pan troglodytes* | AC192641.3 | 2.00E-42 | 98.99% |
| 222T | 221 | | *Pan troglodytes* | AC188594.3 | 1.00E-108 | 99.10% |
| 223T | 121 | | *Pan troglodytes* | AL954246.1 | 2.00E-53 | 98.35% |
| 224T | 150 | | *Pan troglodytes* | AC184876.2 | 1.00E-54 | 89.01% |
| 225T | 195 | | *Pan troglodytes* | AC223435.5 | 7.00E-88 | 97.44% |
| 226T | 114 | | *Pan troglodytes* | AC185967.2 | 2.00E-45 | 96.61% |
| 227T | 91 | | *Pan troglodytes* | NG_046914.3 | 3.00E-36 | 97.80% |
| 228T | 222 | | *Pan troglodytes* | AC226173.4 | 2.00E-97 | 96.85% |
| 229T | 150 | | *Pan troglodytes* | AC226173.4 | 2.00E-62 | 96.10% |
| 230T | 125 | | *Pan troglodytes* | AC161276.2 | 6.00E-51 | 97.60% |
| 231T | 148 | | *Pan troglodytes* | AC145882.3 | 2.00E-46 | 87.19% |
| 232T | 217 | | *Pan troglodytes* | AC142289.1 | 3.00E-102 | 98.62% |
| 233T | 184 | | *Pan troglodytes* | AC160566.2 | 1.00E-89 | 98.91% |
| 234T | 104 | | *Pan troglodytes* | CR937045.1 | 1.00E-43 | 98.08% |
| 235T | 140 | | *Pan troglodytes* | AC194569.3 | 2.00E-58 | 92.85% |
| 236T | 150 | | *Pan troglodytes* | CR937026.1 | 8.00E-72 | 99.33% |
| 237T | 226 | | *Pan troglodytes* | AC188603.5 | 7.00E-114 | 99.12% |
| 238T | 272 | | *Pan troglodytes* | AC201862.3 | 9.00E-133 | 98.90% |
| 239T | 194 | | *Pan troglodytes* | AC187629.3 | 7.00E-73 | 93.30% |
| 240T | 109 | | *Pan troglodytes* | AC190218.2 | 1.00E-47 | 100.00% |
| 241T | 151 | | *Pan troglodytes* | BS000098.1 | 3.00E-67 | 99.34% |
| 242T | 161 | | *Pan troglodytes* | AC195070.2 | 2.00E-76 | 100.00% |
| 243T | 147 | | *Pan troglodytes* | AC270731.1 | 9.00E-68 | 99.32% |
| 244T | 113 | | *Pan troglodytes* | AC279127.1 | 6.00E-46 | 98.23% |
| 245T | 113 | | *Pan troglodytes* | AC204684.4 | 3.00E-52 | 100.00% |
| 246T | 145 | | *Pan troglodytes* | AC147078.2 | 6.00E-63 | 98.62% |
| 247T | 205 | | *Pan troglodytes* | XR_001714532.2 | 2.00E-95 | 99.02% |
| 248T | 226 | | *Pan troglodytes* | AC194989.3 | 3.00E-94 | 94.69% |
| 249T | 246 | | *Pan troglodytes* | AC198805.2 | 7.00E-116 | 98.78% |
| 250T | 104 | | *Pan troglodytes* | AC188667.3 | 1.00E-43 | 100.00% |
| 251T | 172 | | *Pan troglodytes* | AC186553.1 | 3.00E-80 | 100.00% |
| 252T | 167 | | *Pan troglodytes* | AC152353.1 | 3.00E-68 | 92.21% |
| 253T | 231 | | *Pan troglodytes* | AC191699.3 | 9.00E-89 | 92.64% |
| 254T | 177 | | *Pan troglodytes* | AC184055.3 | 6.00E-77 | 97.75% |
| 255T | 197 | | *Pan troglodytes* | NG_029917.1 | 2.00E-63 | 88.34% |
| 256T | 150 | | *Pan troglodytes* | AC197160.3 | 5.00E-58 | 95.33% |
| 257T | 225 | | *Pan troglodytes* | AC158369.5 | 2.00E-104 | 98.67% |
| 258T | 158 | | *Pan troglodytes* | AC192128.3 | 5.00E-71 | 99.37% |
| 259T | 98 | | *Pan troglodytes* | AC190422.2 | 1.00E-37 | 98.96% |
| 260T | 135 | | *Pan troglodytes* | AL954234.1 | 4.00E-58 | 98.26% |
| 261T | 205 | | *Pan troglodytes* | AC214850.4 | 2.00E-95 | 99.02% |
| 262T | 223 | | *Pan troglodytes* | AC197403.3 | 3.00E-102 | 97.76% |
| 263T | 54 | | *Pan troglodytes* | AC275851.1 | 5.00E-17 | 100.00% |
| 264T | 282 | | *Pan troglodytes* | XM_054358264.1 | 3.00E-143 | 100.00% |
| 265T | 255 | | *Pan troglodytes* | AC192167.3 | 4.00E-61 | 77.39% |
| 266T | 159 | | *Pan troglodytes* | AC183378.2 | 8.00E-65 | 95.24% |
| 267T | 205 | | *Pan troglodytes* | AC188675.3 | 2.00E-81 | 94.15% |
| 268T | 100 | | *Pan troglodytes* | AC192137.3 | 7.00E-40 | 99.00% |
| 270T | 125 | | *Pan troglodytes* | AC198664.3 | 3.00E-53 | 99.20% |
| 271T | 285 | | *Pan troglodytes* | AC163769.4 | 9.00E-122 | 91.50% |
| 272T | 197 | | *Pan troglodytes* | AC163769.4 | 9.00E-122 | 91.50% |
| 273T | 144 | | *Pan troglodytes* | XM_054677932.1 | 3.00E-55 | 95.14% |
| 274T | 176 | | *Pan troglodytes* | XM_001155344.5 | 1.00E-75 | 97.16% |
| 275T | 239 | | *Pan troglodytes* | AC183528.2 | 4.00E-106 | 96.09% |

Table S9. Quantification of DNA Extraction for Petroleum Samples

| Bench number | Sample ID | Extraction phase of petroleum oil well | Oil volume to DNA extraction | $OD_{260}$ | pDNAmix （ng） |
|---|---|---|---|---|---|
| 1 | NY201905 | First oil recovery | 120 mL | 0.082 | 205 |
| 2 | | | 120 mL | 0.103 | 257.5 |
| 3 | | | 120 mL | 0.115 | 287.5 |
| 4 | | | 120 mL | 0.087 | 217.5 |
| 5 | | | 120 mL | 0.113 | 282.5 |
| 6 | NY202102 | Secondary oil recovery (using hot water and surfactants) | 120 mL | 0.051 | 127.5 |
| 7 | | | 120 mL | 0.215 | 537.5 |
| 8 | | | 120 mL | 0.093 | 232.5 |
| 9 | | | 120 mL | 0.089 | 222.5 |
| 10 | | | 120 mL | 0.067 | 167.5 |

Table S10. Results of DNA Extraction from Petroleum and Illumina NovaSeq Sequencing

| Bench ID | Sample ID | Extraction phase of petroleum oil well | Oil volume to DNA extraction | $OD_{260}$ | pDNAmix (ng) | Total reads amount | Sequences from non-indigenous species |
|---|---|---|---|---|---|---|---|
| 1 | NY201905 | First oil extraction | 720 mL | 0.618 | 3708 | 3,159,002 | Multiple |
| 2 | NY202102 | Secondary oil recovery (using hot water and surfactants) | 720 mL | 0.675 | 4050 | 390,932 | None |

Table S11. The Length of pDNAs and Blast Results

| Taxo | Minimum (bp) | Maximum (bp) | Mean | SD | Number |
|---|---|---|---|---|---|
| *Bacteria* | 50 | 290 | 159.73 | 48.12 | 1620178 |
| *Homo sapiens* | 50 | 290 | 150.55 | 46.05 | 1047532 |
| *Fungi* | 50 | 290 | 152.24453 | 48.2 | 121735 |
| *Virus* | 51 | 290 | 150.3744 | 22.26 | 16290 |
| *Primate** | 50 | 290 | 130.64 | 43.00 | 35108 |
| *Mammalia** | 50 | 290 | 129.24 | 48.69 | 4785 |
| *Actinopterygii* | 50 | 290 | 147.76691 | 41.01 | 4835 |
| *Insecta* | 50 | 290 | 161.15654 | 49.99 | 2945 |
| *Archaea* | 53 | 288 | 154.88 | 47.67 | 851 |
| *Arthropoda** | 50 | 285 | 139.90 | 43.87 | 1094 |
| *Spermatophyta* | 50 | 290 | 151.97439 | 45.78 | 15188 |
| *Aves* | 51 | 288 | 149.28199 | 10.42 | 9252 |
| *DAASIA* | 55 | 248 | 167.32 | 46.30 | 6988 |
| *Viridiplantae** | 60 | 276 | 156.03 | 48.86 | 117 |
| *sauria* | 61 | 288 | 157.40476 | 62.39 | 42 |
| *amphibia* | 95 | 181 | 140.75 | 38.35 | 4 |
| *Cyclostomata* | 83 | 272 | 181.50 | 80.47 | 7 |
| *Chondrichthyes* | 78 | 221 | 159.50 | 63.61 | 4 |
| *ABPNNR* | 59 | 267 | 187.17 | 88.41 | 752 |
| *PCM* | 68 | 298 | 197.20 | 85.64 | 187 |
| *EHT* | 77 | 219 | 186.34 | 92.41 | 49 |
| *HRC* | 61 | 247 | 163.72 | 80.79 | 15 |
| *Unnamed* | 88 | 296 | 157.26 | 49.87 | 271062 |
| Total | | | | | 3159020 |

Please note that the data in Viridiplantae* does not include Spermatophyta, the data in Arthropoda* excludes Insecta, the data in Sauria* does not include Aves, the data in Primates* does not include Homo sapiens, and the data in Mammalia* does not include Primates*.

Table S12. Appendix table of the pDNA sequences used in this research

| ID | Bench ID | Sequences |
|---|---|---|
| 1T | A00583:504:HM2HWDSXY:2:2677:15781:24095 | CATCCTGGCTAACATGGTGAAACTCTGACTCTACTAAAAATACCAAAAATTAGCCAGGCGTGGTGGCAGGTGCCTGTAATCCCAGCTACTCGGGAGGCTGAGGCAGGAG |
| 2T | A00583:504:HM2HWDSXY:2:2673:19443:6825 | CACTGCACCCTAGCCTGGGCAACAGAGCAAGACTCCATCTCAAGAAAAAAGAAAAAAAAAGAAAAGAAAAGGAAAAAGAGAAGAGAAGAGAAAAGAAAAGAATAAAGAAAAGAGGCCTTTAGGAAATCC |
| 7C | A00583:504:HM2HWDSXY:2:2672:18900:12555 | CACACCCACTGCACTCCAGCCTGGTAACAGAGCGATACTCCGTCACCAAAAAAAAAAAAAAAAAAAAAAAGGAAAAGGAAATATCACCAGAACGGATATCACTGCACTCTTCACGAAAAAGAAAAAGAAAATACCAGA |
| 28P | A00583:504:HM2HWDSXY:2:2672:18900:12555 | ACACACCCACTGCACTCCAGCCTGGTAACAGAGCGATACTCCGTCACCAAAAAAAAAAAAAAAAAAAAAAAAGGAAAAGGAAATATCACCAGAACGGATATCACTGCACTCTTCACGAAAAAGAAAAAGAAAATACCAGA |
| 3T | A00583:504:HM2HWDSXY:2:2672:14823:26694 | GGGTGACTTCGTTTTGATCAAAAATCCAAGGAAACAGAGGACTTGTTGGGTTGTGGGTTTCCTGATGGCTTGGTGCCAGCTATGGATTTAGTTGATAAGGCCCAGTTTCT |
| 14A | A00583:504:HM2HWDSXY:2:2667:23845:4460 | CCAATGCAGGACAATCAAGCTGCGTAATGCAAGCAGGAGAGCAGCACTACGCCCAATAATGGAAGAATGATGAACAAGGCAGAGCCATTTCTCTGCTCTTCCTGCAAGGCATGTACTCACTCTGCTTTCATCTTTTCTAATGGAAAAGGAAGCA |
| 4T | A00583:504:HM2HWDSXY:2:2664:25147:4210 | CTCCAAAGCTAACTTCAGATAGTGTCAGAATAGGATTGAGTTGTTGTACACCCAGTTGGTGTCCTAAGAAAGGGTACCCCACATTTGGTGTTAGAAGTAACGTCAGAAAGCATCACACAATTGTATG |
| 25H | A00583:504:HM2HWDSXY:2:2660:32199:32581 | CAGGGTTTGTTAAGATGGCAGAGCCCGGTAATCGCATAAAACTTAAAACTTTACAGTCAGAGGTTCAATTCCTCTTCTTAACAACATAACCATGGCCAACCTCCTACTCCTCATTGTACCCATTCTAATCGCAATGG |
| 6M | A00583:504:HM2HWDSXY:2:2651:30309:23422 | ACACACACACACACACAGAAGGGTAGAGCCCCAGGTGACCATGGTCTTCTCAGTTCACATGAAAGAGAATTAGAGAGAGCGGGAGGGGGGGAAGCAAAATATGGACCCATTGTGGAGTAAAGTTTTACAAACATGGACTGGATTCAAATAGAATGCAATAAGAGGAGAAATGAAAGAAATAATAGAAAAAACCTAAAAAATAAACTAAAAAATAAATTTGAATTTCTACCATTTATAGGCCAATTTATACTTC |
| 5T | A00583:504:HM2HWDSXY:2:2645:31747:28354 | GGGGCGGTTTCCCCCATGCTATTCTCATGATAGTGAGTGAGTTCTCACGAGAGCTGATGGTTTTAAAGTGTTGTACTTTCTGGTGCTCTATCTCTCTCTCTTGCCTGCCCCATGTAAGAATTGCCTTGCTTCCCCTTCACCTTCCACAATGATTGTAAGTTTCCTGATGACTCCCCAGCCAT |
| 6T | A00583:504:HM2HWDSXY:2:2643:12174:30718 | TTATGTTTAACTGATATTTGGGGATGATTAAAATACAAAGCAATACGTTTAAAAGTTTTATCTGTATTTAAAAGTCTTAGTATTTAGAATAAATGTATTCTGTGTGTTGCCTAAAGGCATAAATCAAACTGTACATGTGTATATAGATACATAGATATTTTCAAGCGACAACTTGATTGATTTGTTACAGTTGCCTACAATGATGTATAT |
| 7T | A00583:504:HM2HWDSXY:2:2641:19678:29496 | AAGGGGTGCTGGAAACAAAGCAAACCATTTTCTCATTCTTATACAGAGGTACAAATGGGAGGGACATGTGGACTGCTGCCTGCATCACTAGTCACTGTGCCATGGAGTAGGCCCACAGGAGTAGGAGGGGTCCTGAGATGGGAAATCAACATGACCATGAAGAAGGAAGGCTTTGGAGTGATGTCATCCAACTTGGTTCTCA |
| 21P | A00583:504:HM2HWDSXY:2:2638:18023:8218 | CCACCCACTCACGCATCCACCATTATCTACCGTCCAGATACCCAGTGTTTCGCAGTGATTAATATACATAT |
| 8G | A00583:504:HM2HWDSXY:2:2631:1931:27477 | TTATTTTACCAGAAGGTTTTGAATTAGCTCCTTCTGATCGAATTCCTCCTGAAATGAAAGAAAAAATTGGTAATCTTTTTTTTCAACCTTATAGTAATGATAAAAAAAAATATTTTAGTAATATGTCCAGTTCCAGGAAAAAAATATAGTGAAATG |
| 8T | A00583:504:HM2HWDSXY:2:2628:3296:25488 | TCTCGTTTATTAGTTCTAGTAGTTTTTTTTTTTGTAGATTATTTGGACTATTTTTTAGAGATCATCTCATTGTGTGGTAAAGACAGTTTTACCTACAACAAACCCCCATGACACAAGTTTACCTATGTAATAAACCTGCACATGT |
| 9T | A00583:504:HM2HWDSXY:2:2627:10113:1344 | GTCTGGTTGTGTACTCTGGTGTACATGTAATGTTTGGTGGAAATTGTATTGTTTATATTCTAGGACATGTTTCTCTACTCACCTCTCTCATCCCACAAATTGCCTTTCCAGTGATTAGCATGTAATAAAGGATTTATTCATTTGACATGTATTTGTTGAGTGTT |
| 11T | A00583:504:HM2HWDSXY:2:2625:28302:10488 | AACACACACTCACATTCACACACACCACACACATTCACACACCACACATTCACACACCACACATTCACACACACCAACACACTCACACACACACCACACACATTCACACACACACCCCACCTGCTTAAAATCCTCCAAAGGCTTCTCATTATGCAAAAACCTTCTG |
| 12T | A00583:504:HM2HWDSXY:2:2622:15456:1548 | TGTTCCACTAATTCTATTATTTTGGCAACGATCTCTGACACAATAAAAAAATGAACAGAAAGATCAAATTTGTGATCTTTAACATT |
| 13T | A00583:504:HM2HWDSXY:2:2621:4752:20337 | GACCTGCCTGGGCTCAGGTGATCCTCCCACCTCAGCCTCCTGAATAGTTGGGACTACAGGTGTGCACCACCACACCTAATTTTTTTTTTTTTTTTTTTTTTTTTTTTTAGAGATGAGTTTTAACCGTGTTGCCCAGGCTGGTCTTGAACTCCT |
| 14T | A00583:504:HM2HWDSXY:2:2614:18855:35556 | TGTCCCCAGCCAGGTTGCAGGTGTGACCAGGCCTGGCCAGACTCTTGGGTTTTGCCCCTTGGACCAATCAATCCTGGAGGTCAATTTGAAGGGGGTAGCAGCCATGTAAGGGAAAATACCCTGGTAAAGCTACACTCTCTATGAAATGGC |
| 15T | A00583:504:HM2HWDSXY:2:2608:23050:6621 | ATGTGCAAAGACTTGGCAGACACAGCCATTGAGCCAAAGGCTATTTTCCCCAAAGATGAATTCTTGTACTGAAAGCCCTTCATAAGGCAAGGAATAGCAGCAGGCATTTACATAAAGCGAATAAACTGGCACTGTTAAGTTCCAGGACACGTCC |
| 9A | A00583:504:HM2HWDSXY:2:2605:23863:4836 | TGTATAGCAGAATCTGTAGCCATAGACTGCTGCTAGCTGCCATATCTGCAATAAACAAACTGACACACATAACTCCTTCCATTAAGACAGATTTGAATGTCCATAGATTAAACCTGCAAAAATTTGAGGGGCATGGGTGAATTTTACATCC |
| 17T | A00583:504:HM2HWDSXY:2:2603:3748:18004 | GTAATGTTCCCCAAAACAAAGAAAAATGTTTCAAATAGTGGATATCCCAATAACCCTGATTTGATCATTGTACATTGTATACATGTATCAAAATATCACAAATGCCCCAAAATACATACAACTATAATAAATTGATTTCAAAATCT |
| 18T | A00583:504:HM2HWDSXY:2:2602:21070:36166 | TTCTAACATGAGAGTCATAATATCTGTTCTCTGTTTCCCATTGAATCCAGAAACTGCACATTAAATTCCACAAAAATTTGTATTAAAGATCTCTCACAATTTTGAGGTAAAGCAACGACATCTCTCATCACTCTATCCCAGTATCCCAGTTTTTATTTTTTTTTTTTTTTTTTTTTTTTTTTTTTTTTAGACAGAGTCTCGCTCTGTCGCCCAGGCTAGGGTACAGTGGCGCAATCTTGAGTCACTGC |
| 32H | A00583:504:HM2HWDSXY:2:2577:21504:30060 | AGCCAACCCCTTAAACACCCCTCCCCACATCAATCCCGAATGAAATTTCCTATTCGCCTACACAATTCTCCGATCCGTCCCTAACAAACTAGGAGGCGTCCTTGCCCTATTACTATCCATCCTCATCCTAGCAATAATCCCCATCCTCCATATATCCAAACAACAAAGCATAATATTTCGCCCACTAAGCCAATCACTT |
| 2A | A00583:504:HM2HWDSXY:2:2575:12310:12007 | CAAGGCAGTGATGGCGCCCTTGTCTTGGGGCTGGGGTCCTCGATTTTCTCTTGAAACCCAGTGCCATCTGCGAGTCCCTAGACGCAGCACTCATACAACCACGAATGGCATGCATGCGTTGTCCAGGCTTGTCGCCAAGAACGACACGG |
| 19T | A00583:504:HM2HWDSXY:2:2573:9724:28667 | TCCCTTCCCCTCTCCAGGGACACAATCAGCCTCTGGCTTCAGTCTTGCTACTTCCTTCGCTTGGAAAGTTCTTACCCAAGAGGGCTCCATTCTACCTTTTTTTTTTTTTTTTTTTTTTTAGACAAGGTCTTACTCTGTCACCCAGGCATG |
| 20T | A00583:504:HM2HWDSXY:2:2570:32497:34256 | ACTTGGCAGACACAGCCATTGAGCCAAAGGCTATTTTCCCCAAAGATGAATTCTTGTACTGAAAGCCCTTCATAAGGCAAGGAATAGCAGCAGGCATTTACATAAAGCGAATAAACTGGCACTGTTAAGTTCCAGGACACGTCCTGGGGCTTAGCCTCCTTTCATTTATTCAGCCTAATTCCCACCGGTGACAGCTACCACCATGTTTAGGGCTGTGATGCCAGC |
| 21T | A00583:504:HM2HWDSXY:2:2567:27760:22608 | CCAAGAAGGAGCCACATTTTGCTCCTAAGTATTGTGATTCCCAGTGAATCCTGCCCACACACCTAGGCGTGGCATGGCCCAGAGGCTGCGTGTCTCCATGCCTGTGCAGATCAAGGCTATCTGGCCAGGTCCAGCTCAAAGGCCTTCTTGGTTGATTCTTTGTTTGAGTTTTTGTTGTTCTCGGATTTTTTTTTCAGACGGAGTCTCCTTCTGTTCCCCAGGCTGGAGTGCAGTGGCACAATCTCGGCTCACCACAACCTCTGCTTCCCGGGTTCAAGTGATTCTC |
| 35H | A00583:504:HM2HWDSXY:2:2566:7428:29575 | GGTATGAAGTTAAGAAGAGGAATTGAACCTCTGACTGTAAAGTTTTAAGTTTTATGCGATTACCGGGCTCTGCCATCTTAACAAACCCTGTTATTGGGTGGGTGTGGGTATAATACTAAGTTGAGATGATATCATTTACGGGGGAAGGCGCTTTGTGAAGTAGGCCTTATTTC |
| 22T | A00583:504:HM2HWDSXY:2:2562:15537:19695 | GATAACCAGCTTTAGGCTTTGTTCATTGTTTTAAATTTATCTTTATTTTCTATTTCATTAATATCTGATATCATACTTATAGCTTCTATCT |
| 23T | A00583:504:HM2HWDSXY:2:2561:30337:4993 | GCCTGGAACAGAACCCACTCCCCCAGGTGAGCATCTGACAGTCTGGAACAGCACTCTGCACCCACATGTGAGCATCTGATAGCCTGGAACAGACCCCACACCCCCAGGTGAGCATCTGAC |
| 10T | A00583:504:HM2HWDSXY:2:2561:17761:20134 | CTCACCTCTCTCATCCCACAAATTGCCTTTCCAGTGATTAGCATGTAATAAAGGATTTATTCATTTGACATGTATTTGTTGAGTGTTTCCTATAAATCTGGTACTGAAAAGTACAGAGAGAATGGTATTAGCTGCCATTTATCAGT |
| 22P | A00583:504:HM2HWDSXY:2:2559:31702:35383 | TTAGTTTAATCTACTGAAAGAATAAAATGAAATTCTATGAAAATAATATAACAAATACTGAGAGAAAGTAATTTGTCTGTAACCCACTTACCACTCTTCAGAGTCCTTGATTCTGTGCCACCAGAAAGCCTCAGGAGTATTCAGGG |
| 24T | A00583:504:HM2HWDSXY:2:2559:11758:24267 | TAAACCCTCAAAGCCTCAGTTTCCTCATCTGAACAAAAAAAAAAAGGATGGTATTCTGTACAACACATGTTTGTGGTAATGACTGAGTGGTATAGTGTGTCAACAGGTTCTGGAACATGGTTAGTGATCAATA |
| 23P | A00583:504:HM2HWDSXY:2:2556:7102:8594 | TATTCTTAAGGAAAAAAACCACACTTATGAACTATGATGAAATCTGTTGTTTTGTCTGTGTAAGTTTCATCCACGCTTCCAGTAACACCAAGTAAACTTTTTTTGAGGAAACATCCTTCTCCCACTCACAGTCCTTGTGAT |
| 33H | A00583:504:HM2HWDSXY:2:2553:5710:31140 | GTAAGGGCGCAGACTGCTGCGAACATAGTGGTGATAGCGCCTAAGCATAGTGTTAGAGTTTGGATTAGTGGGCTATTTTCTGCTAGGGGGTGGAAGCGG |
| 25T | A00583:504:HM2HWDSXY:2:2552:16179:26694 | TCTTTTAACAGTTCTACAATGCTGAAACTGAGGAACAGCAAGGTTATGTCACCCAAAGACATACAGGACAGGGAAGGGGCAAACTCAACTTGACCTGTCTCCAAAATCTATGTTTTTTAAACTCAGCCACGCTTCTGCCCTCCTTTGAAAGGAGTTTCTTTTACTTGATGAATTTTCTAACTTACTGTTTAACCATATTTCTCTGTATTATAAAAAT |
| 26T | A00583:504:HM2HWDSXY:2:2549:32172:6355 | GAAACCTCTTTGCTCCTCTGGTAGGCATTTGTAAAGGTGGCTTGGGTCAGGCACAGGCGGCCCCCAACCCCGCGGATCCCTTGTGTTCTTAGTTTTAACATGGCGTTGATGGAAAGGTCACCCGTCTCCCCTTCCACCGGCACATGCCTGGACCCC |
| 28T | A00583:504:HM2HWDSXY:2:2545:18285:23077 | TGAGCCAAGATTGTGCCACTGCACTCCAGCCTGGGCAACAGGGCGAGACTCTGTCTCAAATAATAATAATAATAATAATAATAATAATTTTTTTCAGAGCTTACCATCTATTCATCCTCTCTGTTTGTTATTATTTATATAAATTATCTTTTGCAAACTTAATATTTGAGT |
| 29T | A00583:504:HM2HWDSXY:2:2540:7455:31626 | AAACCTCTCTTTTGGCACTTAACTATGAAAAGAAAAGACATAGGAATCAAGAATAATTTATAAACTTTTGATCTATTTATAACATCCAATTTTTTTTTTTTTTTTTTTTTTTTTTTTTTTTTTAAGATGGAGTTTCACTCTTGTTGTTGCCCAGGCTGGAGTGCAATGGTGCGATCTTGGCTCACTGCAACCTCTGCC |
| 30T | A00583:504:HM2HWDSXY:2:2537:13006:21261 | ACAGATATACACACAGCCTGTGGATCAAGAGATGGGGTATAGGGGCTGCACTGTACCTAGAGGGACCAATTCTTTGTCTTTTCATGAGGAAAAAAAAGAAGAAAAATAAGAAAGAAAGAGAAGGAGGGAGGAAAGGGAGGAAAAGAGGGAGGGAGGAAAAGAGGGAGGAAGGAAAGAAGGAAGGAAGGAAC |
| 26H | A00583:504:HM2HWDSXY:2:2536:18918:31594 | GAGTCATGTAGTTAGTATTAGGAGGGGGGTTGTTAGGGGGTCGGAGGAAAAGGTTGGGGAACAGCTAAATAGGTTGTTGTTGATTTGGTTAAAAAATAGTAGGGGGATGATGCTAATAATTAGGCTGT |
| 5M | A00583:504:HM2HWDSXY:2:2534:32624:28401 | GATAGCACGTCTTACAGACGCTTTGCCACTGGCAGCATCGGTAAATGACGAAACACATATTCAATCTGGCCGATCAATGGTCGAATATCAAAATCAAGCCAAACAACTATTTAAGAAAATGAACCAGAATTCGCCGATTAGAGGCACTCTCGAAACCGGACCATATCTGTTCCAGTAATACACT |
| 6P | A00583:504:HM2HWDSXY:2:2534:26422:1658 | AGACACTCCAAGGCACTTCCCAAAGCCAAACTTGCACCATTATAGGTCATGGTCACTGTTTGATTGTCTGCTGCCGGTCTGATTCACTACAGCTTTCTGAATCCCAGTAAAACCATTACTTCTGAGAAGTACACTCAGCGCATCAATGAGATGCACCAAAAACTGC |
| 31T | A00583:504:HM2HWDSXY:2:2533:27389:23563 | GGGAAGAGCCAGAGGAGGAACCCAGCTCTTCTGAGCCAAATGGCAGCAGTCTCAGTCCTGGGTATATTGAGGGTGTGACCAGTGGCACTGTGGCCATTTCCAAGTGGA |
| 32T | A00583:504:HM2HWDSXY:2:2527:26051:8813 | TTAAAATATATGTAAATATAGAAGATGGTTTTAAAGAGCTTTTTGTTGTTGTTTTGAGACGGAGTCTCGCTCTGTCGCCCAAGCTGGAGTGCAGTGGCACAATCTCAGCTCACTGCAACCTCTGCCTCCTGGGTTCAATAAGTTCTCTGC |
| 33T | A00583:504:HM2HWDSXY:2:2524:3079:29622 | TCAGATCCTCTGGCGCCCGGGCGCCCCCTCCCGGCGACACCTCCCGGCTGTGAGCGCGCGGGAAGAGGCCAGAGGCAGGGCCTCTCACCTGCTCGGACTGCAAACCACAAAGCGTCTCTTTTTTTTTTTTTTTTGAGGGGCGGAGGTGGTCATTTTTATGTTCCCTGGGCAAAGCGAGCAGCCGTTTGGCTTATAAAATA |
| 36T | A00583:504:HM2HWDSXY:2:2524:23457:33692 | TCCAGCCTGGGCAACAGAGCGAGACTCTATCTCACACACACACACACACACACACACACAATTTTGGTAAATATACATAAATAAAATGTACTATGTTAACC |
| 37T | A00583:504:HM2HWDSXY:2:2523:19054:3114 | ACTGCACTCCAGCCTGGGTGACAGAGTGAGACTTCGTCTCAAAAAAAAAAAAAAAAGATACAGGTTCTCGGTTATACTTTGCAGATTTCAGTTGTTCCTAATCTCTCTAACTTCATGTCAATCTTTCCTTATCCATCTCCTGCACTC |
| 39T | A00583:504:HM2HWDSXY:2:2522:8431:15562 | TTGAGGCTACAGTGAGCTATGATCATGCTACTACCTGGGTGACAGAGCAAGACCCGGTCTCAAAAAAAAAAAAAAAAAAAAAGAGCAGAGCCAGGCGTGGTGGCTCGTGCTTATAATCCTGGCACTTTGGGACACCAAGGTGGGTGGAATGCTTGAGCTCAGGCGTTCAAGACCAGCCTAGGCAATATAGCGAAAC |
| 40T | A00583:504:HM2HWDSXY:2:2520:4182:13620 | CATCTATACAGGGGGCATGTGGGGCAGGGCTGCTTTGGGGTCAAGGGGTCAGATGTAGAGTTGTTTTGTTGTTGTTGTTTTTTGAGACGGAGTCTCGCTCTGTTGCCCAGGCTGGAGTGCAGTGGCGCAATCTTGGCTCACTGCAAACTCTGCCTCCCGGGTT |
| 7P | A00583:504:HM2HWDSXY:2:2517:9742:6433 | GAAAAAACTGTCCATGGTTAAAAGCTGTTCTGTTACGTTGGATCTTCCATGTTGCTGATGCATACCAACATTTACTTAGTAAACAGGTATTGAATACCTACTCTGTGCCAGGCACTGTGCCAGGCAGAGGACCTATAAGATAGTTCCCAGCTCTCAAGGTGTTCATGGTCTGAAGGGTTTAGACACA |
| 8P | A00583:504:HM2HWDSXY:2:2514:9308:13448 | TGTCTATGATTTATAATGCCTCATGAAATCAAAATTTTTGGGAAAAGTGTTTTACCTTAGAAATTTCTGAAAAGTCAGGCTTAGTTAATTTAT |
| 41T | A00583:504:HM2HWDSXY:2:2512:18412:16877 | TTATTGTCATTACAATAATTAAAAGTATCTCTTCTCTGTCTATCAAGTGTTATTTTCTTCATTTATTTTCCAATACATTTTCATTAGCCTGAAATTATTTGTGTATCTGTTTTTGGCAACTCATCATAAATAAGAAAAAAAAAATCATTCTCAATGTTTCTCCATAGATAATACATTTTATTC |
| 19H | A00583:504:HM2HWDSXY:2:2505:28592:5603 | ACCATAGCCGCCTAGTTTTAAGAGTACTGCGGCATGTACTATTGACCCAGCGTTGGGGGCTTCTACATGGGTTTTTGGGAGTCATAAGTGGAGTCTGTAAAGTGGTATTTTTACTATAAATGCTATTGTGTATGTTAGTCATATTAAGTTGTTGGCTCTGGTTTTTTATTTTTTTTG |
| 3A | A00583:504:HM2HWDSXY:2:2478:25563:24064 | GTGGATTAAGGGCTGAGACTGGGGTGGAAGTGAGCTTGGTGGGGATGCAGCGCAGTCACATGTGTACAAGGCTGTGACGCCCAGGGGGATTGTATCCACAGCTGGTCCTGCGGGTGCTGCCTTCTAAATATGTAGAATCCACGG |
| 42T | A00583:504:HM2HWDSXY:2:2477:12020:3427 | TACCAATGAAGCTAGCTAGCAGTCAGTGAGTCAGAGTTGCATGACATTGGCAATGGGTAAGAAACAATTCATAAAATAGTTTGAGCACATTACAAGGAGTT |
| 4G | A00583:504:HM2HWDSXY:2:2475:24605:28228 | GATTTCGACTTCCATGACCACCGTCCTGCTGTTTATATGAATCAACACCCTTTGTGGGATCTAGGTTAGTGCGTAGTTGGGCG |
| 43T | A00583:504:HM2HWDSXY:2:2474:15293:12790 | TTATGACATCTCTTTATGCTCCTCTTGGATGAGACATTTTCTTAGACATTCCTTGTTTTTGATGATCTTGACAGTATTGAAGACACTTGGCAAGTATTTTCTAGGATATCCCTCAGTTAGGATCTGTCATTTTTTTTTTTTTTTTTTTTTTCATGATCAGACTGGGCTTACAGGTTTTGATGAGGAA |
| 9P | A00583:504:HM2HWDSXY:2:2473:22752:35947 | TACCAGGAGCCAATCCAGGAGCTGATGTCGTAGCGTGGACTTCACACTGACATCCAGGCTTACTCCACGTTCTTTTTAAAGCATTCTTCAACCCTGTCTTCTGACTACTCATTTCTTAAAATGAAAAAGAAACATTTTACTTGTTAAATAAAGCCTTGCTTTTTTTGGTGCTATAGTA |
| 44T | A00583:504:HM2HWDSXY:2:2472:3938:28228 | GATAACACCGATGGGAAAAGAAAAAACGAAACAGTGTCCTTCTAAGCAAAATTTCCTTCCTATGAAATTGAATCTACATGCTCAAGTACAGTGAAGTG |
| 45T | A00583:504:HM2HWDSXY:2:2471:32497:16188 | CCTGTTCTTGGGGAGGAAAAAAAAGGCCCATCAGTGAAAGAGCCAAAGGCATGCCTCCTCCAGTGTGTCAGCACAGTTGGCCACGGTCAGCCACTCTTACCACATGGCCTATTGGTGGAAACAGTTCCCTCTTGTCTCTCTCCTTCCCTCCTACCTTTCTCTTACTGGGAATATTGCGTTCATAGACTACACAGTC |
| 47T | A00583:504:HM2HWDSXY:2:2471:16116:4570 | ACGCCGACGGCACCACACTCAGCTAATTTTTTAAATTTTTTTGTAGACATGGTCTCACTTTGCTGCCCAGGTTGGTCTAAAACTCCTGGCTTCAAGTGA |
| 10P | A00583:504:HM2HWDSXY:2:2471:14380:24392 | CATGCCCAGCTAATTTTTGTATTTTTATTAGAGAGGGTTTCACCATATTGGCCAGGCTGGTTTCAAACTCCTAACGTCAAGTGATCCGCCTGCCTCAGCCTCCCAAGTGTTGGGATTACAG |
| 48T | A00583:504:HM2HWDSXY:2:2471:13422:1658 | GCCTGGAAACAGCTATACATCCACTTGTATAGTAAGGGTCTTTTGGTTGAAAGTGACAGAAACAAACCCAACTCAAAGTGACTTAAAAAAAATAGTGTGTTTTTCTGTGAATGTTTTGGTTTATGTAAATGAAAAGTCCAGATGGTTTCAACTTCAGGAATGGCTGGATTCAGGTGTGCACGTGTTGTTTAGGAATAATGTTGTTTCCACT |
| 2M | A00583:504:HM2HWDSXY:2:2468:13304:35869 | GGAGAGTAATTATGTGGCTAGAGGGACAATAGGTAACTGTTCTTTCTCTTTGATGTCATTCATTCCGTAATGCTTTTATG |
| 12A | A00583:504:HM2HWDSXY:2:2463:32461:35509 | TGTTTTTCATACAATACAATTAAAATCCTGATTAGCAAACCTGACCACAGTTCCTCTTCCATGTACTTTATGGAGGCAGCCTCCAAGCAGGACTGCATTTTAAATAATAAACAGTAACTTTCCACTCCACTAGCTAGAGCACATTAATTTTTACATTATCTATTTTAAGCAAGTGGGCATATGCAATTCACACTTTTCTTAATAGCTTTGGATTTTCTCTAGGAAGTGTTATGCAAAGATTGCTCTGTGACAGATCAGCTGTGTTTATGGTATAAATTAGTGTGAATG |
| 24H | A00583:504:HM2HWDSXY:2:2463:1172:21809 | CGTGAGGAAATACTTGATGGCAGCTTCTGTGGAACGAGGGTTTATTTTTTTGGTTAGAACTGGAATAAAAGCTAGCATGTTTATTTCTAGGCCTACTCAGGTAAAAAATCAGTGCGAGCTTAGCGCTGTGATGAGTGTGCATGCAAAGATGGTAGAGTAGATGACGGGTTGGGCC |
| 49T | A00583:504:HM2HWDSXY:2:2461:7527:17002 | GAGCCGAGATCATGCCACTGCACGCCAGCCTGGACGACAGAGTGAGACTCCATCTCAATAAAATAAAATAATAAATAAATAAATAAATAAATAAATAAATAAATAAATAAATAAGCATCACAGGTCAATCTTACC |
| 50T | A00583:504:HM2HWDSXY:2:2461:17833:14497 | GTAGCATGGTTACTTGCTTGTTAGGAGACCCTTAGAAAATAGCTAAGTTTTTGTTTGTTTGTTTGTTTTGTTTTGATTTTTTTTGAGACGGAGCCTCGCTCTGTCCCCCAGGCTGGAGTGCAGTGGCGCAATCTCGGCTCACTGCAA |
| 51T | A00583:504:HM2HWDSXY:2:2459:11351:22936 | TTCACCATGTTGGTCAGGCTGGTCTCGAATACCTCAGGTGATCTGCTCATCTTGTCCTCCCAAAGTGCTGGGATTACAGGCGTGAACCACCGTACCTGGCTGCTATTGGAGTTTTAAGCAGGGAGTGGCATAATCAATTCTGAGAAATCTTAGAGGTAACATTCTCAGAATTAGGTAATTTCTAAAAATACATTTAAAGG |
| 52T | A00583:504:HM2HWDSXY:2:2458:15573:35446 | AGTGGTGTGCTGGTAAATGCTTAACCCACCCCTAAAAAAAAGAATATAAATATATATATATACATATTTATGATTGTTATAAATTTTACTGCTATAAAGGATGTGTAATGAACACTTTACAAATAATAATAATAAAACAGGCAATACTCTATGTTGTAAATTCCTTATAGTTGATTGATTCTAACAGAATGCTTTCACTGATTTTTGCCA |
| 53T | A00583:504:HM2HWDSXY:2:2455:2727:22060 | TTCTGAACTTGGTTTTCTCATTTGTAAAATAAAAGAAGTTGAACTATGATCCCCAAGGCCCCATTTGCTTATCAACAGTGCTGATTCTTAACTACAGTAGCAATAAGATCATCAAAAAAATTCACTT |
| 54T | A00583:504:HM2HWDSXY:2:2452:29695:33129 | GTTTTTTAAAAATTTATTTGTTTGTTGTTGTTGCATTTTGTTTGTTTTGAGACAGATGCTCACTCTGTCACCCAGGCTGGAGTGCAGTGGCACATTCTTGGCTCACTGCAATCTCTCCCTCCCAGGTTCAAGCAATTCTCCCACCTCAGC |
| 55T | A00583:504:HM2HWDSXY:2:2448:25292:11725 | CTCCATTTTATTTATTTTCTTTCTAATCTTTATTATTTTCTTCCTGATGCTTATTTTTGGGTTTAGTTTGTTTTTCTTTTTCTAATTTCTGAGGATGGAATGTTAGATTATTAATATGAG |
| 56T | A00583:504:HM2HWDSXY:2:2447:32244:28964 | TCCCACAGAGGATCGCTGAGAAAGACAGAGAGTAACTGCCCGCAGGGTCACAGAGAGGGGATGAAGTTCTTTTTAAAAAACACTATTTTTTATTGTGGTGA |
| 33P | A00583:504:HM2HWDSXY:2:2445:12536:27618 | AGCTCCACCTGCAGCCCCGGTGTGGGATCCTCTAGGTGAAGCGAGCTGGGCTCCTGAGTCTGCTGGTGACTTGGAGAATCTTTATGTCTAGCTAGGGGATTGTAAATACACCAATCAGCACTCTGTGTCTAGCTCAAGGTTTGTAAACACTCTAATCAGCACCCTGTGTCTAGCTCAAGGTTTGTAAATGCACCATTCAGTGCTCTGTG |
| 57T | A00583:504:HM2HWDSXY:2:2442:21314:16172 | GGCGTCCAGCCTATCCGCGGGGACTCCACGTGCACCCCTTCCTCATTGTCCTTGTCTAGGGCCGCGGCAAGGTCCTCGCTCCCATGGCGCGACTCCGGGGGTGTAGGAGCCTGGGCTGAGCAGGTGGAGTAGGGT |
| 58T | A00583:504:HM2HWDSXY:2:2442:19235:1767 | ACCTCATTTAACCTTGATCACTTCCTTACTCCAAATCCAGCCACACTGGGCACTAGGACTTCAACACATGAACTGGAGAGCACAGTTCAGTCCATAGCATTCCGGCTTTTCCTTTTTTTTTTTTTTTTTTTTTTTTTTTTGAGAAAGGGTCTTGTTCTGTTGCCCAATGTGGAGTGCAGTGGTACAAT |
| 31P | A00583:504:HM2HWDSXY:2:2438:9372:15718 | ATCTATACATGTATATGTGTGTGTATATATATGTGTGTATATATATAAAAAACACACATGCACGGATTCACCACCACCAACTCAGAAATTACCGTCTCCCTCTATTCTAAGGAATCATTTTTCATTTTGCCATCT |
| 11P | A00583:504:HM2HWDSXY:2:2438:30843:15076 | TGAACAGCTGTAGAATTATGTCATTGATAGATAAAAAGAATAAGCCCCAGAAATAGATGAATGCTCCCATTCATGTAAACAAACAAGTAAAAATTTGTCTGCATACAAAGACCTCTCAACGAGTAGAAGATGGCCTGGGAGTCTTGCGGG |
| 5N | A00583:504:HM2HWDSXY:2:2429:32262:27774 | GGGCCATTTGGCTTTATTGTTTTGCTTTCCTTAACCACCCTTTACGCCGTGGTGTTATCAACTGGGGCCTCTCGTTTAACCGCGGTGGCTGGCACGAGTTTTACCGGCCCTCT |
| 38H | A00583:504:HM2HWDSXY:2:2426:1280:22748 | CACTCGAATAATTCTTCTCACCCTAACAGGTCAACCTCGCTTCCCCACCCTTACTAACATTAACGAAAATAACCCCACCCTACTAAACCCCATTAAACGCCTGGCAGCCGGAAGCCTATTCGCAGGATTTCTCATTACTAAAA |
| 59T | A00583:504:HM2HWDSXY:2:2424:3857:19225 | TTTAGAAAGTAAGTATAGATGGAATAAGTATGCCTAGTTTTTGGCAACTTTCTCCTAATCATGAATTTATCTTTAAGCTTTTTTTTTTTTTCCTTCCAGAGGTGATTCCTATGCACATGCTCTAATTTCCTTTTAATTATTAG |
| 60T | A00583:504:HM2HWDSXY:2:2422:14769:3427 | AGTGTGATCATGGCTCACTGCAGCCTTGACCACTGGGGCTCAAGCATTCCTCTTGCCTCAGCCTCCCCAGTAGTTGGGACCACAGGCCGGCACCTCCATGCTTTTTTTTTTTTTTTTTTTTTTTTGTGATATGATGTGTCAATGTGTTGCCCAGGCTGGTCTTGAACTCCTGGGTTCAAATGATCCTCC |
| 61T | A00583:504:HM2HWDSXY: 2:2422:10800:19977 | CTACTCTCTCCTTCTATGAGATCAGGTTTTTTTTTTTTTTTAGCTTCCAACATGTGAGAACATGTGTATTTGT CTCTCTGTGCATGGCTTATTTCACTTAACATAATGT |
| 62T | A00583:504:HM2HWDSXY: 2:2421:2826:27993 | CATTCCATTCCATTTCATTCCATTCCATTCCATTCCATTTCATTCCATTGCATTGCGTTGCATTCCATTCCATT CCATTGCATTCCATTCCAGTACATTGCATTACTCTCGGGGTGATTCCTTTCCATTCCACTCCATTCCATTCCA CTGG |
| 63T | A00583:504:HM2HWDSXY: 2:2416:4128:15812 | CCTTTGGCCTCCACCTGTTCTTCTGGGTGTTGACTATTGTCCCTTTTCCTTAATCACCACTTTTGTGGTAAGG TGCCCCCCTCTAAGAGACTTC |
| 67T | A00583:504:HM2HWDSXY: 2:2416:13648:23970 | TAAGATACTACTACACACCTATTAGAATCTAGCCTGAAATTAAAAAGACTGACCATTCCGAAGTGTTGGCA AGGGTGTGGAGCAACTGAAACCTTCATACATTGTTGCCAGGAATGTAAAATGGTACAGCCACTTCGAAAA ACACTATGACAGTTTCTTAAAATTTAAGCAGACACCTACCTTATGACCCAGCCATTCTACCCCTAGAAGTTT ACCTAAGAAAACAGGAAAATGTATG |
| 68T | A00583:504:HM2HWDSXY: 2:2415:5475:2049 | GGGTGGCCTGGCAGAGGGGCTCCTCACATCCCAGACAATGGGCAGCCAGGCAGAGACACTCCTCACATCC TAGACGGGATGACGGCTGGGAAGAGGCGCTCCTCACTTCCCAGACTGGGTGGCCGGGCAGAGGGGCTCCT CACATCCCAGACGGGGTGGCGGCCGGGCAGAGGCTGCAATCTCAGCACTTTGGGAGGCCAAGGCAGGCGG CTGGGAGGTGGAGGTTGTAGCGAGCCGAGATCACGCCACTGCACTCCAGCCTGGGCA |
| 5G | A00583:504:HM2HWDSXY: 2:2413:25572:3756 | TCTCCTGGTATCTCGAGCTTGGTTTTCTTCCACAATCCTTCGGTACTGGCGAATACAAGAGGTAGGCAGAG ACCACTTGCGGCACCAAAACATAGCAACC |
| 30P | A00583:504:HM2HWDSXY: 2:2412:28619:3865 | TGGGTGACAGAGCAAGACTCTGTCTCAAAAAAAAAAAAAAAAAAGAGTAGAAAAAGGAGTTTGATCTGT AACTGACTATGAACAATCAACTGAGATA |
| 29P | A00583:504:HM2HWDSXY: 2:2412:27453:4288 | TGGGTGACAGAGCAATACTCTGTCTCAAAAAAAAAAAAAAAAAAGAGTAGAAAAAGGAGTTTGATCTGT AACTGACTATGAACAATCAACTGAGATA |
| 69T | A00583:504:HM2HWDSXY: 2:2410:6325:10755 | AGATCTGATTGGATGGGTTAAGTAATATCTATCTATGTCTCTCTCTCTCTCTCTCTCTCTCTCTGTCTCTC TGCTCTCCCCCTCT |
| 1M | A00583:504:HM2HWDSXY: 2:2408:26250:30608 | GCTCAGTCTGGCGGCGCGTCTACGACCGCGCGCCCCTGCAGCGGGGATGAAGCGGGGATGAAGCGGGCCT GAAGCGAG |
| 6H | A00583:504:HM2HWDSXY: 2:2405:2917:2253 | TAGATGGAGACATACAGAAATAGTCAAACCACATCTACAAAATGCCAGTATCAGGCGGCGGCTTCGAAGC CAAAGTGATGTTTGGATGTAAAGTGAAATATTAGTTGGCGGATGAAGCAGATAGTGAGGAAAGTTGAGCC AATAATGACGTGTAGTCCGTGGAAGCCTTTGGCTACAAAAAATGTTGAGCCGTA |
| 2H | A00583:504:HM2HWDSXY: 2:2404:22426:21762 | AACACCTCTTTACAGTGAAATGCCCCAACTAAATACTACCGTATGGCCCACCATAATTACCCCCATATTCC TTACACTATTCCTCATCACCCAACTAAAAATATTAAACACAAACTACCACCTACCTCCCTCACCAAAGCCC |
| | | ATAAAAATAAAAAATTATAACAAACCCTGAGAACCAAAATGAACGAAAATCTGTTCGCTTCATTCATTGCCCCCACAATCCTAGGCCGACCCGCCG |
| 70T | A00583:504:HM2HWDSXY:2:2403:7600:2566 | CTCACTGCATTCTCCACCACCACCCAGGCTCAAGTGAGCCTCTTGAGTAGCTGAGATGTCCAGTTAATTTTTATTTTTTATTTTATTATTATTATTATTATTATTTGGTAGAGACAGGGTTTTGCCATGTTGGCCAGGCTGGTCTTCAATTGAA |
| 27H | A00583:504:HM2HWDSXY:2:2378:15501:31469 | ACAGCCTAATTATTAGCATCATCCCCCTACTATTTTTTAACCAAATCAACAACAACCTATTTAGCTGTTCCCCAACCTTTTCCTCCGACCCCCTAACAACCCCCCTCCGAATACTAACTACCTGACTC |
| 71T | A00583:504:HM2HWDSXY:2:2377:12689:7560 | CACAATAGCTGCCGAGAGAAGGAAGGGAAGGGAAGAGAAGGGCAGGGCAGGGCAGGGCAGGGCAGGGCAGGGCCTAGGAATACATCTAACCAAGGAGATGAACTACAAGTTCTACAAGGAGATCTACAAAACACTGCTGGAAGAATCA |
| 1A | A00583:504:HM2HWDSXY:2:2376:16694:28933 | AAAAATATTGGGTTTTGACAACATGGAAGGCAATAGTGACTTTGGCAAGAAAAATGTGGATGGGATTCTAGTGTCTGGAGTCAAATTAAGAGTCCGAAGAGTGGGTGGAAGCTGAGG |
| 72T | A00583:504:HM2HWDSXY:2:2375:26151:29465 | AAATATAGAGAAATAATGGCAGAGACAAAATATTTATCAGGATACAGAAAGAGCGAAAGTATTACTTTCCCCACACCTTCCAAGACTGGAATAGTCTCCTTCATACTTGAGAGAGTGAAGACATCAGTTGGGAAAGTAGCATCTGCAGACACG |
| 31H | A00583:504:HM2HWDSXY:2:2374:32316:21668 | CCGACCACACCGCTAACAATCAATACTAAACCCCCATAAATTGGAGAAGGCTTAGAAGAAAACCCCACAAACCCCATTACTAAACCCACACTCAACAGAAACAAAG |
| 73T | A00583:504:HM2HWDSXY:2:2373:3323:23406 | AAAGGCTAAAAGAAAGTGTCATATTACACATGTAGATTTTGAGAAATATTTGTTATTTTTCTTATGATAAAGGTTATGCATACACACTCACACACACACACACACACACACACACACACACAACTAGAAATTACAGACAAGAAAAAAGATTAGCCAGAATGT |
| 74T | A00583:504:HM2HWDSXY:2:2373:12038:29324 | CTGCTATAACAAACTACCCTGGACTACATGGCTTATAAACAACAGAAATTGATTTCTTTTATTTTTTTTTTTGAGACAGAGTCTCGCTCTATCGCCCATGCTGGAGT |
| 75T | A00583:504:HM2HWDSXY:2:2369:5032:3975 | GAAATTATCAGAGACTGGGTCACAAATTTATATAAAAGTTCAGTGCTGCATTATTAAGAATAGTACCAAAAATGAAAACCTAAGCGTT |
| 76T | A00583:504:HM2HWDSXY:2:2368:32660:15718 | TTTTCTGGTCTCCTACTATTTATCATTTACTTTCCACCTTGTCTTAAGTGGTAATCAAGAAAGGTTTGATTTATTTTTAATTTAATTTAATTTAATTTAATTAATTAATTTTTTTTTTGCT |
| 77T | A00583:504:HM2HWDSXY:2:2364:3748:24737 | GTGCGTTGCCTGCTGTCCCTGCTGCCCAGGCCCTCCCCAACATGTCAGGACTTCTGGCTGATTCCTGCAGGGCAGACTGTCTGTTCATCCTCAGGGGGCGTTTGTTGAGTGA |
| 78T | A00583:504:HM2HWDSXY:2:2361:21739:4914 | ACAAAGAAGAGAAGAAAACTTCTGGGATTTAAGCTTGAAGGGTGAAATGAGACAACTCTAAAGTTCTGTTTCAGTTAGAGTCTACTCTGCAAATTTAATCAAGGTCTTCCGCAGGCATCAGCTTGACATTAAATTTAAGTGCCCCTTAAAGTGAGAGTT |
| 64T | A00583:504:HM2HWDSXY:2:2354:8748:9846 | AAGGTTATCTCCCCTCCTACCTGCCTCCTGTCCCTTCTTTTACCATAAACTTCCGTCATCAAATTTACATCCCAATAACTGCCTTAGAATGGCTGCCAGACTCTCTCATTCCATGTGGTATGCCAGTGTAAATTTGCACAGTCTACACACAGTGTTGCCTCCCACTGTCATCTTCCAGAAACTCCTTTTCATCTTCAACATCCAGTCCTGCATGCCCTGCCACCTCCAAGTCAATCCACAGACTCAGGAATCTTT |
| 12P | A00583:504:HM2HWDSXY:2:2354:14299:19617 | TTGAAAATTCTTTAGCCTAAACTAATTATTCTCAATCCTGGCCAAACATTAGAACCACTTGGGGTACTTTTAAAAATTGCTGATGTCCAAGCCCCATTCCTGAATCAGAATCCTTGGGCTTGGGACCCTAGCATCAATATTTTCTATTCTATTCTATTCTATTCTACTCTATTTTATTTTATTTTATTTATTTTTTTGAGACAGAGTTTCACTGTCATTGCCCAGGCTGGAGTGCAATGGTGCAATCTCGGCTCACTGCAGC |
| 80T | A00583:504:HM2HWDSXY:2:2354:10086:23500 | AACTGTAATCTGTCACAATACTCTTTTGTGCCATCTGTATCTGTTTGGATGTTTTGTTATTGCAGGATATTTTGATATTGAGAGTTTGTCCTTGCAATATCATACTAGATCATAATACCATCATGTTGATATCTAGGTA |
| 81T | A00583:504:HM2HWDSXY:2:2353:29541:24565 | AATACAAGGTTAATGTAAAAAACATTATATATACACATATTCATATATGCACATATACATATATATATATATATATATAGAGAGAGAGAGAGAGAGAGAGAGACAGAGAGAGGGAGAGAGACACTAGCAGCAAACCATTGGAAAATAAAATAATTTTTTACAAT |
| 13P | A00583:504:HM2HWDSXY:2:2352:18584:31266 | CAACTGCCAATTAGAAAAATCTTTAAATCCACCTATACCCTGTAAGCCACCCACTTCAAGATATCTCACCTTTTCAGATGGAACCAATGGTCACTTTCCATATTTCGATTTATGATGT |
| 82T | A00583:504:HM2HWDSXY:2:2352:11885:21104 | AATACAAAAAAAAAATTAGCCAGGCGTGGTGGCGGGCACCTGTAGTCCCAGCTACTTGGGAGGCTGAGGCAGGAGAATGGCGTGAACCCAGGAGGCAGAACTTATAGTGAGCCGAGATCACGCCACTGTACTCCAGCCTGGTGACAGAGCAAGACTCTGTCTCAAAAAAATAAAAAAATTAAAAAAAAATTTTCAGTTGGCAGAACTTTGCACTACGCACCTTGTCATTCACTTTGCATGGA |
| 11H | A00583:504:HM2HWDSXY:2:2350:13250:20462 | GGTCTTTGGAGTAGAAACCTGTGAGGAAAGGTATTCCTGCTAATGCTAGGCTGCCAATGGTGTGGGAGGTTGAAGTGAGAGGTATGGTTTTGAGTAGTCCTCCTATTTTTC |
| 83T | A00583:504:HM2HWDSXY:2:2347:14055:11146 | TATCAATCATGGGATTTGCAGTTGGTTCGTGGGGCACATGCATAGTGTATGTGCAGCATAATTATTTTTATTGCATAAACAGAATGACCATATGTGTATATCTCACCACTAACTAACACAACTTAGCATGGTGCTTGGCACACAGTTGGTAATCAATAAAATTATATTTACTAAGTAAATAGAATACTGAAGATGTTAACTCTGATTTAACTATACTGAAAAATACATCACTCATGTA |
| 84T | A00583:504:HM2HWDSXY:2:2347:11867:15781 | TACACAGAAGCATTCTGAGAAACTACTTTGTGATGTGTCCATTGAACTTACAGTGTTGAACCTATCTTTTGCAGAATCTGCAAGTGGATATTTGGAGCACTTTGAGGACTACT |
| 85T | A00583:504:HM2HWDSXY:2:2343:4173:28260 | TGCGCATTGTTCACCTGCGTGGGGGCTTGTGGGGGCTGAGGGGGGCTTGCTGTCATTTTTGTTACACGAGAAGCCCCAGTGCAGTGGGCACGGCCAGTGAGAATGAGTGGAAACGTGCGAGGACCCTGCACATGAGCTTCTATCAGGTCAGCAAACGGCCGCTCGCTTGGGGCTGGGTAAC |
| 86T | A00583:504:HM2HWDSXY:2:2339:4535:9283 | GACAGTTCCGTGCTGTCCCCTGTTTCCCTCGGGGTCTTCCCTTCGGAGGCTGCAGCTCGTCTGAGCATCCAGCCTTGAAGGCACCAAGCAGGCAAGGAGGTAGGGCTTTTCTCTGGGGAGGCCTGCGTTGCAGTACGGCTTCCTCATCCCCACCAAGGGCAGGGAAGGGCAGGAGTCTAGGACTTACCTGAATACAGAACACCTGGG |
| 7M | A00583:504:HM2HWDSXY:2:2337:13404:11898 | CGTTAATTCCCATATATTCCCGTTAATTCCCATGGAAAGTTTCCAGCTTTGAAAATTCCCGGAATTTTGCAACCCTAGGAACAATACATTCACAATGCAAATCATGTTTTATTCATAACTGTAAAATTACTATTGTTACGATTATAAGTTACTCTGTTGTACCTTTCAGTTGTTGGTGTTGTTTTGTGTTTGACACACTGTTA |
| 5A | A00583:504:HM2HWDSXY:2:2335:3206:35884 | TTCTGCCTTCCCCGGAACAGCCCTGGCCCCTGTGCATCTAGGCTGAGATAAAAGGAGAAGCCAGAGAATTCTTTAAATCCTCTTCCCCAGGATGAGCCTTTAGAAGTCCCCGCAAAGGCTGTGTGTGTATTGGGGTGAGTCAGTCTGGGAAGACTGGACAGT |
| 87T | A00583:504:HM2HWDSXY:2:2333:25762:19899 | ACCATACACTTTATCTTTGAATTACCACAAATATTTGTAGAATTTTTTTAATGTCTTTGCCCTTTTAGTTTTTGGATATTCATTAAATTGAGTTTCTAAGTCTGTGTGTGTGTATGTGTATGTGTGTGTGTTTGTGTTTTTTGAGATGGAGTCTCACTCTGTCGCCCAGGCTGGAGTGCAGT |
| 89T | A00583:504:HM2HWDSXY:2:2332:2709:15953 | ATGGCGCGTGCCTGTAATCCCAGCTATCCAGGAGGCTGAAGCAGTGAACCTGGGAGGCGATGGTTGTAGTGAGCCGAGATCGCACCATTGCACTCCAGCCTGGGTGACGACTGCGATACTCTGTCTAAAAAAAAAAAAAAATAAAAAAAAAAAAAAAAGCATGCAGTAGATGCTTGAATGAATGGTTGAAGAAGCACAGGAT |
| 90T | A00583:504:HM2HWDSXY:2:2331:12355:7513 | GCCCCCAGTCTCATTGCTTGCCTACCAAACTGATTGTAGGAGCCGCATGATGCACATGACCTTGCTTCTAACCTACCTGGATGTTTTCACGCTCTGCTGTCTCTTTGAAGTGACTTCCTCTTCTTTCTTCCCTCTTCTTCCTTGCAGTGGCTGTTATGATCTTTAAAGATCTGGTTCAAACA |
| 29H | A00583:504:HM2HWDSXY:2:2327:31729:8563 | GGTGAGCATCAAACACAAACTACGCCCTGATCGGCGCACTGCGAGCAGTAGCCCAAACAATCTCATATGAAGTCACCCTAGCCATCATTCTACTATCAACATTACTAATAAGTGGCTCCTTTA |
| 91T | A00583:504:HM2HWDSXY:2:2327:16062:35446 | ACAATGAAGATGCTTCCTTATCTACCACAGGCTGTGAACCGCTCTAACTATCCACTTCAAATTCTCCAAAAAGAGTATTTCAATACTGTTCTATCGAAAGGAAAGTTCAACTCTGTGAGTGGAATGCACATATCACAAAGAAGATCCTGT |
| 34P | A00583:504:HM2HWDSXY:2:2325:15772:27430 | TGCATGTAAATGTTTTAAATGGCTGATACAACCTTGGTTATCTTCCCTGCTGTTTCAATCTCTTATGAAAGTTTCTTTACTTCCTCTACCAATTCACTTGTGCTCCAAATGTGCTAAAATCATTCTTGCCACAGCCACCACTGTGCCTTTACTCTTTACTGCA |
| 92T | A00583:504:HM2HWDSXY:2:2320:14443:36213 | GGCCAGGACTGAGGTCCGTGAGGTTGGAGGCTCTCTTCTACTTTCCTTCTGGCCGGGACTGAGGTCCGTGGGGTTGGAGGCTCTCTCTTCTACTTTCCTTCTGGCCG |
| 2C | A00583:504:HM2HWDSXY:2:2314:11035:19257 | GGACACCATTGAGATAAATGGAACTTTTAAACACCGCAAAATGACCCTGGTGGAGGAGGGCTTTAACCCT<br>GCTGTCATCAAAGATGACTTGTATTTCTTGGATGACACAGCAAAAATGTATTTGCCTATGACTGATGACAT<br>CTATAATGCCATACGTGCTAAAACCCT |
| 15H | A00583:504:HM2HWDSXY:2:2313:7600:17566 | TGACGTTGACAATCGAGTAGTACTCCCAATTGAAGCCCCCAGTCGTATAATAATTACATCACAAGACGTCT<br>TGCACTCATGAGCTGTCCCCACATGAGGCTTAAAAACAGATGCAATTCCCGGACTTCTAAACCAAACCACT<br>TTCACCGCTACACGACCGGGGGTATACTACGGTCAATGCTCTGAAATCTGTGGAGCA |
| 93T | A00583:504:HM2HWDSXY:2:2311:5810:10050 | GCTTTTCCATTTTTTTTTTTGTCTCCAACTGCCAGAACCTGGCTAAATTAGACACACCTAAATTATAATCCA<br>AGATCAGCCATTTAGTAATTGTGAAATCT |
| 94T | A00583:504:HM2HWDSXY:2:2309:21432:10989 | AAAATGTGCAAAAGATATGAACAGACATTTTACTAAAGTAAAAAGCCAGATGGCAAGCACATGAAAAGA<br>CGTTCTTGTATCACTAAATCAGCAGTGAACTGAAAATTAGAACTACAGTGAGATAGCAATGTTCTCTGATC<br>AGAACGGCTACAATAAAAAAATAGTGTCAACACGGAATGCTGCTGAGGGTGAGCAGAGGCTTGATCATTC<br>ACATATTGC |
| 1C | A00583:504:HM2HWDSXY:2:2307:5981:28667 | CGATTTTTGAGTGGTAGGAGGGGTGATGGGTTCTGGGCTTAGTGGGGGGGGGGGTCCCGGGCGACTGGG<br>GGTGAGGCCTCACCTCAAACACGTGCAGCTCCTCCAGGAGCAGCC |
| 38T | A00583:504:HM2HWDSXY:2:2306:12436:11537 | ACTGCACTCCAGCCTGGGTGACAGAGTGAGACTTCGTCTCAAAAAAAAAAAAAAAAAAGATACAGGTTCTC<br>GGTTATACTTTGCAGATTTCAGTTGTTCCTAATCTCTCTAACTTCATGTCAATCTTTCCTTATCCATCTCCTG<br>CACTC |
| 8A | A00583:504:HM2HWDSXY:2:2303:8187:26005 | AGCTTGAATTAAAAGGAATTCACTTCAGAGTCTATTTTGGTTTGTTAGGGCAGATCGTTGGTGTGCTGGGG<br>GCGCAAACTGATGCACTTGAGGTATGCATTAGCTGGTCCTGCCATATGGCCTGTCACGTCTTACTGGTTAA<br>TGTAACCATCTGTCACCTCTTGACAATTTGCATTCACTTTCAGCACTTGGGTTAGAGACATTTTCTTTGGTA<br>TTTTGCTGT |
| 14P | A00583:504:HM2HWDSXY:2:2277:29505:8656 | ATGAGGTGGATGGATCACGGGGTCAGGAGTTCAAGACCAGCCTGGCCAACATGGTGAAACCCCATCTCTA<br>AAAAAATACAAAAAAAAAAAAAAAAAAAAAAAATTAGCCGGGCATGATGGTGGTGCCTGTAATCCCAGCT |
| 95T | A00583:504:HM2HWDSXY:2:2277:16206:31876 | AATTGTCCCTGGCTTGTATGCAGTCCGTGTGTGTGTGTGTGTGTGTGTGTGTGTGTGTGTGTAGCAGGTG<br>GAGTGGGGAGAGGTGCATGAGGTGTCTGTAAAGGGGTGTTCTGGGGCCTGGGCTCACCAAACA |
| 96T | A00583:504:HM2HWDSXY:2:2275:7636:8484 | TGATGTGTATTAGAATGGGTTTCTTATTCGTCCTGTTTGAGATTCATTAAGATTTCTCAGTCTGATAATTGA<br>CATGTTTTATAAATTCTAAAAAAAAAAAAAACCAAGCTACTATGTTTTTTGAATATTGCTTTCTCTCATTCGA<br>ATCT |
| 98T | A00583:504:HM2HWDSXY:2:2272:9462:17628 | GGATGGGTGGGTGGATGAGTGAATGGATGGAGGAAGGGTGGATGGGTGAATGGATGGAGGAAGGGTGGA<br>TGGGTGGTGGGTGGGTGGATGGGTGGGTGTGTGAGTGGATGAATGGATGGGTGGGTGGGTGGGTGAGTGG |
| | | ATGGATGAGTGAATAGATGGATAGATGAGTGAACATGTGAATGAATGGGCGAGTAGGTGGATAAGTGGATGGATGGATG |
| 99T | A00583:504:HM2HWDSXY:2:2262:21522:10394 | TAATGTAACAGAATGAAAAGTATATTGTATGGTTTTAGATTCTACTCAATGACTTAACCTACCACATGTCAAGTTTTGGTATAGCATAAAAGAACATCCACAGTTATCGG |
| 100T | A00583:504:HM2HWDSXY:2:2258:23068:11318 | GGTGATTGTGATGGAGGTGGTGACAGTGGTGGAGGTGGTCTTGGTTGTGTTGGAATTGGGGAGCTGATGGAGCTGGTGGTG |
| 101T | A00583:504:HM2HWDSXY:2:2257:16586:15687 | GGAGGGCTAGACAGGTTTAAATTTTTTTTTTTTTTTTTTTTTTGGAGATGGATAAGCAATATGTTGCCCAGGCTGGAGTGAAATGGGGCAATAAAGGCTCACTACAACCACCGCCGCAAGGGTTAAAGCGATTATCATGTAACAGCCG |
| 102T | A00583:504:HM2HWDSXY:2:2257:11505:6073 | CTCCAGCCTGGGTGACAGAGTGAGACCCTGTCTCAAAAAAAAAAAAAAAAAAGAAAATGAAAATGGAAAAAATTTGTTTGTGAAGAAAGACCCCTGAAAAGAAAGTCAACTGCCAGTATTAATTATGAAAGCAAAAAAAGAAAAGTCTATAAAAGTAATGGAATGTGAATAAAGGTAGAAGGGATGGACACACTTTTGCATATATTAGGTGGAACGTTAGCCTCAGTAATGAAGAGCCAGGACACAT |
| 36H | A00583:504:HM2HWDSXY:2:2256:3106:5087 | GACTGATAATAAAGGTGGATGCGACAATGGATTTTACATAATTGGGGTATGAGTTTTTTTTGTTAGGGTTAACGAGGGTGGTAAGGATGGGGGGAATTAGGGAAGTCAGGGTTAGGGTGGATATAGTAGTGTGCATGGTTATTAATTTTATTTGGAGTTGAACCAAAATTTTTGGGGCCTAAGACCAATGGATAGCTGTT |
| 103T | A00583:504:HM2HWDSXY:2:2250:10809:5431 | CAGAGCCTGGAAAACAGTGAGCACTGAATAAGCATTATCTATTGTTATTTGATCATCTTCAGATGTAATTAAAAGCTTACGATGAAACATTTGTCAATTTCATGCTAACTAGAGAAATCTTTGTAATTTCATGTGTAAAATGTGCTAGTGTTTTCAC |
| 104T | A00583:504:HM2HWDSXY:2:2246:3450:9032 | GTGAAAACACTAGCACATTTTACACATGAAATTACAAAGATTTCTCTAGTTAGCATGAAATTGACAAATGTTTCATCGTAAGCTTTTAATTACATCTGAAGATGATCAAATAACAATAGATAATGCTTATTCAGTGCTCACTGTTTTCCAGGCTCTG |
| 27T | A00583:504:HM2HWDSXY:2:2244:9263:26490 | GCCGGTGGAAGGGGAGATGGGTGACCTTTCCATCAATGCCAAGTTAAAACAAAGAACACCAGGGACCCGCGGCGTGGGGGGCCGCCTGTGCCTGACCCAAGCCACCTTTTCAAATGCCTACCAGAGGAGCAAAGAGGTTTCTGCAAAATTCGCAACACCCCCA |
| 106T | A00583:504:HM2HWDSXY:2:2244:11116:17926 | TTATCCAAATAATTAGGTATAAATATTGCTTTTTAAAAAATATTTGTCTCTATAGGTTTTTGGGTTTTTTTCCATTCTTCCTGTTCTCCTTTTAGATTTATTAAATATTTTAGGGTCTCATTTTTCTCTCATTGACT |
| 107T | A00583:504:HM2HWDSXY:2:2243:27362:35352 | CGTGGTGGCATGCCTTTAATCTCAACTTACTCGGGAAGCTGATGCAGGAGAATCACTTGAACCAGGAGGTGGAGGTTGCAGT |
| 109T | A00583:504:HM2HWDSXY:2:2233:9263:32127 | AGAGCAAGACTCCGTCTCAAAAAATAAAATAAAATATTAGTTTTGGCTGGCTGGGCTCAGTGGCTCAGCCTGTAATCCCAGCACTTTGGGAGGCTGAGGTGGGAGGATTGCTTGAGACCAGGAGTTTGAGACCAGCCTGGGCAACATAGCAAG |
| 110T | A00583:504:HM2HWDSXY:2:2233:29089:24659 | AGGGTATTATATTATAATTGAGGCACTTTATGAATATATAAATAATATTATGTTTTCATGCTAGAGATCATGCCAATGAAGATATTTACTTTGAAAAGGGGAAGATTAGAAGTTTAAAAGCATTTCCATATTGAAGTAAATATTCATTTCCATATCTTCACAGTTATCTTTCTCTGAGTTCTCTGACTCATTGTGAAAAAAAATTCCAACCTTCTTCACAGCTCTACCATCTTCG |
| 111T | A00583:504:HM2HWDSXY:2:2229:30752:12633 | TGCTAACTTTGGTTGTTATGCCATGCCTCTTTTAATCTAAACTGGCCTCTCCTTACTATTTTTATTTCCCATTTCATGAATTCTTTGTGGAAAGTTATTTCTTTTAAAATAGCATGCTCATGAGGTTGAAGGGGTAAAGAACATTGATCTCGTGGTTGCAGTGTGGATGAAAATTCAGCCTGAAATGTGCC |
| 112T | A00583:504:HM2HWDSXY:2:2229:23484:4429 | AGAAGAATTGAATTTAAGAACTGATTGCCAGTAAGCAAAAATAATGATTAATTTGTTTTGTTTTTAATCTTCACCCACTTTTCTGTATACTTTTTTTTATTAGATGTCC |
| 113T | A00583:504:HM2HWDSXY:2:2227:26919:24314 | GGCTGCAGTAAGCCGAGATTGTGCCACTGCACTCCGGCCTCGGCGAAACATAGAGACTGTGTCTCAAAAAAAAAAAAAAAAAAAAAAAAAAGAAAAAGAAAAA |
| 114T | A00583:504:HM2HWDSXY:2:2225:13503:21778 | GTGCCACTGCACTCCAGCCTTGGTGACACAGTGATATCCTGTTGAAAGAAAGGAATGGAATGCAATTGAGTGAAATGGAATGGAATGGAGTGGAGTGGAGTGGAGTGGAATGGAATGGAATGGAATCGAATCAAATCAAATTGAATGGAATGGAATCGAATCAAATTGAATGGAATGGAATTGAATGGGAGCTGAGATTTTGCCACTGCGCTTCAGTCTGTGT |
| 15P | A00583:504:HM2HWDSXY:2:2222:13910:25050 | AAAGTTTTTATAATTTTAGTTGCTATCGATTTAAAGATTACAGAATTCTTTTGTAATCATATGTTCTCACCTATTCTTAATATAATATACATTTATGATCAAATAACATCTATAATATGATATTCATTTTACAATCAAATAAATTGAAA |
| 116T | A00583:504:HM2HWDSXY:2:2219:22155:13620 | TCTGTGCTTAGAGAATTATTTTCTGGCAACTCTCTTTCTGCAGCTCAGTAAGTGTCTCTTGAGCAATTCTATAATTTCAAGAACGATACATGTGCCA |
| 117T | A00583:504:HM2HWDSXY:2:2215:25084:33285 | GGAACATTGTGAATATACTAAAAGCAAGTGCATTGTATGCTTTAAAATGGTTGTTATTAATTTTATATTATGTGATTTTTACCTTAAAAAACAAAAAAGAGAAAATAGCCTTACTCTATATACAATAAATTCAAGATGTGTTACAAATTTATATGTGAAATCCAAAATAGTATAATATTTAAGGAATAGC |
| 118T | A00583:504:HM2HWDSXY:2:2211:23249:11945 | TCTGGCCCAAGTCCCCGGCACCCAGAAGGAGGCCCATAAAAACTTCCCAGACTGAAGGGACCATGAGCTAAAGTCAGCAGGGGAAAAGGTGGCTCCAAGCAAGGAGAGGAAAACCCACAGACAGCAGCCTGTGGGGCCGTCACCGCCACCACACCCTGGAGCCCCGAGAACCATCACTTCTGCTCCCTG |
| 119T | A00583:504:HM2HWDSXY:2:2209:7889:10050 | ATAGGGAGCCAGTGAAGGGTATTGAGCTGGGGAGGTCAGATTTGCATTTTGGAAAGATCTCTCTTGCAACTTTTTTTTTTTTTTTTTTTTTTTTTTAATGGGTCTCCTATCACCCAGGCTGGAGTGTAGTGGTACAATTATGGCTCACTGCAGCCTCTACCTCTTGGGCTCAAGTGATCCTCCCACCTCAGCCTCCCAAGTAGCTGGCCACTACTACACACAGGTGCATACAACCATGTCCAGCTAGTT |
| 120T | A00583:504:HM2HWDSXY:2:2209:23249:27320 | CCCCACTCCTTCCTGAGCGTCTCCCTCCACAGACATCCGTGACTGTACCGCTACTCCCACCCCCGATTCCTT |
| 20H | A00583:504:HM2HWDSXY:2:2203:3314:31219 | ACCGCAAACATATCATACACAAACGCCTGAGCCCTATCTATTACTCTCATCGCTACCTCCCTGACAAGCGCCTATAGCACTCGAATAATTCTTCTCACCCTAACAGGTCAACCTCGCTTCCCCACCCTTACTAACATTAACGAAAATAACCCCACCCTACTACACCCCATTAAACGCCTGGCAGCCGGAAGCCTATTCG |
| 17H | A00583:504:HM2HWDSXY:2:2203:1271:29966 | ACCGCAAACATATCATACACAAACGCCTGAGACCTATCTATTAATCTCATCGCTACCTCCCTTACAAGAGCCTATAGCACTCGAATAATTCTTCTCACCCTAACAGGTCAACCTCGCTTACCCACCCTTACTAACATTAACGAAAATAACCCCTCCCTACTAAACCCCATTAAACGCCTTGCAGCCGGAAGCCTATTCG |
| 4A | A00583:504:HM2HWDSXY:2:2202:28754:29622 | TGATGCCCCGATCGGGTGTTTTTGTTCACAGACTAGTTCAGGGTTCAGGTTGCTCTCTGATCAGAGGTGAGGAACTGGTAGCCCCCGCTGTCTGTTT |
| 121T | A00583:504:HM2HWDSXY:2:2177:22227:1125 | CTGACCTTGTGATCCACCCACCTCTGCCTCCCAAAGTGTTGGGACTACAGGCATGAGCCACCATGCCTGGCCTTCTTATTTTTTTTATAGGTGTGATTACAAAGCAAAGAAAGAGTAGTGACAGAGAAATTGATGAAAAGAGATGAAAATTC |
| 122T | A00583:504:HM2HWDSXY:2:2171:8910:36119 | CCTCCCGGGCGGGGCAGCTGGCCGGGCGGGGGCTGACCCCCCCACTTCCCTCCCGGACGGGGCGGCTGGCTGGGCGGGGGGCTGGCCCCACCACCTCCCTCCCAGATGGGGCGGCTGGCCTGGCGGGGGCTGACCCCC |
| 123T | A00583:504:HM2HWDSXY:2:2171:28917:13182 | CTCTGGGCCTACAACACTGTAAATTATCTCAGCCTCTAGCATGAAATTCGTATTTCATATCTTGTTCTAACTCTCTTTCCCTCTCCTGTATGCACGCGGCATGCACACCAAGCCGTACCGGAG |
| 79T | A00583:504:HM2HWDSXY:2:2170:25174:6511 | GTTGTGTGAAGGAGGTGAAACATTACTGAGGTGGGTGGCTTAAGGTAGTATATACCAAGTTATGTTTCTTAATTTGAAGATAATCATGCAGGAAAATTTATTATAGGGAAAAGGATGATATGGTGCAGTGAAAAAATGGGGAATTAGGAATCAAAAGGTTTGAGTTCCCACTG |
| 124T | A00583:504:HM2HWDSXY:2:2169:13313:24956 | GGCAAGAGAATGGCGTGAACCCAGGAGGCACAGCTTGCAGTGAGCTGAGATTGCGCCACTGCACTCCAGGCTGGGCAACAGAGCGAGACACCGTCTCTAAAATAACATAACATAACATAACATAACATAAACTAAACTAAACTAAACTAAAATAATAAATCATGATAGGTAAGTCCAGATATTCAAACCTCTCAAGTAACAGAATTATTGTTAGAGCTATCTACAGGCCAACCCAAGTTTGGCAAATCCATATTCTAG |
| 16P | A00583:504:HM2HWDSXY:2:2166:18032:15593 | GGCACTGTTGGCAGTGACCTTGTCATTGAGCTCCTGGCCAGTTGGGACCAGTCAGTGGTTTGGATCAGTGGGGCCTGGTCGGTGGGGCTGTGGTACTGCAGGCCTGGTCTGTGGGGCC |
| 88T | A00583:504:HM2HWDSXY:2:2163:19072:24220 | GCCGGGTGCAGTGGCTCACACCTGTAATCCCAACACTTTGGGAGGCTGAGGTGGGTGGATCACTTAAGGTCAGGAGTTCAAGGCCAGCCTGGTCAGCATGGTGAAACCCCATCTCTACTAAAAAATACAAAAATTAGCTGGGTGTGGTGG |
| 125T | A00583:504:HM2HWDSXY:2:2162:16957:17581 | CACCATCGGCACTCTGAGTCCTTCCCAGGTTCCCTAAAGGGGGAGGAGGGGACACCGTTTTAGATTTAATTTGTACTGTACCTGATATGATTGTGATAATTTTCCAATGAATAATTGTTAATCTTACCTAATTAAAATAATTATATGTTAAATAAAATATGCAGATGTTAAA |
| 40H | A00583:504:HM2HWDSXY:2:2158:3604:34726 | TGCCCCAACTAAATACTACCGTATGTCCCACCCTAATTAACCCCATATTCCGTACACTAGTACTCATCACCCAACTAAAAATATTAAACACAAACTACAACCTAACTCAATCACCAAAGACCATAAAAATAAAAAATTATAACAAACCCTGAGAACAAAAATGAACGAAAATCTGTTCGCTTAATTCATTGCACCAACAATCCTAGTACTAACAGCCGCAGTAAT |
| 126T | A00583:504:HM2HWDSXY:2:2154:14642:28510 | TACTTTTCATACGATGGAATTATAATGTGAAATCATGTTTTATTAAAATCATGTTTTAGAAGAATATTTTACAATGTTATTGGGAAACAATCTTGAAATATGAC |
| 21H | A00583:504:HM2HWDSXY:2:2152:29812:33426 | CAATTAGGGAGATAGTTGGTATTAGGATTAGGATTGTTGTGAAGTATAGTACGAATGCTACTTGTCCAATGATGGTAAAAGGGTAGCTTACTGGTTGTCCTCCGATTCAGGTTAGAATGAGGAGGTCTGCGGCTAGGAGTCAATAAAGTGATTGGCTTAGTGGGCGAAATATTATGCTTTGTTGTTTGG |
| 128T | A00583:504:HM2HWDSXY:2:2149:1244:13667 | ATTTTATTATTATTATTATTATTATTATTTTGAGATGGAGTCTTGCTCTTTTTCCCAGTCTGGAGTGCAGTGGCGAGATATCGGCTCACTGCATGCTCCTCTTCCTGTGTTCGCACCGTTATCCTGCCTCAGCATCCGGAGTAGCTGGGA |
| 129T | A00583:504:HM2HWDSXY:2:2145:15971:3912 | TATTTATAGGATAAAGTGATTCTTTTTTTTTTTTTTTTTTTTTTTTTTGAGAGGGAGGCTCGCTCTGTCGTCCAGGCTGGAGTGCAATGGCGAGATCTCGGCTCACTGCAAGCTC |
| 131T | A00583:504:HM2HWDSXY:2:2144:4499:27289 | ACTACTGGTAGAAGAAAGTGAAAACACAAAGGACAAAAGATGCAAACACATGTGGTTCCTTATTTAGATTGCGACAGGCATTTTGCTACAGTAAGAGTTAAGAAAAATCGGTGTCAGGGAAATTTTAAATGAAGAATTTACTAATTCCAAGATATCCAGACAGCATTACTGACTCAGTTTGGCCTTCAGTTTGACTCACAACTATCTGAATACTTAGAAGTAGCATATTTTTGAGAGAAAGAAGTAGATGAAG |
| 132T | A00583:504:HM2HWDSXY:2:2141:30617:34100 | AAAATATAATCTGCCAAAGATGTTTTGGACAAATTACCGAAAACTATGGAAGAAGCATGACTCAGGGATGAAATGGCTACAAAACCACAGTTCAGTGAAATCCTGTCCCACAGCTTAACTGTGATCAGAGTGATTTCTGTCTAATTTTCTGC |
| 30H | A00583:504:HM2HWDSXY:2:2140:6533:1094 | GTTGTAGGCCCCTACGGGCTACTACAACCCTTCGCTTACGCCATAAAACTCTTCACCAAAGAGCCCCTAAAACCCGCCACATCTACCATCACCCTCTACATCACCGCCCCGACCTTAG |
| 134T | A00583:504:HM2HWDSXY:2:2132:32407:29653 | AGATACTTAATCTCTCTCAGACTCATTTTTACTATCTTTACTCATCTTTTTTTTTTTTTTTTTTTTTTACCACTCCTAAACTAGGAGTACTAATCATTTAGAATTCAAAAAGTTTTAAATTGG |
| 1H | A00583:504:HM2HWDSXY:2:2124:2528:23500 | TCAACTACCTAACCAACAAACTTAAAATAAAATCCCCACTATGCACATTTTATTTCTCCAACATACTCGGATTCTACCCTAGCATCACACACCGCACAATCCCCTATCTAGGCCTTCTTACGAGCCAAAACCTGCCCCTACTCCTCCTAGACCTAACCTGACTAGAAAAGCTAGTACCTAAAACAATTTCACAGCACCAAATCTCCACCTCCATCATCACCTCAACCCAAAAAGGCATAATTAAACTTT |
| 135T | A00583:504:HM2HWDSXY:2:2123:26476:31031 | CAGTGCACTTGCTGGCAGATTTTGAATGTGCATAGCAACTCTGTTCCTGGAATTTTAAAATGTGGAATCTGTGACAGACTCTTACCATTTTATAATCAGTTTCAGTACTTTTTTTTTTTTTTAACTTTTGACTTGTTTTGTGTCTTGTAGCTCAGATAAGTGATTGCAATAATTTATCCAAATTTATCTTTTAAAACTCTGCAGAGAATAGTCTTTCTCCGGGATTATTAACATACCT |
| 136T | A00583:504:HM2HWDSXY:2:2123:24198:6950 | TTTGAAAGCTAGATAACAGTAAAACAATGTCTTTAAAATTCTGAGAGACGTAATATTCAGCTTACAAATTTTTACCCAAACTATCAATCAAATATGAAGGTAGAATAAGGGCATTTTCAGACAATTAAATGTCACAGTAAATTGGCTTGTCTCATCCCCATTCTTAAAATGCTATTGAAATATGTGCTCTTTCAACAGAAAG |
| 24P | A00583:504:HM2HWDSXY:2:2123:16866:11005 | CACCACCACGGAGACCACCTCACACAGTCCTCCCAGGTTCACTTCTTCAATCACCACCACCAAGACCACCTCAGACAGTACTCCCGTCTTCACTCCTTCTATCGCCACGTCCGAGACAAGCTCACACAGTACTCCCGGCTACACTTCTTCAACTGCCACCACC |
| 137T | A00583:504:HM2HWDSXY:2:2121:20582:2096 | CACTGAGACCACCTCACACAATACTCCCAGCTTCACTTCTTCGATCACCACCACCGAGACCACATCCCACAGTACTCCCAGCTTCACTTCTTTGATCACCACCACGGAGACCACCTCACAAAGTACTCCCAGCTTCATTTCTTTGATAAACACGTCTGAGACCCCCTCACA |
| 97T | A00583:504:HM2HWDSXY:2:2119:9200:9502 | TGGCTACATTTTTTTGTTGTTGTTTCTGTATTTTTAGTAGAGACGGGGTTTCACCATGTTGGCCAGGCTGGTCTTGAACTCCTGACCTCGTGATCCGTCCGCCTCGGCCTCCCAAAGTGCTGGGATTACAGGCATGAGCCACCGTGCCCG |
| 138T | A00583:504:HM2HWDSXY:2:2118:9598:20995 | ACGAAAACAACACTAATGAATGTTAATGAAAGCAATGAAGAGAATCTCCAGATCTTTCACATTCATCTTTGACATCTTTGACATAGGCTGTGAACATATGTCAAGGAGATTGAATTTCCAGGTGATAAAATATTGGTAATGTGGAAACAA |
| 139T | A00583:504:HM2HWDSXY:2:2117:15076:18490 | CAAGATTGCGCCACTGTACTCCAGCCTGGGCAACAAGAGTGAAACTCCATCTCAAAAAAAAAAAAAAAAAATACAGAATTAGCCAGGCATGGTGGCGTGTGCCTGTAATCCCAGCTACTCAAGAGGCTGAGGCAGGAGAATCACTTGAACCTGGGAGGCAGAGGTTGC |
| 140T | A00583:504:HM2HWDSXY:2:2116:5990:31344 | CTGTGCCTCACTAAAATTCAGAGACTTGTGGGTCAGTTTAGTTGGGGTAACCCTGTACACACACACACACACACACACACACACACACACACACACACACACACTTTCTCTCTCTCTCTCTCTCTCTCTCTCTCTCTCTCTCTCAGGTAGGGCAACACTTTAAACACACTCTCTTTCTCTTTTTCTCTCTCTCTCTCTCTCTCTCTCTCTCAGGTGGGGCAACCC |
| 37H | A00583:504:HM2HWDSXY:2:2116:32072:6527 | CCCACTCATCCTAACCCTACTCCTAATCACATAACCTATTACCCCGAGCAAGCTAAATTACAATATATACACCAACAAACAATGTTCAACCAGTAACTACTACTAATCAACGCCCATAATCATACAAAGCCCCCGCACCAATAGGATCCTCCCGAATCAACCCTGACCCCTTTCCTTCATATATTATTCAGCTTCCTACACTATTAAAGTTTACCACAACCACCCACCCATCATACTCTTTCACCCACAGCACCAATCC |
| 141T | A00583:504:HM2HWDSXY:2:2115:14534:4116 | GCAAGACTTCCTGACAGTCACATCTGTCCTGTCCTTGTGTTGGTTGGTCTTGTTGCTCTGGTCACTTCCTCTGTCTGGCTTTCTGAGTTCAAGTAAAACATACTGAGCAAGTCTCAACACACACACACACACACACACAACACACATAAAAAAA |
| 142T | A00583:504:HM2HWDSXY:2:2114:3766:32377 | GCAGAGGTTGCAGTGAGCCGAGATCCCGCCACTGTACTTCAGCCTGGCGACAGAGTGAGAGACTCCGTCTCAGAAAAAAGAAAAAAGAAAAGAAAAGCCCCACAGACCCAGCGTGA |
| 143T | A00583:504:HM2HWDSXY:2:2110:5376:14372 | TTATTGCCATAACTTTTTCTCTTTATACCACTTTAGCTGTGTCCTGCAACTTCTGGTATGTTGTTCTTGTATTTTCATTCAGTTCAAAATATTTTTATAACTTCGTTATGATTTCTTCTTTCACCTATGGGTCATTTAAAAGTGTGTGTGTGTGTGTGTTTGAATCCCGATGTTTCCTTTTTTCTTGATATCTTTCTATTACGTATTTCTCGTTCAATTTCTTTTTGGCCAGAGAA |
| 144T | A00583:504:HM2HWDSXY:2:2101:21721:3693 | ACTATAGCTTTGTAATACAATTTTAAATCAAGAGGTGTGATGCCTCCAACTTTTTCTTTCACAGTAATCTGTTGGCTGTTTGGGGTTTTTTGTGGTTCCATACGAGTTTCAGGATTGTTTTTTCTGTTTTTGTTTTTTTTTTTTTTTCTTGAGATGAAGTCTCACTCTGTTGCCCAAGTTGGAGTGCAGTGGCACAACCTTGGCTC |
| 145T | A00583:504:HM2HWDSXY:2:1676:1922:15749 | TCAGCAGTTTTCACAATTTAATTATTTTATTATTATTATTATTATTATTATAGTAGAGACAGGGTCTTGCCATGTTGCTAGGCAAGTGTCAAACTCCTGGGCTCAAGTGACCCTCCCACTTCAGCCTCCCAAAGTGCCAGGGT |
| 8M | A00583:504:HM2HWDSXY:2:1676:18548:15640 | AAATCAACGAGGGGGGCAAACCATTGGAACCACTTGTTAATATAATGCACCCAAGTATGACCACCCACTGAGACATCAATAATAGACACATACTAGGCATATAATTTTTCCTGACAATTAAAACCTTGTAGAAGTATAATTTGTCAGAGCTGCTGTATTGGATTGCATTAACTTGTATGGGTGTGCC |
| 146T | A00583:504:HM2HWDSXY:2:1676:15935:12806 | GCAGACTGACACCTCACACGGCCGGGTACTCCTCTGAGACAAAACTTCCAGAGGAACGATCGCGGTTCACGAAAAACCACTGTTCGGCAGACACTGCTGCTGATACCCAGGCAAACAGGGTCTTGAGTGGACCTCTAG |
| 147T | A00583:504:HM2HWDSXY:2:1675:3432:28322 | AGTTGCCATCTGTGGCTAAGAGCACCGCAGACTGCAGGCCCCGAGGCGGAGCAAAGGTGGGTGTTCAAGCGAGAGGCTGGCATGCAGGTCTGG |
| 148T | A00583:504:HM2HWDSXY:2:1675:15338:16752 | AATATAGATCATATGATACATCACAAAACAAATCTCAATTTTTTTTTTCATATTCAGGCACCTCAAAACATATTTTCTTTATTTCCTCCAGAGAACTGGCTTTAGTTCTGTGGTCTTTGTAAACATTTAACACAAAATAACAAACTGAAGAATTTATATCTTTGCCCTGTGTTGGTGGGTTGTGTTTACTTATGTTTAAGTTG |
| 149T | A00583:504:HM2HWDSXY:2:1669:7970:9064 | GCACTTGCGGTTGCCAGCTTCTCCTGGATCTTTCCCCCACAGGTTTAAAATGAAAAAAAAAAAAAAAAAGAAAGAGAACAAGATCCTAAATGTAACATATTTTTCCCCACTATTTTCTAAGCTTCCTTATTTG |
| 150T | A00583:504:HM2HWDSXY:2:1668:8603:12007 | AGCTTTTGGATTTGATTCTTATTGGAAACAAAGTCATTTATACACAATCATGTTATTTTTCCCTTCCATTAGCCAAAATAGCCAACAACAACAACAACAACAACAACAACAACAAAACAAAACCAACAGTAATAACAAGAATTGAGTGGACATAAATAATGACAAATATATGTAGGCTAGACATATCTATATGTACGTACATGGGCTAGTTGAATTCTAAAACAGAGGTAAAATCTGTTCTTCAACTAATATTAACATCTATTGCTAATCTTCCATGTGCTTTGAAAG |
| 151T | A00583:504:HM2HWDSXY:2:1659:27172:5556 | GAGTTTGACCTCTGACGGCCAGTTGTAATAGCATTAAGTCTTTGAAATTTTGTAGCGGGGTAGAAGGGGCTAGGAAAGGAAGAAAACATCTTTTTAAAAATATAAGCGATCGGCCGGGCGCGGT |
| 115T | A00583:504:HM2HWDSXY:2:1639:3974:9627 | GTGCAATTGAAAGTTGACATGTAATGTGAGCTGAGATTGTGCCACTGCACTCCAGCCTTGGTGACACAGTGATATCCTGTTGAAAGAAAGGAATGG |
| 155T | A00583:504:HM2HWDSXY:2:1638:29604:35853 | TGGAGATGATAGATAACAGATACATAGAGATGATAGGTAGATTAGATAATAGATAGATGAAAGATCAATAGATAGATAGATAGATAGATAGATAGATAGATATGATAAGTAGATAGGGAGAGAGATAGAGAGATGATAGGCAGACAGATGATAGATAATCACACACACATACATCCTATTGGTTCTGTTTCTCTGGTCTCAGGCTGAACC |
| 156T | A00583:504:HM2HWDSXY:2:1637:8232:10300 | GATTGAAGTTCTATTTACTGCTGGCAAAAAGCAACTTCATGGTTTGAAAAAAAAAAAAAAGACATACAAAATACTCATACC |
| 157T | A00583:504:HM2HWDSXY:2:1636:2889:8938 | TCATGCCTGTAATCCCAGCTACTCGGGAGGCTGAGTCAGGAGAATTGCTTGATCACAGGAGGCAGTGGTTGCAGTGAGCCGAGATCACGCCACTGCACTCCAGACTGGGTGACAGAGCAAGATTC |
| 35P | A00583:504:HM2HWDSXY:2:1635:16333:1564 | CGGGGGGCCGTGGCCGGGTGCTGCGGCGGGCGAATTGGGGGCGCGGGCCGCGTGCAGCGGGAGGGGGAGGGGCTGGCCGGGTCCCCTCGGGTGTGGCGGAGTCCTGGCTTCCGCGCTGTGGACCATCGCAACTCCCAGCACACCTGCGGG |
| 25P | A00583:504:HM2HWDSXY:2:1629:26169:26772 | CAGCTGATAGTGAAAGAGACATCAACAGACATGTGGTCAATACCAGTAGATTTTTAGAAGATAGATTTCTCCAAGGTGTACTTTTTATTTTCATATCATTTCAAAAATTAACTCTAGATTCAATTTGCTATTTCTACAGAAGTGCTAC |
| 158T | A00583:504:HM2HWDSXY:2:1625:28537:28369 | TGGTATTTATTCTATTGATATGCTATAATATATTAATTAATTTTTGGCTGTTAAACAAAATTTGCCTCCTAGGATGAATCCCACTTGGCCAAGATATATAATTATTTCAACATGCTGATCAATTAG |
| 3G | A00583:504:HM2HWDSXY:2:1624:32687:9157 | GTGATCTATAATGATACTCAAAAAGAACATTGTGATGCAGTTCTCAGGATCATCCGTAAGCTTGCTAACGCAGGGCTTCAACTTGATTTTGATAAGAGCGAGTTTGAAGCAGGCACTATTAAGTATCTTGGATATTTGGTGAAGACCGGTTGGGGTTTGCAGGCTGATCGCACAAAACT |
| 17P | A00583:504:HM2HWDSXY:2:1621:14886:29246 | TAGTGTCTTAATTATGTCATAAAGTTACCTAGTAACAGAATCCAATAAAATATCAATGTACATTCTAGTATCTTTGCTTTTTGGGATATATATATGTATGTATATATATTATATATATATTATAAATATATTTTATAATATATATTTTTTGGAAACAGGATCTTCCACTGTCACCCAGGCTGGAGTGCAGTGGCAT |
| 6C | A00583:504:HM2HWDSXY:2:1617:30273:9768 | CGGCCCCTGAGACAGCGGGTTCCGCCGAAGCTCCGCTGCAGTACAGCCTGCTCCTGCAGGACCTGGTGGGTGTCAACCGTCAGCCCCGGCTCCCGGAGCCTGGGAGCCTGGGCGGATCCCAAGTCCAGCCTATAGTGAGGTGCAGATGATGTTCTAGTAGGGGTAGGAAAGCTGCGCCTTCAAGGCTGCGCTGGCCTGCGTGGGAG |
| 159T | A00583:504:HM2HWDSXY:2:1617:19877:35540 | GTGTCTCCACCCGAATCTCATCTTGAACTGTAGCTCCCACAATTCCCACATGTCCTAGGAAGGACCTGGTGGGAGGCAGTTGA |
| 9M | A00583:504:HM2HWDSXY:2:1617:18873:15734 | GCTTGTGTTCCATCACCTGGACAACAACAAACAGACTGAGTCAAGTACAAGTCCCAGAGTGCACTGCTGTCAGTGCTGTCACT |
| 160T | A00583:504:HM2HWDSXY:2:1613:11822:29356 | GACAATTACTGGCCAGGCGCGGTGGCCAGAGCGAGACTCCATCTCAAAAAAGAAAAGAAAAGAAAAGAAAATTACTGGCGGCAAGCAGGAACATTGTAGATTTTGAAACTGTCTTGTTTTACAAGATACTGAAGCAAGGTGGTGCAATTATTACGTCCTTCTAAAGCTGATCGGATAAAGGCTTTAATTTTGTAATTTTCAGAGAATATTACCAATGTAGC |
| 16T | A00583:504:HM2HWDSXY:2:1612:9362:11412 | GCTGGCATCACAGCCCTAAACATGGTGGTAGCTGTCACCGGTGGGAATTAGGCTGAATAAATGAAAGGAGGCTAAGCCCCAGGACGTGTCCTGGAACTTAACAGTGCCAGTTTATTCGCTTTATGTAAATGCCTGCTGCTATTCCTTGCCTTATGAAGGGCTTTCAGTACAAGAATTCATCTTTGGGGAAAATAGCCTTTGGCTCAATGGCTGTGTCTGCCAAGT |
| 161T | A00583:504:HM2HWDSXY:2:1612:31385:6057 | TGGCTGGGCCTCACGATCTGGCCGGGTGCCCGTGAGGGCTGGCCCGCTGGCACCCTTGGTAACAAAGGCCAATCATGGACCCTCCGCCTGGTCAGGAACATAAGGAGCCGGTCCTGGGCTGGGCCCTGGGGGTGAGGGACGTCTGACACCCAGAACCAGAGTGCAGGACAGAGGCCGTGGCTGTGGCTGGGCTCTGGGTGCTGAGTGGGACA |
| 26P | A00583:504:HM2HWDSXY:2:1611:4878:6057 | TACTAGTTTACCTTTCAGATTTGTGATTAAGGGCTGGGTGTGTGTGTGTGTGTGTGTGTAAGTGTGTTGGGGGAGGGAGGAGTCTGAGAAAGTGAGCAAGTACGTAGAAACTTTTTTTTAAAAAGTCCAGATTAAGTGATTGAAATCGACT |
| 162T | A00583:504:HM2HWDSXY:2:1609:6686:31892 | GGACAAGTAGAATTGTACAATATGTGATCTTTTGTGGCTTTTTTTTCCCTCTTAGCACAATGTTTTCAAAGTTCCTTTATGTCATAGTGTGTATCAGTATTTCATTTCTTCTGTGGCTGAATAATATTCCATGGTAGAGACACACT |
| 163T | A00583:504:HM2HWDSXY:2:1609:3052:5212 | ATCCTAAAATTCCCTACTGTGTCCCATTACAAACACTACTCAAGCTGCCTTTTAATGCCACAGATAGTATTGCCTGACTCTGAACTGTGTATACAAGGAATTATATGGTAATTCCACATCTGACTTCTTCCACTTAGTATTATGTTTGATTCAGCAATGCTGTTGGTACTAACAGTTAGTACATATTTTAAAGTAAAAAGTAGTCTTGGATGTGTAATATGCAAGTATGTCTGTCAAAAAAGACTAAGATGCTGTTAAGAACAA |
| 165T | A00583:504:HM2HWDSXY:<br>2:1608:19144:11224 | AGGAGTATGATGATGCCTCTGGCTTTATTATTTTGCTTGGGATTGGTTTGGCTGCTTGGGCTTTTTTTTTTTT<br>TTTTTTTTTTTTTTTTTTTTGGTTTCAGATGCATTTTAGAATTTTTTTTTCTAATTCTGTGGAGAATATTGTTTA<br>TGATTTGATAGGAATATCTTTTTTAAAAAGCACAAGTATGAATAAATTTAAGAAAAATGTAAATTAGTTCT<br>ACTCTAGAAACTACAGAATAATGCAGATAGAAATTTTT |
| 166T | A00583:504:HM2HWDSXY:<br>2:1607:26078:36260 | CGGCCAAGGATGACGTTGATTGCGGACACAAATACGTAAAATTTCTTAAAACATTACGAGATTTTTTTTTT<br>TTGCCATTTTTTTCTTTTAGCTTATCAGTTATCGTTAGT |
| 167T | A00583:504:HM2HWDSXY:<br>2:1603:25482:35321 | AGGAGACAAAACCACAAGTTACTTATAAGTAAATTTTAGATCCACATCTCATAAACCAGAGGGATATCTG<br>CATAGTCACAAATATTAACAGAAGAGTATC |
| 168T | A00583:504:HM2HWDSXY:<br>2:1603:17824:23343 | TTCTTTAATTTTATTCTCAGATTTTTCACTGCTAGCATGTAGAAATACAATTGATTTTTTTTATCTTGTATCA<br>TATAATCTTACTATCCTCCTTTATTAGGGGTAGTAGTGTGTGTGTGTGTGTGTGTGTGTGTGTGTGTGTGTG<br>TCTGCGTGTCTGTGTGATTTCTTAGAAGTTTCTT |
| 169T | A00583:504:HM2HWDSXY:<br>2:1603:12228:10394 | GAAGGCACTATCAGAAAGGGTACATTGAAGTTGAGGTAACTTCAAAAGGTAGTGAATACCTGCTATTTGG<br>CAGTCCTCTTTTCTGCCTGCCCCTAGTCCGCCGTTTGAAGGAAACGTCATACAATGAATGCATCCGCGTTAG<br>GAAAAGTACTTTTAAGAATGTGGCCCTGT |
| 4M | A00583:504:HM2HWDSXY:<br>2:1576:29568:8296 | ATGCTTGTCTAGGTAATTGCTTGGATGAGTTCGAAATCGCATGTGAAACTAGGTCACCGGAGCTAAAAATA<br>GAAAAATCTTGTTAACACTCTAGTTGCTACTGTTTTGATCC |
| 170T | A00583:504:HM2HWDSXY:<br>2:1574:29668:34710 | CCCAAGATCATGCCATTGTACTCCAGCCTGGGTTACAGAGTGAGACTTTGTCTAAAAAAAAAAAAAAAA<br>AAAAAAAAAAAAAAAATATATATATATATATATATGTAAATAATAATGGCTGAAATTTGATAGATGTTTTTTAC<br>CCTCTTAATTTCTTAGT |
| 2L | A00583:504:HM2HWDSXY:<br>2:1572:3568:4601 | AAGAAGGAAGCTCTATCTGCAAGCCAAGAAGAGAGCCCTTAGAAGAAACAAACTTGCCAATACCTTAATC<br>TTAGACTTCTAGCCTCCAGAACTGTGATAAAATAAATTTCTGTTGTTTAAGCCTCATTGCTTGTGTCTTTTTG<br>TTATAGCAGCCCTAGCAAACTAATACAGAGGTCATAGTGATCATTAATGGGTATATACCACACAATAGATC<br>TTCTAAACACAGGAAAGAATAAAAGAGTTAGAAGGA |
| 3M | A00583:504:HM2HWDSXY:<br>2:1570:3884:16297 | AATGAAAGAATATAATGTCCCTGTCGGGATTCGAACCCACAGCGGTGAGGGGCAAGTGGTTTGAAGTCAA<br>CGACCTTAACCTCTCGGCCACG |
| 10M | A00583:504:HM2HWDSXY:<br>2:1568:15257:11068 | GGGATTTAAAAAATCAACTTAGGATTTTCATAAATTTATTCATTTTAGGGTTAGGGTTAGGGTTAGGGTTA<br>ATATTAGGATTAGGTTTCTAAATCCCAAGTTACGGCAAAAAAATCTCGAGATACGGCAAA |
| 5H | A00583:504:HM2HWDSXY:<br>2:1565:24469:28087 | TGAGTGGTTAATAGGGTGATAGACCTGTGATCCATCGTGATGTCTTATTTAAGGGGAACGTGTGGGCTATT<br>TAGGCTTTATGGCCCTGAAGTAGGAACCAGATGTCGGATACAGTTCACTTTAGCTACCCCCAAGTGTTATG<br>TGCCCGGAGCGAGGAGAGTAGCACTCTTGTGCGGGATATTGATTTCACGGAGGATGGTGGTCAAGGGACC |
| 171T | A00583:504:HM2HWDSXY:2:1559:25943:11005 | CCTAAAACCTCATGGCTTGCCTGCCTCCACAGTCTTTTTCCTTTTCAATAGCCTATCCATCACCTGGTCCTGGCGAACCCACCTCAGAAATGTTTCCCAGATAGGTCCAGACCTAAAGAGCCAAC |
| 9C | A00583:504:HM2HWDSXY:2:1551:14922:6230 | GTTGCGTGTCCTGTGGATATTGAGGTAACGTTGGTGAGAGGAAGAGGGGTAGCGTGACCTGTGGACTCTGAGTAAGCGGCGGTGACAGGAAGAGTGGTGGCGTGACCTGTGGACTCTGAGGAAGCGTCGGTGACAGGAATAGGGG |
| 127T | A00583:504:HM2HWDSXY:2:1549:27579:26428 | AAGGTTGCATGGTCAACCTGTTGACCACGGCTGGTGCTGCCATGGGAGAAATGGTGATAATGGTGGTGGTGATGGTGATGGTGATATGATGTTGATGGTGATGTGATGATGATATAATGGTGATGGTAATGGTGATGGTGATGTTGGTGATGGTGGTTATGTGATGGTGATAGTGATGGTGATGTA |
| 172T | A00583:504:HM2HWDSXY:2:1548:31611:9518 | TGGCCTCTCAAAGTGCTGGGATTACAGTCATGAGCCACTCTGCCCGGCTGATAATTTCTAATCTTTTAGACTACTATATCCCAGGGAAAAACCTCTCATCCAGCCTTCCTTGTTTTAAAACAGTTTCCTGAGCCTATTTCATCACACACACACACACACACACACACACACACACTTCTCTTTACTCATGCTTTTCCATTGTTAATAAACATGCTTATTAAAAAAC |
| 173T | A00583:504:HM2HWDSXY:2:1541:18665:31062 | AAATCCCAAGGGAAAATTAAAACTCAAGTGGACCACTCAGATATCTGTCTTATAAATTGACTTCTCATCTCATCACTACTTTGCCTGACGAAGTCTCAATTTTCGAACAAATTGTAATATATTCTCCATGGGAAAAAGCTACAAAC |
| 174T | A00583:504:HM2HWDSXY:2:1538:23185:33849 | CCAAGCCGAGCATCAGCCCGTCGTGGCCCATGGACTCTGGGTTCCCCAGGCAGGGGAAATGTGCCAGAGCCACGGGGAAAGCTGAGCTCTGCGGCCCTGCGAGCGCGTTCACAAAGGTTTATCAGTTTGTAAAACTTGCAGAAAACTCAGCCATCCGTGTTTGGGACAGGCGGCCCCTTCCACTCAGGGTTCATCCGTCACGCTCCTTTCGCCT |
| 4N | A00583:504:HM2HWDSXY:2:1536:9851:21684 | ATGTTAAAAGTACTAATTCCTACCATTATATTATTCCCAACAATCTGATTAACTTCCCCCAAATGACTATGAACAACCACAACCACACATAGCCTCCTAATTGCCCTTATTAGCCTCACATGATTAAAATGAACATCAGAAACCGGATGGACTATATCCAATTCATATTTAGCCACAGATCCACTATCAACCCCTCTCTTAGT |
| 175T | A00583:504:HM2HWDSXY:2:1536:25554:21887 | TATTTTTAACTTTTAAAGTAACTTTCATACTGGCTACACCAACTTACATTCCCACCAAGAGTGTTCAAATTTTCCACTTTCCTTTACTCCACATCCTTGCCAACAAATCTATCTATCTACCTATTATATATCTATTTATCTCTCTCTCTCTCTCTCTCTCTACTTATCTAATAGCCAACCTAACAGGTGTGAGGTGATATGTTATTGTGTTACAGTAAGGACTTTTTGCTTGAAAAT |
| 176T | A00583:504:HM2HWDSXY:2:1535:23475:13119 | CCTTGCAAGAGTTTCTCTTGTTTATTTGTTTGTTTGCATTTTTATTTATTTAAATAGTCAAATCTGTGACTGAGATAGTTTTTTATGCTTTTTTACCC |
| 177T | A00583:504:HM2HWDSXY:2:1533:8287:28964 | CCTCTCACCATTTCCCAGTGTCCCTCACCATTCCCTAGTGCTCTTCACCATTTCCCAGTGCCTCTCACCATTCCTCAGTGCCTGATGCCCTCTCTTCCCTGGACATGCACACATATTGTTATCTTGGATGGCCACCCTTC |
| 178T | A00583:504:HM2HWDSXY:2:1533:25075:8907 | GTTGCACAAGTCCATAAATTTACTAAGAATCATTAAATTGTATGCTTACAATGAGTGAATTTTGTAAAATGTAAATCATATCTCAATGAAATTGTTAAGAGAAGAAGCTGGGGACATTTCGACTCTGATTAGTCACCCACCTGACTAGTCATAG |
| 179T | A00583:504:HM2HWDSXY:2:1530:31114:23437 | GTGAACCCAGGAGGCGGAGCTTGCAGTGAGCTGAGATCGCACCACTGCACTCCAGCCTGGGCAACAGAGCGAGACACTGTCTCAAAAAAAAAAAAAAAAAAAAAAAAAAAAGGAGTCTTTAGGGTAAGGAATATTAACAAGGATAAAGTCAGACATTCCATCATGATAAACAGTCAATTCATCAAGAAGATATGCAACAATCCTAAATCTGATGGACTAAATAA |
| 180T | A00583:504:HM2HWDSXY:2:1525:15167:18897 | GGAATGGAGTGGAATAGAGTGGAGTGGAATGGAGTGGAATAGAATGGAATGGAATGGAATGGTGAAATGAAATGTGAGCTGAGATTGTGCCACTGTATTCCAGCCATTTTTGACACTGTGAGATCTTGTCAAAAGAAAGGAATGGAATCGATTGGGATGGAATAGAATGGAGTGGAATGGAATGTTGTGGAGTGGAGTGGAGTGGTGTCGAGTGGAACGGAGTGGAATGGAATGGGATG |
| 181T | A00583:504:HM2HWDSXY:2:1522:22752:32033 | GCAGACTTAAAAGTTATGAATAAAAATACCTTCTGTGATGAACAGAAGTTTCTCATTGATCACACTGACAAGAAACCTTTCGATGAACAGTCCGGGTAGAAGACATGTCACTCTAAACAGTGTCTGTCTTCTTTCCTGGGCCTAACTGAACTTTGCAGAGTTCATTTGGGACATAAGGAAACAGAAGGTGTCAACTTGGCCCTTGTGCTCACTGGGAAAATGGGGAGAGACAGAACTCATTCAGGAGACCT |
| 182T | A00583:504:HM2HWDSXY:2:1520:24578:20134 | GAGAATGGCGTGAACCCAGGAGGCGGAGCTTTCAGTTATCTTTTATCGGGCCACTGCACTCCAGCCTGGGCGAAAGAGCAAGACTCCGTCTCAAAAAAAAAAAAAAAAAAAAAAAAAAAAAAAAAAAAAAAAGAGTACAGGGCCAGGGTCCTTTTCCATGAATCTCCT |
| 183T | A00583:504:HM2HWDSXY:2:1519:18123:21512 | GTGAGAGTTACCTAAAGCTCAGCGTCCATGATGGTCTACGGGGCTTCTGAGGTGATCGGGCAGCATAAGTCTTCAGCCGCTAAGCTGAGAAGATCTGGGAAGGAGTCAGTCAGAGAGCCTTGGGGCAGAGTTCCAGGGGCTCTGGGAGTG |
| 184T | A00583:504:HM2HWDSXY:2:1515:29866:35744 | CGGGAGGTGGAGGTTGTAGTGCCTAGATTGTGCCATTGCACTCCAGCCTGGGTGACAGAGTGACAGAGTGAGAC |
| 23H | A00583:504:HM2HWDSXY:2:1515:29731:25207 | CCTCAAGATACTCCTCAATAGCCATCGCTGTAGTATATCCAAAGACAACCATCATTCCCCCTAAATAAATTAAAAAAACTATTAAACCCATATAACCTCCCCCAAAATTCAGAATAATAACACACCCGACCACACCGCTAACAATCAATACTAAACCCCCATAAATAGGAGAAGGCTTA |
| 186T | A00583:504:HM2HWDSXY:2:1514:2772:15781 | GATCACGCCATTGCACTCCAGCCTGGGAGACAGGGCAAGAGTCTGTCTCAAAACAAACAAACAAACAAACAAACCCAATGGTGGTTTTTGAAATACAGAAAAATTCATTCTCATATTCAGATGGAATCTCAAGGAACCCTTAATAGCTGAAACAAACTTGA |
| 187T | A00583:504:HM2HWDSXY:2:1513:4101:26537 | TGTTTACGATATGACTTTCAGCTATTGAATTTTCCCCCACAGATGTGAAGGGATTAAATAAATCAGACAAGACATTATTAATGTCAATGCAGTTGGCCTTGTAGAATGACATTTGTCCAGAAGGCCTCCAAAGTTATCCTCAGTCACAGC |
| 188T | A00583:504:HM2HWDSXY:2:1510:3079:23359 | GCCATGGTTTACCAGATGCACAAGGCACTGTTCTTGAAGTGCTTACCTTCCAGTTGGGGGATATAAAAAAGGCACAAACAAGAGAATGAGTTAGAGAAACAGACCATGTGAAATCACACAAACACATGGACAGCTGGATTGACCATTTGGTCCCATTTTTCACTGTCATTTAACTTCTTAGTCTTCCTCATCTGTGAAAGAGAAGAAGAAGATGTGACTTACCTGTTT |
| 189T | A00583:504:HM2HWDSXY:2:1510:26485:13228 | TGGAATTTCCCTAGTGATCAGGGTGAGAGGACTGTTTCTGTTATTCATAGTAAGCC |
| 190T | A00583:504:HM2HWDSXY:2:1506:9290:15201 | CCTCTGAGGAAGTCTTCAAAAAGATACATAAAATACCTGTGACTCTATTAAAGATTCCTCTTAAGGAGGCAAACCCTGACCTTCGAGTGACTGAGTCTGTGACAT |
| 3H | A00583:504:HM2HWDSXY:2:1502:4598:29121 | TATACTAAAAGAGTAAGACCCTCATCAATAGATGGAGACATACAGAAAAAGTCAAACCACATCTACAAAATGCCAGTATCAGGCGGCGGCTTCGAAGCCAAAGTGATGTTTGGATGTAAAGTGAAATATTAGTTGGCGGATGAAGCAGATAGTGAGGAAAGTTGAGCCAATAATGACGTGTAGTCCGTGGAAGCCTGTGGCTACAAAAAATGTTGA |
| 1P | A00583:504:HM2HWDSXY:2:1473:5095:19398 | CAATTTATGGTTGCCAGGAATATAATCACTAAAGACTTGCTTCAGACCAAAGTCTTGAGGTCTTTTTCTGATGAAGAAGGGAGTGCCGTTGATCTGCTGTTGCTGTCAGACGTTGACATTGATTCATCTGACATTAGGTTAATCATGCCC |
| 191T | A00583:504:HM2HWDSXY:2:1471:6262:9862 | CAAAAAAGTAAAATAAAATAAAAGAGAAAGAAGAAATGTGCTATTCAAGATGACAGCAGGTATTGAATTTAAAACTTTTTTTCAAATAAGTTTTGGAAACTATTTTTTTTGTGACAGCTACTTATAATGTCTTTC |
| 192T | A00583:504:HM2HWDSXY:2:1471:30526:30405 | GCCTGGGTGACAGAGCAAGGCTCTGTCTCAAAAAAAAAAAAAAAAAAGGTATGATTTACTTTTCATTTAACTGGGTGTCTGGTGAGACATTTCTCACCTGAGGCATGGATAAATATC |
| 1G | A00583:504:HM2HWDSXY:2:1469:20347:7670 | ATTAAAAATAGATGTATTCTTACTGGTCGTACTCATTCAGTATTACGAATTTGTAAGAATACATCTATTTTTAAT |
| 193T | A00583:504:HM2HWDSXY:2:1468:26883:16705 | CGGCCTCCCAAAGTGCTGGGATTACAGGCGTGAGCCACCGTGCCCGGCTTTTTTTTTTTTTTTTTTTTGAGATGAAGTCTCGCTCTGTTGCCCAGGCTGGAGAGTGCAGTGGTACGGTCTCAGCTCACTGCAACCTCCAC |
| 194T | A00583:504:HM2HWDSXY:2:1467:11912:4805 | AAAATCTATGCACACAAATCAGTAGCACTGCTATATACCAACAGTGACCAAGGTGAGAATCAAATCACGAACTCAACCCCTTTTACAATAGCTGCAAAAAATAAAATAAAATAAAATAAAATACATAGGGATATACCTACCAAGGAGGTGG |
| 195T | A00583:504:HM2HWDSXY:2:1466:28971:22921 | GGAGGCCTAGGAGGAAAAAATGGTTTTATGGGCCAGGCCCAGGGCCCCCTTGCTCTGTGCAGCCTAGGGACTTAGTGCCCTGCATCCCAGCCATTTCAGCCATGGCTAAAAGGAGCCAAGGTATAGCTTGGGCTTTTGCTTCAGAAGGAT |
| 196T | A00583:504:HM2HWDSXY:2:1466:19904:19617 | CGCCATCCTCACCATCCTCACTGTCCTCGCCATCCTCGCCATCCTCAGAGTCCTCGCCATCGTCGCCACCGCTTCGCCCTTCGTCGGCCTGTTCGGCACCGTCTGGGGCATC |
| 197T | A00583:504:HM2HWDSXY:2:1465:23339:35806 | AGGTTTACTGATTTGGCTTTGATGAGGACTCTGAGCTGTATAACATGTACATTCATTCATTTAAATTATACATATATATATATATATATATATTTTTTTTTTTTTTTTTTTTTTTTTTTTTTTGAGATGGAATCTCACTGTCACCCAGGCAGGAGTGCAGTGGCACAATCCTGGCTCACTGCAACCTCTGCCTCCTGGGTTCGAGTGATTCTTGTACCTCAGCTGCTCAGCAGCTGGCATTACA |
| 198T | A00583:504:HM2HWDSXY:2:1461:8938:13495 | TGGACAAAAATGACCACAGATAACTAGCTCTTAATATACAAAATTTGCAAAAAAAAAAAAACATTGTGA |
| 199T | A00583:504:HM2HWDSXY:2:1461:22679:24549 | GACAGAGTGAGACTCCATCTCAAAAAAAAAAAAAAAAAAAGAAAAAGAAAAAGAAAAGAAAAAGAAAAGCCTAGGCCAAGTATTTGGTAGAATGTTCTTCAATTTAGGTGTGACTGATGTTTCCTCATGAGG |
| 133T | A00583:504:HM2HWDSXY:2:1460:4707:5447 | CGTGTGCCCAGGGGTCCTCTGTGTGCCCAGGAGTCCTCCCCTTGCTTCTGACATGATTATTTACGCCCATGGCCTGCATGCCCATGCATGAGACATCATGGTTTTCCCTGTAATCGTTATCTCAAAACCAGCTTTCCCCTGTGGCAACCCTCCCTCCTGAGGGCATTTTCAAT |
| 200T | A00583:504:HM2HWDSXY:2:1457:13060:7889 | GTTTTTATCTTCATGTTCTATGTTGTTTGTAAGTACAGTGACCTGGTCAGTGGGCTGAGGGAGTCTGAGAAGTTGCTGGGAACACTTTTTCTTGCCT |
| 201T | A00583:504:HM2HWDSXY:2:1453:30463:3552 | CATTCAACTCACAGAGTTGAACTTATCCTTTCATTGAGCGGTTTTGAATCTCTCTTTCTGAAGAATCTGCAAGTGGATATTTGGAGCCCTTTGCTCCGTATGGTGGAAAAAGAAATATCTTC |
| 202T | A00583:504:HM2HWDSXY:2:1451:5773:30592 | TTCTCATATAGATTTATTCTAGTTTTCATCCTGGGATATTAGCTTTTTCACCATAGGGTGCAATATACTCCCAAATATCCCTTCACCAATCGTACAAACACAGTGTTTCCAAAATGCTGAATGAAAAGAAAGGTTTAACTTTGTGAGGTGAATGCACACATGACAAAATGGTTTCTCAGATAGCTTCGTTGTAGTTTT |
| 3N | A00583:504:HM2HWDSXY:2:1451:26033:17832 | CCCGCAATAATTCACCGACTCCCCCCTAAACTTAACCTAACCTTAGGGCAATCAGTCGCCACCAAACTCGACCAAACATGA |
| 203T | A00583:504:HM2HWDSXY:2:1448:4345:5948 | GAATGGAATGAAATGGAATTGAATGGAATCAACGCGAGTGCAGGGGAATGTAATGGAAAGAAATGCAATTGAATGGAATCATGCTGAATGGAATGGAATGGAATGGAATGGAATGGAATCAACCCGAGTGCAATGGAATGGATTGGAATGGAATGGAATGGAATGGAACAACCCGAATGGAATGGAATGTAATGGAGAGT |
| 2P | A00583:504:HM2HWDSXY:2:1447:4969:12508 | TCTGTTTCTAGAGAACCCAGCCTGCAACACAGTCCTACGTGATCTAGCTCCCTGTAGCCTCTCTGAATTCCTGCTCACTCTGTGCCAAGCACACTGGCCTTTGATATTTCCAGAGGATTGGC |
| 14H | A00583:504:HM2HWDSXY:2:1444:3468:32174 | ACCCAACAATGACTAATCAAACTAACCTCAAAACAAATGATAGCCATACACAACACTAAAGGACGAACCTGATCTCTTATACTAGTATCCTTAATCATTTTTATTGCCACAACTAACCTCCTCGGACTCCTGCCTCACTCATTTACACCAATCACCCAACTATCTATAAACCTAGCCATGGCCTGCCCCTTATGAGCGGGCGCAGTGATTATAGGCTTTCGCTCTAAGATTAAAAATGCCCTAGC |
| 164T | A00583:504:HM2HWDSXY:2:1439:8793:23735 | GTAATAATAGTAGCTATCTTGTACTATGCATTTACAATGGTCCAGGCACTCTTCCAAACTCTAGATAGATAGATATATACATACATACATACATACATACATACATCATACATACATACATACATAGCTAGGTAGATAGATAGGCAGATGAAGTTC |
| 204T | A00583:504:HM2HWDSXY:2:1439:19822:17879 | CCAGCTTTCAGTCCTGGCTCTGTCTCATCCTAGCTGCCTGCCCTGGGGCCACTCTCTGCTTCCTCATCCATAAACCAGTGGCAGGAATGCCCACCCCATAGACTCGTGGTGAAGATGAAATGAGCCACATATGCGGAAGCAGATGGCACC |
| 205T | A00583:504:HM2HWDSXY:2:1437:19931:30248 | CAGCTCCTCCGGAGTTTGAGGCAGGTGTTTGTGGCTGCAGTGGGCCGTGATCATGCCACTGCACTGAAGCCTGGGTGACAAAGTGAGATCGTGTCTCTAAAACATTAAAAAAAAAAAAAAAAAAAAAAGGAAAAATGAGAGTGAGAACTGAAACAGC |
| 206T | A00583:504:HM2HWDSXY:2:1436:20446:28260 | CAGCTGGTGGCTTTCAGGGCCAGGGCAGCACCACACAGCACACCTCAGATGCCCAAACACGCTAACCCTAGAAACAGAGCCTCAGACAACTTGGTGCCAGAACTGCTCCTTTGGGCTTTTGTTTTGCTGTATTTTATTTTATTTATTTATTTTTTTGAAACAGAGTCTCGCTCTGTCACCCAGGCTGGAGTGTAG |
| 10A | A00583:504:HM2HWDSXY:2:1434:13196:5901 | AGGAAACAAAAATGTACAGTTTTTGTCGTGCTACACGCTTGAGTTCCTAGGACTGCAAACTATAGCTTTAGCAGTTCTGTTTCTTGTCAAATACAAGCTATGTCCAAATCTGCTGTG |
| 207T | A00583:504:HM2HWDSXY:2:1430:32389:9612 | GGTTGGAACTTGCAGTTTGAAGCATACACCTAAATCTAAGATAGTTTTTTCTTTCTTTGTTTTTCTTTGTTTTTGTTTCATTTTTTAAAAGTCTGAGCATAAATAAGAGAGGTGGAAGTAAAATTTATGTT |
| 208T | A00583:504:HM2HWDSXY:2:1429:2121:25175 | GACAAATTCCTTTGGCTTTTACTTGTCTGGGAATGTCTTTTTCTCTCCTTTATTTCTGAAGGATGGCTTTGCTGGATACAGTATTCTTGGTCAGCAGGGTTTTTTGTTTTTATTTTTTTCTTCTTCAACCCTCTGAATATATTTTCTAAATTCCTAGAAGCCTATAAAATTTCTGCTGAGAAATCTGCTGCCA |
| 105T | A00583:504:HM2HWDSXY:2:1427:18041:17018 | TCCACGGCTTCTTCAAGACTGTACCCCCATCCAAGGGCAGCCGCTTGTCTTTTTACAGCTCACCCCAGCCCACACACACATACAAGCACACATAAACACACAGGCACACAATTCTG |
| 13H | A00583:504:HM2HWDSXY:2:1424:1967:34898 | CTCAAGGACTTCAAACTCTACTCCCACTAATAGCTTTTTGATTACTTCTAGCAAGCCTCGCTAACCTCGCCTTACCCCCCACTATTAACCTACTGGGAGAACTCTCTGTGCTAGTAACCACTTTCTCCTGATCAAATATCACTCTCCTACTTACAGGACTAAACATACTAGTCACAGCCCTATACTCCCTCTACATATTTACCACAACACAATGGGGCTCACTCACCCACCACATTAACAACATAAAACCCTCATTCACACGAGAAAACACCCTCATGTT |
| 22H | A00583:504:HM2HWDSXY:2:1421:28565:10661 | CTCTTCTTAACAACATACCCATGGCCAACCTCCTACTCCTCATTGTACCCATTCTAATCGCAATGGCATTCCTAATGCTTACCGAACGAAAAATTCTAGGCTATATACAACTACGCAAAGGCCCCAACGTTGTAGGCCCCTACGGGCTACTACAACCCTTCGCTGACGCCATAAAACTCGTCACCAAA |
| 209T | A00583:504:HM2HWDSXY:2:1414:11984:36276 | GAAGACTATGGAAAAGCATAGTATCTGGGCCAGAGTGCACCATTCCTCACAGCACAGTCCCTCAGGGCTTCTCTTGGCTAGGGGAGGGAGTTCCCCG |
| 210T | A00583:504:HM2HWDSXY:2:1411:7554:31234 | CAGCAGGAGAATCGCTGCCACAGCGCGGCGGTGGCGATATTTAAAGGGGACGCAGCCTATCTGGCAGGAGCAGCGCGCCAGTCGGCTCAGCCAATGCGCATGCGCGAGGCGCGAGCGGTTTCTCCCATCACAGTGGTTCCCACGG |
| 211T | A00583:504:HM2HWDSXY:2:1411:22083:36824 | AAACTCAGACATGGTAGTGATCCCTGCTGCACCTGGCTTGATGAACTAATGCAATGTGGGCTATAAACAGTCTTTTAGATAATGTTTAAAGCTCTTCTTCTAAGGCTTATTTTTGGGACAACATGTGATTTTTTTTTAATCCGAAGACACCATCCCGAGTGCTACACTGAAATATATGTGCTTCTGTTCTCTTTATTTTCTAAATGTGCTCATTATATTGCTAAAACAAATGCGAGGAGATA |
| 7H | A00583:504:HM2HWDSXY:2:1410:19045:31908 | GCCTCACTCATTTACACCAACCACCCAACTATCTATAAACCTAGCCATGGCCATCCCCTTATGAGCGGGCGCAGTGATTATAGGCTTTCGCTCTAAGATTAAAAATGCCCTAGCCCACTTCGTACCACAAGGCACACCTACACCCCTTATCCCCATACTAGTTATTATCGAAACTATCAGCCTACT |
| 212T | A00583:504:HM2HWDSXY:2:1410:18774:32878 | CACGCCTGTAGTCCCAGCTCCTCAGGAGGCTGAGGCAGGAGAATTGCTTGAATCTGGGAGGCAGAGGTTGCAGTAAGCCGAGATTGTGCCATTGCACTGCAGCCTGAGCAACAGAGCAAGATACTGTCTAAAAAAAAAAAAAAAAAAAAAAAAATTAATAAACTATTAGGACAGTCTTGGAGGAAAAAGTCTTGGAAGAAAAAGTCTCTTGGGTAAGGATTAATTAACTGGTACACAGGCTGGCAAGGTG |
| 18P | A00583:504:HM2HWDSXY:2:1406:28556:9361 | GCCTGTCTTTGTAAATAAAGTTTGATTGGAACAGAGTCATGCCCATGTATCTACACATTGTCTGTGGATGCCTTTCTGCTATAATGGCAGAGTTGCAGAGTTGCCAAAGAGGACATATGTCCCACGAAGCCTAAAATATT |
| 213T | A00583:504:HM2HWDSXY:2:1404:14868:25081 | AGCATCCCGCTTTCCATTGACAAGGAGGTCAGCAAAAAATCACTAGCTTTTTTTTTTTTTTTTTTTTTTTGAGATGGAGTCTTGCTCTGTCCCCCAGGCTGGAGTGCAGTGGTGTGATCTCCACTCACTGCAACCTCCGCCTCCCTGGTTCAAGGGATTCTCCTGCCTCAGCCTCCCGAGTAGCTGTG |
| 214T | A00583:504:HM2HWDSXY:2:1375:18222:13698 | AGTCTGGCCCCTCCCTTTCTCTCTCTCTTTTGCTCCCGCTCTCATCATGTGACATGCTGTCTCCCTGTCACCTTCCACCGTTATTGAAAACTTCCTGAAGCCTCACCAGAAGCCAGATGTTGGTGCCATGCTTCCTGTACAGCCTGCAGAACCCAGAGCCAATTAAACCTCTTTTCTTTGTAAGTTACCCAATTTCAGGTA |
| 216T | A00583:504:HM2HWDSXY:2:1374:29903:9878 | GAAAGAAAGAAAGAAAGAGAGAGAGAGAGCCAGGTGTGGTGGCTCACGCCTGTAATCCCAGCACTCTGGGAGGCCAAGGCGGGCGGATCACAAGG |
| 217T | A00583:504:HM2HWDSXY:2:1373:25554:21104 | TCAGGTGAGGGAGGCAGGAGGCAGGGTCTGGGGACTGGGGGCAGCGGCTCGACCTCCGCTTTGGAAGGGAGAAGGGTTAGCTGGGTCCCACCCTGCTCTCCCTTCCCTTTCCTC |
| 218T | A00583:504:HM2HWDSXY:2:1372:28510:21872 | ATATTTTTTACTATGAAAATCATTATAATGAAAAAAATCAAAATTTAGTAAGGCAAAAAAAAAAAAAAAAAAAAAAAAAAAAAAAAAAAAGAAGGTGACTTCATGGCTATGAACAGAGTTAAAAGGCCAGCAGATAAAGAGAAATGTCAATCTCT |
| 1L | A00583:504:HM2HWDSXY:2:1372:24957:21480 | GTGGTTCTCACCACCACCACCACCACACCGACCCGGCTCTGCGTGCTGCTGACGGTCCGTACCTGCAGATCCTGGAACAGCCGAAACAGCGTGGTTTCCGTTTCCGTTACGTTTGCGAAGGTCCGTCTCACGGTGGTCTGCCGGGTGCTTCTTCTGAAAAAAACAAAAAATCTTACCCGCAGGTTAAAATCTGCAACT |
| 219T | A00583:504:HM2HWDSXY:2:1368:27877:19460 | TAAGCCTGGGAGTTTGAGACCAGCCTGGGCAACATAAGGAGACCTTGTCTTTACAATAAATAAATAAATAAATAAATAAATAAATAAATAAATAAATAAAATTAGCCAGGTGTGATGGTG |
| 220T | A00583:504:HM2HWDSXY:2:1368:17372:10786 | TGACTCAGCTCCTGGGTGTGGGCCACGTATCCAGATGAGCCAGTTTATGGGTTTAGGTGGCACCCGCTGATCTGTCAGGATGCAAGTTATGAAAAATACCTCAAACACCAATCTTAG |
| 221T | A00583:504:HM2HWDSXY:2:1366:7247:24283 | AATGGTTCCTTCTGTACAAGTTTTTCTCAGCTCTCATAAATTGCACAAAACAAAGAAATGTGATTCCTTCTATTTCGGCATTTGGTAGACCAATTTATA |
| 222T | A00583:504:HM2HWDSXY:2:1366:20175:34898 | ATATGTTTTATAGTGATAAATATGTCTTGTTTTATTATTCTTTAGGCTCCATATTTCTGAGATGTGTGTTTTTCTTTTTCTTTTTTTTTTTTTAACACTTCTATTCTCACATCTTAATACCTTCTTTCATCTTGGAGGTGGAGCTATGTATGTGACCATAGGAACGTGAAGATATCTTATGATTTGGTAGGGAGGGAATGAAATTTGGATGTGAATGTGAC |
| 39H | A00583:504:HM2HWDSXY:2:1364:2736:22795 | CACCAATGGTACTGAACCTACGAGTACACCGACTAAGGCGGACTAATCTTCAACTCCTACATACTTCCCCCATTATTCCTAGAACCAGGCGAACTTCGACTCCTTGAAGTTGACAATCGAGTAGTACTCCAAATTGAAGCCCCCATTCGTAT |
| 15A | A00583:504:HM2HWDSXY:2:1362:11406:27132 | CTTATTTATAGATGCGCTAGACTTCAACATCTGTTACGAGTGCTACTAACTAGTTAGAAGTTGTCCCATGCTTTCGCATTGAACTTACCCCACCAACCTACCG |
| 273T | A00583:504:HM2HWDSXY:2:1359:11379:20823 | GATCCACGGGGACGGGACCCAGGAGGTGAGGATCCACGGGGATGGGACCCAGGAGGTGAGGATCCACGGGGACGGGACCCAGGAGGTGAGGATCCACGGGGATGGGACTCAGGAGGTGAGGATCCACGGGGATGGGACTCAGGA |
| 223T | A00583:504:HM2HWDSXY:2:1356:8883:18161 | CCAGGCTGGCCATGCCACGTGGGAGATGGAGTTATTACTCAAATCAATCTCCCCCAAGAGACGGAGTTATTACTCAAATCAATCTCCTCCAAAATTCTCAGGCTCGGGTTTTTCAAGGATA |
| 28H | A00583:504:HM2HWDSXY:2:1355:24442:1141 | GCCACATATCCCTCGTAGTAACAGCCATTCTCATCCAAACCCCCTGAAGCTTCACCGGCGCAGTCATTCTCATAATCGCCCACGGACTTACATCCTCATTACTATTCTGCCTAGCAAACTCAAACTA |
| 2N | A00583:504:HM2HWDSXY:2:1348:32488:25066 | CCTTTAGCTGGAAACTTAGCCCATGCAGGAGCTTCAGTTGATCTAACAATTTTCTCCCTACACCTTGCAGGTGTATCATCAATCCTAGGGGCTATTAATTTCATTACCACAATTATTAACATAAAACCTCCCGCAATGTCTCAATACCAAACACCCCTGTTTGTCTGATCAGTACTAATCACAGCCATACTAC |
| 224T | A00583:504:HM2HWDSXY:2:1348:20826:5368 | TCAACAATCTCCTGGACAGTTTGAAGTTATTCTGCAGGCATAATTCCCTTTTCCTTGGGTAATGTCAGTCTTTTGTTTCTCTCTCTCCTTTTTTTTTTTTTTTTAGTCAGAATCTCGCTCTGTCACCCAGGCTGGAATTCAGTGGCATGA |
| 16H | A00583:504:HM2HWDSXY:2:1346:2302:18630 | GGGCCATACGGTAGTATTTAGTTGGGGCATTTCACTGTAAAGAGGTGTTGGTTCTCTTAATCTTTAACTTAAAAGGTTAATGCTAAGTTAGCTTTACAGGGGGCTCTAGAGGGGGTAGAGGGGGTGCTATAGGGTAAATACGGGCCCTATTTCAAAGATTTTTAGGTGAATTAATTCTAGGACTATGGG |
| 18H | A00583:504:HM2HWDSXY:2:1343:6045:34475 | AGTCAATACTTGGGTGGTACCCAAATCTGCTTCCCCATGAAAGAACAGAGAATAGTTTCATGTAGAATCTTAGCTTTTGGTGCTAATGGTGGAGTTAAAGTCTTTTTCTCTGATTTGTCCTTGGAAAAAGGTTTTCATCTCCGGTTTACAAGACTGGTGTTTTAGTTTTTACTTCATGGACAGGCCCATTTGAGTATTTTGTTTTCAATTAGGGAGATAGTTGGTATTAGGATTAGGATTGTTGTGAAGT |
| 7A | A00583:504:HM2HWDSXY:2:1343:4878:18176 | GCATGGAAGAGGCACACGCAAGAGGGTGCATGGCAAATGGGGAAGGTGTAGAGCCAAAGGAAGCAAGAGGTGGGAGTTAGCAGCAAACTGAACACATAGCTAAGACTTATTATTTTGCCAGATCTGAGGCTTAATCTGCT |
| 9H | A00583:504:HM2HWDSXY:2:1342:30843:10473 | ATTTACCGAGAAAGCTCACAAGAACTGCTAACTCATGCCCCCATGTCTAACAACATGGCTTTCTCAACTTTTAAAGGATAACAGCTATCCATTGGTCTTAGGCCCCAAAAATTTTGGTGCAACTCCAAATAAAAGTAATAACCATGCACACTACTATATCCACT |
| 4C | A00583:504:HM2HWDSXY:2:1339:8621:32487 | CGGGCGTAGTACATCTCCAGTCGTTCCATCTCCCGCGCAGTAACAGTGTCCGGATTTACTCGAGACTCAAGGGGGATGGCGTATGCTGTCACCACCAGAAGAATCCAACTCACCAGGAAAAACAGGGCGGAAACAAAGCAATAATCCACC |
| 4H | A00583:504:HM2HWDSXY:2:1332:10447:13197 | TATTGACTCACCCATCAACAACCGCTATGTATTTCGTACATTACTGCCAGCCACCATGAATATTGTACGGTACCATAAATACTTGACCACCTGTAGTACATAAAAACCCAATCCACATCAAAACCCCCTCCCCATGCTTACAAGCAAGTACAGCAATCAACCTTCAACTATCACACATCAACTGCAACTCCAAAGCCACCCCTCACCCACTTGGA |
| 6A | A00583:504:HM2HWDSXY:2:1330:20112:31000 | TGTGTCTCCTGCATTGCAGGTGGATTCTTTACCACCGAGCCACCACAGATAATTTGTGTAAAAGTTCTTTTAAGTTGTGGTGTGCTCAGCAAAGATGAACTCTTACTTCTGGCGATTCA |
| 3P | A00583:504:HM2HWDSXY:2:1324:30626:15107 | AGCAGTGTAGTGACTGTGTAGTATAAATATGGAAACCAGCATTTTGAGGCCTTTGGTCAATAT |
| 225T | A00583:504:HM2HWDSXY:2:1322:13331:4570 | GGAGGTTGCAGTGAGCCGAGATCACACCACTGTAATCCAGCCTGGGCAGCATAGTGAGACTCTGTCTAAAAAAAAAAAAAAAAAAAAAAAAAAAGAGAAAGAAAGATTAAACTTTTGTCTTTACCTTTGTCTTTATTACAAATTATTTTCCTGACAATCATTTATCCTTTGACTATGTTTATGTCTTTTGCCATA |
| 226T | A00583:504:HM2HWDSXY:2:1321:12852:25285 | CCCTCCCTTTTGTGCCTGCTCCCTTCAAAACACACACACACACACACACACACACACACAGCCTTAGCTTTTTTAAAACATAGGGGCTAGGGAGAAAAAAATAGAGAGCCAT |
| 227T | A00583:504:HM2HWDSXY:2:1319:15076:23062 | AACTATTGGGAGATTACGGCTCCATATGATAAAATAGCATGCCCTAACTTTCTTGGTAGGTTAAAGAATGTCCATTACCATCAAGAATTCA |
| 152T | A00583:504:HM2HWDSXY:2:1314:4354:9345 | TACAGGCATGAGCCACGGCGTTCGGCCAGCTGTTTCTAGATTCTGTCACTGCCCTGACAGTTTCTGCACTGAATACATATACGTACATACATGCGTACATACATACATAGATTTTTTCATTTAAAATGTTTAGGCCGGTTGCGGTCCTGTAATCGCAGCACTTTGGGAGGCAGAGGCAGGAGGATCACTTGAGCTCATGAGTTCGAGACCAGCCTGGGAAACATGGCAAGACCCTGTTGCTACAAAAAAAACT |
| 66T | A00583:504:HM2HWDSXY:2:1311:24306:23140 | GAATCTTTTTCCTCAATTTTTGGAAGATTTTCCTATTCAGAGCTCTCTTGGGCTTGCAGAATTCTATTTATGCTTAACAGTATCAAAGAAATAAGATCAAGAAACACTCAATAAAACAAGGGAGTTTATCTCCACTGAACATGGG |
| 3C | A00583:504:HM2HWDSXY:2:1310:23357:5869 | GTGTAATGAATGTGGGAAGACCTTTGGTCAAAATTCAGATCTTTTAATTCATAAGTCAATTCATACTGGGGAGCAACCTTACAAATGTGATGAATGTGAAAAGGTTTTCAGTCGTAAATCAAGCCTTGAGACACATAAGATAGGTCATACTGGAGAGAAACCATACAAATGTAAGGTTTGTGACTTGGTTTTTGCGTGTAATTCCTTTCTGGCAAAACATTCTAGATTTCTTAGTGTAGAGAAACCTTACTAGTTTAATG |
| 228T | A00583:504:HM2HWDSXY:2:1310:13313:9361 | CAGATGTGCAGAGAAACTGAGGTTTCTTTACTTCCTATGTATGACCTGACCTTCATGAGCTCACCCTGCTGGGTGATGGGGCAGTTTGTGTGTGTGTTTGTGTGTGTATGTATGTATGTATGTATGTATGTATGTATGTATGTATGTATGTATGTAAGGGAGAGTGTATTAGTCTGTTCTCATGCTGCTAATAAATATATAGCTGACACTGGGTAATTTATA |
| 230T | A00583:504:HM2HWDSXY:2:1308:29116:23422 | CATGTGACAAACTATATTAGTCAGGGTTCTCTTAGAGGAACAGAACTATATATATATATATATGTACAGGAGTTTGGCCGGGCATGGTGGTCACACCTATAATCCCAGCATTTTGGGAGGCCAAG |
| 231T | A00583:504:HM2HWDSXY:2:1307:16857:15029 | CCGCGGGGTCAGGGTGGAAAAAGCCACAGCGGAGAAAAAGCCGCAGCGGCGGGGGGGCAAAAAGCCACGGCGGCGGGGGGGCAGAAAGCCACGGCGGCGGGGGCTAAAAAGCTGCTGCGGGCAAAAAGTCGCGGCGGCGGGATCGGTT |
| 232T | A00583:504:HM2HWDSXY:2:1306:18367:29575 | TGACGTGACATGATCAATAGTTATATAGTTATATATCCTTGATGTGCAGGTCAGGGTATAGACAGCACATAAGGAATAGCAAGGGCAGAATGTCTACCACATGTTTGACCACAAACCATTTTTTAATTCCTGGGCACACAAGAACGTGTACTACCTTTCCCAGTCTCTTATGACCAGTGGGAACTAGAACTGAGTTCTGGCCAATTGAATGTGGCAG |
| 233T | A00583:504:HM2HWDSXY:2:1277:28158:13338 | GGAGCTAAAAACATCACCTGTGCATTGGGGTTATATCTGTTTTACCATGTTTACGCTCATGAATGCCTTACAAATTCTGTGTTTGCAGTTTGGGTAAGAGTAATGCAGAGATGCTCAAGTTCAGATTCTTCTTGAGAACAATTACTAAAGTCAACAGAGTCTGGAGGAAAGCAATTGCTTAATT |
| 234T | A00583:504:HM2HWDSXY:2:1277:23357:1924 | GCTGGAAGCCATTATCCTCAGCAAACTAACGCAGGAACAGAAAACCAAACACCACATGATCTTATCATAAATGGGAGCTGAACAAGGAGAATGCATGGAAATGG |
| 235T | A00583:504:HM2HWDSXY:2:1276:3812:3740 | AGAGAAAAAAAAAAAAACAGCAAGATACTATCTCACACCCATTAGGATGGCTACTATATATTTAAAAAATTGCACCTGTAGTCCTAGCTACTTGGAACACAGAGGCAGGATTGCTTGAGCTCAGGAGGTCAAGGCTGCAG |
| 13A | A00583:504:HM2HWDSXY:2:1276:29487:24565 | AGGTCTTTAGTGCTTGATATGACAAAAAATTACACTGCAGTCTTCCTTCAAGGAGCTATGAGTTGCATCTTTGCTCAACAGATACCTATTTC |
| 236T | A00583:504:HM2HWDSXY:2:1273:4146:7012 | GGCACTGCACCTGGCTGAAGAATCACAGCTTCTAAAGGCAGCAAAGATCTTTGCCGAATTGTATTTATTTAACTTCATTAATTGAAAAGAGGCATTGGTGGGTCAGTTCTCTTTCTTCTTTTTTCTTTTTTCCAGTCAGTGTATTCTATA |
| 237T | A00583:504:HM2HWDSXY:2:1273:27905:33849 | TGAAGGGGCCTCAGAGCAGAGAAACAGAAAGAGCTGAGAAGCAACGTTGCCCTCTTGAACATTTCTGCTCATGGCCGAAGCTGACTGAAGCGGTTTTTCAAAGAACATCTTCCCAGAAGGCTGGTGGACAACTTGGAGGTCTTTTATGAGTCAGAGACAGCTAGGCTTTCCTAAGGAGCTGGAAGTCAGGGCATGTCCACCCAGACAGAAGAACAGCTACCCTGGG |
| 238T | A00583:504:HM2HWDSXY:2:1263:29839:26115 | TGACTTCAGGTGATCCACCCACCTTGGCCTCCCAAAGTGCTGGGGTTAAAGGCATGAGACACTGCGTTCATCCACCTCCTATTTTACTTGGGAGAAATGCACAGATTCTGGGTGCCATGTGCATTTGTTTTGGGAGTGATAATTGATCTAACTTATGGAAATAATACTAAATAGTTAGCGGATGGATTCTGCATCTGATGAGAGTTTTGGGCAAAACGAATTCCTAGTTTCTGAGTCTTATTTTTCCCCTGATTCAAGAAAACTGTGAATTA |
| 215T | A00583:504:HM2HWDSXY:2:1260:21531:31641 | TGTGAATGTACTAAAAACCATTAAATTGTACACTATAAGTGGGTAAATTATGTAGTATGTAAATATATCTCAATGAATCTGTTATTTAAAAAAGATAATAGTAGCCTAAATATGAATGATATTCAGGTGTTTAACTATTCTCAAATATCATTTTGAATCTTCTGCTTTAAAATATCTTCACAGAATTTGGAAACTCTTCTC |
| 19P | A00583:504:HM2HWDSXY:2:1254:15212:24048 | AAAGGCAGCAACAGAATTCCTGGAGAAAGTAAGACATTGGAGAACTGCACAGCACCTCTTCTTGGGGTCTGGGGTTACCCAGACTTAGGGAGAGTTCACCTTCTGGGAAAAACTAAACTTCGGTTCTTGTCCTTTTTTGTTCTTATGGCAAGACCAAAATTTCAGAAAACCAAGATGAAGCACCAAGCTAAGGTGAAGTACTCTCTCTCTCTTTTCTTTTTAAGAATATCTAATACCAAGGATGATCCAT |
| 10H | A00583:504:HM2HWDSXY:2:1253:1190:14575 | TATGTCGCAGTATCTGTCTTTGATTCCTGCCTCATCCTATTATTTATCGCACCTACGTTCAATATTACAGGCGAACATATTTACTAAAGTGTGTTAATTAATTAATGCTTGTAGGACATAATAATAACAATTGAATGTATGCACAGCCGCTTT |
| 239T | A00583:504:HM2HWDSXY:2:1250:11071:6825 | ACAGGCAGGTCGTGTAGATAGAGTGGATCTGCTGGACAGGCAAGTAGTATAGATAGGGCGGGTCTGCTGGAGAGGCAGGTAGGGGAGATTGCGTGGAACTACTGCCCAGGCAGGTAGTGTAGACAGGGTGGATCTGCTGGCCAGGCAGGTAGTGTAGACATGGTGGACCTGCTGGACAAGCAGGTAGTGAAGAC |
| 32P | A00583:504:HM2HWDSXY:2:1249:11831:15374 | ATCCTGGAAATGATTATGTGGTTTTTGACCACAGGAAAAAAGGTGGTGATTGGGGATCTAACAATAGATTTATTAAGAAAAAATTGTGTCAAATAAACACAGA |
| 240T | A00583:504:HM2HWDSXY:2:1248:31123:32002 | GAAGTAACCCTTTCTACCCCATCCCATCTCCACCCACACGTCTACAAAAAGAATTTTATGCAAGACACAGATTTTGAAGCACATAGAGGAGGAGCTAACATGAGTTGCT |
| 130T | A00583:504:HM2HWDSXY:2:1244:27868:13495 | TCTCATATCCCTAGCTCTCAGTAGAGGTGGCATTATCTTCATATAATCCCTGTCCCTAGGTCACTATTTTATATGTCTACATT |
| 27P | A00583:504:HM2HWDSXY:2:1244:15600:27978 | AGAAAGTAGTACCTACCAATGCAAAACATTTCTGTGCTTGTGCCAGAGCAGACTCAGGTCTTAACCTAATACATTCATCAAAATCTGCCACTGCTTCTTCAACTTGATCAAGGAGTATTTTCAGCTAAGAAAAAAATACGTAAATAAATGTAATTCAACTAGGTGCACCAGTT |
| 241T | A00583:504:HM2HWDSXY:2:1238:26069:32988 | TGCAAGCTGGAGATAAACTTATTCTGAGGCATAGGCATGCTATGTAAGTATTTCATATTTGTTCAGAAGGCCTATGCTTTCATACAGGTATAGATACGTACATATCTATCAGCCATCTATCTATCTTTCTCTATTCTGAGTGAGAAATAGA |
| 3L | A00583:504:HM2HWDSXY:2:1237:22064:30091 | TTATTTTCATCGTAACATGAACAGTTACTCAATTTTCCAGTTGCAATTCCCATTTGGCTTGTGTTCCATGTTATCCTTCCTCCTTTCTAGACACTCTGCTTTTTTTTTTTTTGTAAAAAGGACAGCTCATATTCAGCCTTTTCACTTTTGTTATTTGCCTTGCAAGAACAAACTCAAAATTGCTTGATCCAATATGATTAACAACGTGCATCCCAAATAAAGCAAATTCCACTCTGTGGTGTGTGACAGAATGCAAAAACAGGTTAGCATCCT |
| 34H | A00583:504:HM2HWDSXY:2:1236:31114:1830 | GGTTAGCGAGGCTTGCTAGAAGTCATCAAAAAGCTATTAGTGGGAGTAGAGTTTGAAGTCCTTGAGAGAGGATTATGATTCGA |
| 7G | A00583:504:HM2HWDSXY:2:1235:5773:17002 | TGAAGTAGTTCAACATATTGTTGTGCGAAAAGTAGATTGTGGTACTCTTTATGGTATAAATGTCAATAATTTATCAGAAAAAAAAAAAAGAATTTTCAACAAAAATTAATTGGGCGGGTAATTGCAGAAAATATATATATAGATCATAGATGTA |
| 242T | A00583:504:HM2HWDSXY:2:1232:14127:30405 | AAGGGCGGTGCAAGATGTGCTTTGCTAAACAGATGCTTGAAGGCAGCATACTCGTTAAGAGTCATCACCACTCCCTAATCTCAAGTACCCAGGGACACAAGCACTGCAGAAGGCCGCAGGGCCCTCTGCCTAGGAAAACCAGAGACGTTTATTCACATGTT |
| 8C | A00583:504:HM2HWDSXY:2:1230:3568:21543 | GGCACCAGCTGCTGCCCAGGGTCCCTCACGCCGCCCCCTCCCCACTGTGCAGAGTCCCACACACCACACACACCACCAAGTGGGTGTCCAGGCTCTGGGCCTTCACGCCCCACCCCACCCCACGAGTAAGAGGCAGATGGTGTCACTGCAGGGCCCAGGG |
| 4P | A00583:504:HM2HWDSXY:2:1230:3568:21543 | AGGCACCAGCTGCTGCCCAGGGTCCCTCACGCCGCCCCCTCCCCACTGTGCAGAGTCCCACACACCACACACACCACCAAGTGGGTGTCCAGGCTCTGGGCCTTCACGCCCCACCCCACCCCACGAGTAAGAGGCAGATGGTGTCACTGCAGGGCCCAGGG |
| 46T | A00583:504:HM2HWDSXY:2:1230:25943:33739 | GATTGCTGAGAATAATTAATTGGTTAGTACTCGGATCCTTATGCATTTCTCCTGCCTCTTCCCCCTCACTTTCCTCAAATGCTATAGAGATATCATAAAAGTAAGTTACTGAAATAAAGGCCCTATTATGTCTGGGTATTTCCTTTTCAAAGCGCTGGTTAGTGCC |
| 243T | A00583:504:HM2HWDSXY:2:1228:31412:6198 | GAAGCTGAGGCAGTAGAATTGCTTAAACCCTGGAGGCGGAGGTTGCAGTGCACCAAGATTGTGCCACTGCACTGTAGCCTGGGCAAAAAAGAGCAAAACCCCGTCTCAAAAAAAAAAAAAAAAAAAAAAAAATTGAAGTAATAATAGC |
| 185T | A00583:504:HM2HWDSXY:2:1226:19388:5415 | TGGGATGCTCCTCTAGTACCAGGTACTCAGGAGGCTGAGGTGGGAGGATTGTTTGAGACTGGGAGGCTGAGGCTGCTGTTAGCCGTGATCACACCACTGCATTCCAG |
| 244T | A00583:504:HM2HWDSXY:2:1223:3794:27070 | GAGAAAAGTTTACTTTTCCCAACAAAGTTGAATAGCTTTTTTTTTTTTTTTTTTTGGAGACGGAGTCTCACTCTGTCACCCAGGCTGGAGTGCAGTGGCACCATCTCAGCTC |
| 245T | A00583:504:HM2HWDSXY:2:1221:31159:15405 | CAGAACAGTCACCAAAAATTATAATATATGAGGCTTTGACCCAATTACACTACCAAACATCCCCCATTAAATTTTTCCTGTGGATTTAAAACTCTTTATCTGTAGGAATCCTG |
| 11A | A00583:504:HM2HWDSXY:2:1214:7392:4492 | ATCTTAAAATATAAAAATGGGGATTTTTTTTCCCCCCAGCCACTAGGCACTTCTGGATACTTTCACTTCAGAGATGCTGTTTCAGTGTCAGTTGGATGTTTTGCCGACTTCATTATTTATCTGCATTAACGCTGTGCTTATCCTAAACAATCTCACGTCACTGCTCAGAAGCACGCCAGACTGAGCTGCCCTCCCCCT |
| 153T | A00583:504:HM2HWDSXY:2:1212:10782:10895 | ACTGGCCAGGAGAATAAGTTTCTGCCAAGTGAGTGGTTAAAAAAAGGCAGACTGGAAAAATAACTGTGGGATGGTAAGTATTTCTTTATTTACAAGGCTAACATAAGTTCTCTCCATGTGTTGGGTTTGGAAGAGGGGATGGATTGGTTA |
| 246T | A00583:504:HM2HWDSXY:2:1208:31259:20682 | GTGTGTCGTTTCTAGGCTTTTTCTTTTCTTTCTTTCTTTCTTTTGACAGGTTCTCACTCTGTCATCCAGGCTGGTGTGCAGTAGCACAATCACAGCTCACCTCAGCCTCGACCACCCAGGCTCAGGTGACCCTCCCACCTCAGCC |
| 247T | A00583:504:HM2HWDSXY:2:1205:16107:32894 | CTGGTATCTGCTTCTGGTGAGGCCTCAGGAAGCTTACAATCATGGCAGAAGGCCGCAGGGAGCCAGCATTTTACATAGGAAAAGCAGGAGCAAGAGAGAGAGTGGGGAGGTGCCACACACTTTTAAACAACTAGATCTCATGAGACCTCACTCAGTATCATGAGGACGGCACCAGGTCATTCATGAGGGATTCACTCTTATGACC |
| 5P | A00583:504:HM2HWDSXY:2:1204:8540:20791 | ACCCTCCGGCCAGGACGGAGGAGATGCGGCTGGAGCCGCCCCCGATGCCGCAGGAGCCCTTCATGGAGCTGGAGGAGGTGAACTGGCGGCTGCAGGTGCTCATGGTGCCGAGGAGGGAGGTGAGTGAGCGAGCAGTTGGCTGAGTGAAGAGAAGGTGCTCAGGT |
| 248T | A00583:504:HM2HWDSXY:2:1202:28058:15295 | GGTTTCAGGGAGATGATAATCCACACTTTGTAAAGCAAATGTCATATAAAATTGACAAAATAGAAATGAAAGCAAAAACAACCCTCAAGAGCAAAGCTGTTGTGTTTTGTTTTGTTTTGTTTTGTTTTGTTTTGTTTTTAATGAACATAGTAATTCAGGTCATTTTCTGGTGCCTAAAGAAAAATTGAGAACTTCAGATTAGGCTGGAGGGGGGAAGTTCATCACC |
| 249T | A00583:504:HM2HWDSXY:2:1175:23827:11287 | GCGGTATACAATATGTAAATGAATGAGTGTGGCTGTATTCCCATAAAAGTTTATCTTTGGATGCTGAGATTGGAATTTGATAGGTTTTTCATGAAATAGTCTTCTTTTTAATTTTTCTTCAACCATAAAAATGTAAAAAACATGATTAGCTCACAGGCAGTGCAAAACCAGTTTGTTAACACTTGCCCTAGGGCTTTGGAATCCTTACTTCCAGCCAGAGGCTGGAGAAAGGATCTTGCATGGGAA |
| 250T | A00583:504:HM2HWDSXY:2:1171:7835:11365 | TGCCTCAGCCTCCCGAGAAGCTGGGACTACAGGCGCACGCCACTGCACCTGGCTAATTTTTGTATTTTTAGTAAAGACGCGGTTTTGCCATGTTGGCCAGGCTG |
| 251T | A00583:504:HM2HWDSXY:2:1169:10248:19774 | CAGAGCAAGATTCTGTCAAAAAAAAAAAAAAAAAAAAAAAAAGAAGTTATAGCACCTCTGTATTTCTCAAAATTATTTTTGCCTTCTCAATAGCACTTGTGCCAACTTAGCTCTGATTTGGCCTATGAAACTCTCTTTTAATATAAATCTAAAAATATAATTTAGCCAGTACC |
| 65T | A00583:504:HM2HWDSXY:2:1168:28628:25895 | TAGGCACCTTCCCCCTACTTTCCTTTCCTAAGGATCCACTTCTGAAAAGCCTGGGATTTTGCCTCCCTACCGTAATCCTATGGGTTGGCCTTGATAAGAAGCATGATACCAACTTCCCTGAGATCTGATAAAACCCCATTTGGGGAACTCTGATTTCCTCCTTCCCCTTTTCTGCATCGGAC |
| 1N | A00583:504:HM2HWDSXY:2:1166:28022:12759 | ACCTAGAGGAGCCTGTTCTATAATCGATAATCCACGATTCACCCAACCACCCCTTGCCAGCATAGCCTACATACCGCCGTCGCCAGCCCACCTCTAATGAAAGAACAACAGTGAGCTCAATAGCCCCTCGCTAATAAGACAGGTCA |
| 252T | A00583:504:HM2HWDSXY:2:1165:33003:5134 | GTTGATATATATATATATATATATATATATATATATATATATATATCCCCTAGATAGTGTGTTATTGATTTTTAACATATTCTGTTCATGTTTAGGTATGGAGGAAAGTTTTCAGTCCACAGCAGCTTTGCATTTGACAGATTTTCAAGTAACCATTATCTGAGCGAA |
| 253T | A00583:504:HM2HWDSXY:2:1164:30798:29935 | GAATAATATTCCTGATTTTTCTTTTCTATTAACTGTGGATTATCCACATTTTAGGAAAATCAAAGATTGAGTTTAAAGATTAAAAAAAGTTATTAGCAGAAACTTTGAATTTTTAAATTCATCAAATTCATAGATACACAAAACTGGTAATAATGCAGTAAAAACAAATTTTATTCAATAACTAATTTAGTATGGAAAGGAGCTGGGCTCAATTCCAATATGTTCAGAGGT |
| 254T | A00583:504:HM2HWDSXY:2:1162:30861:1548 | CAGCCTTTTCTTTTCTTTTCTTTTTTTTTTTTTTTTTTTTTGAGATGGAGTTTCACTCTTGTTGCCCAGGCTGTAGTGCAATGGCAGGATCTCAGCTCACCGCAAACACTGCCTCCTGGGTTAAAGCGATTCTCCTGCCTCAGCCACCTGAGTAGCTGGGATTAGAGGCATGCACCAC |
| 255T | A00583:504:HM2HWDSXY:2:1161:2148:6308 | CGGGAGGGTGAGGCAGGAGAATCACTTGAGCCCATGAGTTGGAGGCTGAAGTGGGCTATTATGGAGCTACTGTATTTCATCCTTATTAACAGAGTGGTACCTTGTCTCCAAAAATTAAAAATGAAAAAAAAAAAAGAAATTACAAAATTTTTTAGTTATCAATATATATTTGGCAATAACAAATTTAAAAGCATTCG |
| 256T | A00583:504:HM2HWDSXY:2:1158:8666:4570 | AAATCACACACTCCAGGCTGGGGGTGAATGCACAGAGATCTGTCCACAATGAAATATTAAATAATTACTGTTAAATAGCTCATCAAAGCTTTGGGGGTGGGGGTAAAAAGTTAAGCTTCTTTAAGGAGAGTCCTGCTGCAAACAAGAAAT |
| 257T | A00583:504:HM2HWDSXY:2:1154:7798:33379 | GTGGCGGGCGCCTGTAATCCCAGCTACTGGGGAGGCTGAGGTGGGAGAATCGCTTGAACCCAGGAAGGCGGAGGTTGCAGTGAGCCGAAATTGCACCACTGCACTCCAGCCTGGGCAACAAGAGTGAAACTTGGTCTCCAGAAAAAAAAAAAAAAAAAAAAAAGAAGGCCAGGACTTGATCACACCATTCTTTGACTACTGTGTGGGTTCTAGGAGGGAAAGGAGA |
| 258T | A00583:504:HM2HWDSXY:2:1154:16179:8688 | GAGAAAGGCATGAACACAGGAGGCGGAGCTTGCAGTGAGCCGAGATGGCGTCACTGCACTCCAGCCTGGGGGACAGAGCGAGACTCCGTCTCAAAAAAAAAAAAAAAAAAAAAAGCAGAAGGCAAGGCCCTATCCTGGTAGCTGGGGAGAGGAACCTGA |
| 259T | A00583:504:HM2HWDSXY:2:1147:28420:27445 | GTGCACGTGTAGTCCCAGCTACTCGGGAGGCTGAGGTGGGAGGATCCCTTGAGCCCTTGAGCCCAGGAGGTTGAGGCTGCAGTGTGCTATGATGGCAC |
| 260T | A00583:504:HM2HWDSXY:2:1147:1108:34851 | AATACAATCGAGGAGCAGAAAAATGCAATAAAGGTCATCAACAGTAGACCTGATAAGGCAGAAGAAAGAATCTGTGAACTTAAAGAAGAAATTATATAATTAGAACAGAAAAAAAGAAAGAAAGAATGAGAAAAT |
| 261T | A00583:504:HM2HWDSXY:2:1140:9842:4977 | TAGACCGAGACTCCGTCTAAGAAAGAAAGAGAGAAAGACAGCAAAGAAAGAAAGAGAAAGAGAGAAAGAGAGAGAGAAAGGGAGAAAGGGAAGGGAAAGGAAGGGAAGGGAAGGGAAGGGAAGGGAAGGGAAGGGAAGGGAACGGAGAAAGAAGGTGTTCTGCGAGAGGGAATAATAGCATAGTAAAATTTCAAGGAGTTTTGGG |
| 154T | A00583:504:HM2HWDSXY:2:1139:21504:28651 | GGCCACCATTTTATTTTCAACTCCCTTTCATGGAAGAACGTTTAGCCTTTGGTCTCTTTCTTGATTTATGACAGCTCGCGGCTTTCAAGAAAACTACCT |
| 6G | A00583:504:HM2HWDSXY:2:1138:9932:3944 | CCAAAAACCGATCAGGAGATAAGAGCACATGCCCACAAGTTCCCAGAACACGTATACTTGGACCAAATTAGGGCTAAGAACCAGACCTAGCATCGAGGCGGTAAACAAGCTAAGATAAACAAAAAAGCGGACGTATCCCTG |
| 4L | A00583:504:HM2HWDSXY:2:1136:27082:4366 | GCTCAAGTCATACTGGTCAACCCTGGACTACAAGATTGCCAGTCAGCAGAAAGAGGAAGACTCCACCAGATGAGTTGCAGGCCCCTGGTAAGCCCAATCTCTAACTCATATTCATAGATGGTCAAGTGACTGTCCATGTGTTTGTGGTCATCTGTACATTGTCCATACTTGCTATATTGTTCTTCATGAAGGATTGCCCATGAAAGAGGTTGC |
| 20P | A00583:504:HM2HWDSXY:2:1133:3784:7138 | TGTGCGTCCATTACTGCTACACTGACACCGGGTGAGTGCTCCTGGGCTGATGCTCCGGGTCCAGTGCTCCCAGGTGGTCATAGTACACGCCCCTGCCTCAGGCAGCAGGGCTGCAGGTGGC |
| 262T | A00583:504:HM2HWDSXY:2:1129:2953:36198 | GGTGAAGCAAACTTTGATTGTTAAATTATTATTGTTGTTGTTGCAGAGGTTCTTTTTAAAAACTTTGTTTGGTTTGGTTTATACACAGAAATATCTAGAAATGTTCTGGGACTAGTTGAGTTGTATCTTTAGTATTCAGGTTGTGAAAAATTAAGATGTTTGGCTATGCACAAAATAGTTTGTATGCTTTGAACTTTAGTTTACTTGGAGTTTGATATGGTTT |
| 108T | A00583:504:HM2HWDSXY:2:1128:16052:24596 | TAAAAAACTAAAAAAAAAAAAAAAAAAATTCCAAACTCTCCACAATCTAATTGTAAATTGAAGAGTCTAAATTAAAATGCCGTTTATGAAGCTGGGTGAGATGGCACACACCTGTAATCCCAGCTACTTGGAAGGCTGAGGCAGGAGAATTGCTTGAGCCCAGGAATTTGAGGCTGCAATGAGCTATGATCATGCCACCACACTCCAGCCTGGGCAACATAGTGAG |
| 263T | A00583:504:HM2HWDSXY:2:1125:25183:10723 | AACCTTGGCAGAAGCCAAAACGGTCAGACAGTGTTGTGTAAAAATGATCACCCA |
| 264T | A00583:504:HM2HWDSXY:2:1124:28339:10927 | GGGAGGGGAGTGGGGGTTACCTGCTGGAGACTGTGTGCTATTTCACTAAAGGTGGTGTTGGCTTGGGGCAGGATACTGGCCAGTAAAGGTTTTGATGCCTTCTCTGTGCCCCCCAAGAAGGAATGATTGTTCAGAGTGTGGGAGGATACCCTGTTCTCCGCACAGTTTTACCACAAAGGCCAGGGTGGGGCTTTCTGGCTCTCTACCCACCA |
| 2G | A00583:504:HM2HWDSXY:2:1120:8386:24596 | AAGCTTCATCTACAATGGCAATTGCTGGGAGTGGCAGGGGCATACTACATTTCCATTTGCTGGTGGGGCAGAGGGTCAAGATTCCCTCACCCGGACAAGGCGTCAGACGGTGACGTGAGCGAGCCCGGAGACGCCGGCGTCGGCCCTGGGAAGAGTTTTCTTTTCTTCTTGAC |
| 265T | A00583:504:HM2HWDSXY:2:1120:28682:27305 | GCTCTTATTTTTGCAGCCTTTGTGGAAAAATCAAAACTCATTCATTTGCAAAGTTATTTCCATTGACTCCTTGCAATGTGCCTGGTACATTCTGCTCAATAAATGATGAGAGAATGAGTGAGTGAGTGCATGAATGACTGAATTAGTGAGTGAAGGAATGCGTGACCAAGTGAATGAATGAGTGAGTGAACAAGTGAATGGGTCAGTGAGTGAATGAGTGATTGAATGAGTGAGCAAGTAAGGGAATGAGTGAGT |
| 274T | A00583:504:HM2HWDSXY:2:1120:19090:27665 | CTTTTAGAGTGTCATGAATTAAACTAAAAATCTCAGGAACTATCTCTCTCTTTCTCTCTTTCTCTCTCTCTCTCTCTCTCTCTCTCTCTCTCTCACACACACACACACACACACACACACACACACACACACAACAAACAAATAAAAAACACACATTTTGACTTGAAGTTTTACTTTA |
| 266T | A00583:504:HM2HWDSXY:2:1120:16721:21183 | AGGTGGGTTTTACTCAGACTGCATACTGTGAAAAGGATAAATTAGTGTGGTTTGTGAACTAGATGTGGAAATTGTGTGTGTGTGTGTGTGTGTGTGTGTGTGTGAGAGAGAGAGAGAGACCAATCCCACCATGAGGACCCGGAAATGGAGTTTGATTTGGG |
| 8H | A00583:504:HM2HWDSXY:2:1118:21377:12618 | GTAAGGTCAGCTAAATAAGCTATCGGGCCCATACCCCGAAAATGTTGGTTATACCCTTCCCGTACTAATTAATCCCCTGGCCCAACCCGTCATCTACTCTACCATCTTTGCAGGCACACTCATCACAGCGCTAAGCTCGCACTGATTTTTTACCTGAGTAGGCCTAGTAATAAACT |
| 267T | A00583:504:HM2HWDSXY:2:1115:1452:29997 | CCAACTGAAGCTGACCAGGCTGCCTCCCCTCCTCCACAGCGGATGAGACCACATGTATCCAGTGTCTCTCACATTCAGTATACACACACACACACACACACACACACACAGACACACACACCCCTACCTTAGAGATTCGGCATTTATCATTGTGAAGAACTGTCCTACAAAATACAGAGGCTTTGAATACCTGTGGTCCACACTC |
| 268T | A00583:504:HM2HWDSXY:2:1111:2682:12931 | TGTAACTAACCTTCACAATGTGCACATGTACCCTAAAACTTAAAGTATAATAAAAAAAAAAAAAAAAAAGAAAAGAGAAGAAACTCATCAGTGGGCTGCT |
| 269T | A00583:504:HM2HWDSXY:2:1110:9715:22482 | CTGATTTTGTTCTTTGGCCTTGAACTTGTTCAAGTTGAATTGTGTGTCCTGAACACTGATCTAGTTGTTCACAAACTAGGGTCAGCCAACTGTAGTGGTATGTTGATGTATTCTTGGGAATCGAGAGTTCTTTATTATAATCTTTTAATATTATGGCTAAAAAACCCAAAATATCTATGTGTAAATATATAATGTAAGCATGGATTCGTTCAT |
| 270T | A00583:504:HM2HWDSXY:2:1110:32696:10269 | AGTAACAAATATCATGGTAAAATGAAAGACCAGTAATCTGAAATACTTATATTCTAAGTGGTCATTTAATTGTCCCCCCAATTTTATTATAAAATTAAGAATTGACAGATGATAGATATTTTTTC |
| 34T | A00583:504:HM2HWDSXY:2:1110:14335:5431 | TCAGATCCTCTGGCGCCCGGGCGCCCCCTCCCGGCGACACCTCCCGGCTGGGAGCGCGCGGGAAGAGGCCAGAGGCAGGGCCTCTCACCTGCTCGGACTGCAAACCACAAAGCGTCTCTTTTTTTTTTTTTTTGAGGGGCGGAGGTGGTCATTTTTATGTTCCCTGGGCAAAGCGAGCAGCCGTTTGGCTTATAAAATA |
| 36P | A00583:504:HM2HWDSXY:2:1110:12807:9330 | TTTTTCTCCACAATCTTGCCAGCATCTATTATTGTTTTTTTTTTTTTTTTTACTTTTAATAAATAGCCATTTGGACTGGTGTGAGATGGTATCTCATTGTGGTTTTGATTTGC |
| 35T | A00583:504:HM2HWDSXY:2:1109:21884:24956 | CAGAAGGAGTAGCCTTTTGATCTAGTGCACAGGTGTCCAGTCTTTTGGCTTCTCAGGGCCACATTGGAAGAAGAATGCTCCTGGGCCGCACATAAAATACACTAATGCTAACAACAGCTGATGAGGTTAAAAAAAAAAAAAAAACAAGGTTTGTGCATAATTTTCATGCTACCCACCACCACAGATAGGTGGAAAAGTCCTTGTAGTC |
| 12H | A00583:504:HM2HWDSXY:2:1107:9480:32534 | ACTCACGGGAGCTCTCCATGCATTTGGTATTTTCGTCTGGGGGGTATGCACGCGATAGCATTGCGAGACGCTGGAGCCGGAGCACCCTATGTCGCAGTATCTGTCTTTGATTCCTGCCTCATCATATTATTTATCGCACCTACGTTCAATATTATAGGCGAACATATTTACTAAAGTGTGTTAATTAATTAATGCTTGTA |
| 5C | A00583:504:HM2HWDSXY:2:1106:5611:13839 | CGCAAGGGCAACTAATCGGAGCGCGTGGGCGCCGGCGCCCCGGTCTATCTTGCCGCGGTGCTCGAGAACTTGACTTCCTAGATCCTGGAGAATGACGGCAACGCGGCGCGCGACAACAACAAGACGCGCATCATCACGCGCCACCTGCAG |
| 229T | A00583:504:HM2HWDSXY:2:1104:23077:27211 | ACCTCTTTTTCTTTATAAATTACCCAGTGTCAGCTATATATTTATTAGCAGCATGAGAACAGACTAATACACTCTCCCTTACATACATACATACATACATACATACATACATACATACATACACACACAAACACACACACAAACTGCCCC |
| 271T | A00583:504:HM2HWDSXY:2:1103:31430:34882 | GCGGCGGCAACGCCACCCTGCAGGTGGACAGCTGGCCGGTCCACGAGCGGTACCCGGCAGGTCGGCGGCGCGGGCCCGCGGAGGAGGCGCGGTGGCGGGGCGGGTGGCGGCCTCTGATAGGAGTGTGGCCGCACCGGGGCGGGGCTAAGCGAGAGACGAAGCTTGCGAGGCGGGACTAGGGCAGGGGCGTGGCTAGGAAAAGTCGGGATTTGTGGCGGCGGGGCGAGACCGTGGTGGTGCGAATGCGCCCCCTATGGGGGAGGGTGCAGGGCTGAGCTGGGAAGG |
| 272T | A00583:504:HM2HWDSXY:2:1103:25229:7138 | GTCCTTTCCTCTGTCGAACTCAATCCACCTGACCTTTCATCATCTCACCACACCCTTCATTGACACGTGGTCCCCTGTGTTTCCAAGGTAGCAGCACTTCTATGCCCTGCTTCTTCCTTGGCCTCCTCCCACCTCTCTTTCATTCCTTGCTTATCTTACCCACACTATGCTTTGTCATCCATGAACTGCGTCACACA |
| 275T | A00583:504:HM2HWDSXY:2:1103:21224:1799 | ACCACCGCTCTTCTGCCAGACAGCGGGCAAAGTGCTGGCCCATGTCGATGGGGTGAAGCTGGAGGGCAAGGAGCCCATGCACAAGCTGTTCTTGGAGATGCTCGAGGTCATGATGGACTGTGGCGAGGGGTGGGACTGGTGGGGGTTCTGGCAGGACCTGCCTAGCATGGGGTCAGCCCCAAGGGCTGGGGCGGAGCTGGGGTCTGGGCAGTGCCACAGCCTGCTGGCAGGGCCAGGGC |